\documentclass[11pt]{iopart}
\pdfoutput=1
\usepackage{graphics}
\usepackage{longtable}
\usepackage{graphicx}
\graphicspath{{figures/}}
\usepackage{amsfonts}
\usepackage{epsfig}
\usepackage{color}
\usepackage{amssymb}
\usepackage{subfigure}
\usepackage{caption}
\usepackage{geometry}
\usepackage{layout}
\usepackage{multicol}
\usepackage{arydshln}
\usepackage{float}
\usepackage[normalem]{ulem}
\expandafter\let\csname equation*\endcsname\relax
\expandafter\let\csname endequation*\endcsname\relax
\usepackage{amsmath}
\usepackage[dvipsnames]{xcolor}

\usepackage{tikz}
\DeclareRobustCommand\full  {\tikz[baseline=-0.6ex]\draw[thick] (0,0)--(0.5,0);}
\DeclareRobustCommand\dotted{\tikz[baseline=-0.6ex]\draw[thick,dotted] (0,0)--(0.54,0);}
\DeclareRobustCommand\dashed{\tikz[baseline=-0.6ex]\draw[thick,dashed] (0,0)--(0.54,0);}
\DeclareRobustCommand\chain {\tikz[baseline=-0.6ex]\draw[thick,dash dot] (0,0)--(0.5,0);}
\newcommand{\tikzcircle}[2][red,fill=red]{\tikz[baseline=-0.5ex]\draw[#1,radius=#2] (0,0) circle ;}

\usepackage{multirow}
\usepackage{parskip}      % space line between paragraphs
\usepackage{indentfirst}  % indentation at first paragraph

\usepackage{afterpage}    % blank page
\usepackage[font=scriptsize]{caption} % figure script small

\usepackage{pdfpages}   % insert a pdf page

\usepackage{graphicx,subfigure}
\usepackage{longtable}    % table divided into two pages

\usepackage{threeparttable}

\usepackage{threeparttablex} % for "ThreePartTable" environment
\usepackage{booktabs}        % for well-spaced horizontal rules

\usepackage{hyperref}
\hypersetup{
    colorlinks=true,
    linkcolor=blue,
    citecolor=blue,    
    urlcolor=black,
}

\usepackage{cleveref}

\usepackage{cite}

\begin{document}

\title[]{Zero-dimensional and pseudo-one-dimensional models of atmospheric-pressure plasma jet in binary and ternary mixtures of oxygen and nitrogen with helium background}
\author{Youfan He$^1$, Patrick Preissing$^2$, David Steuer$^2$, Maximilian Klich$^1$, Volker Schulz-von der Gathen$^2$, Marc B{\"o}ke$^2$, Ihor Korolov$^3$, Julian Schulze$^3$, Vasco Guerra$^4$, Ralf Peter Brinkmann$^1$, Efe Kemaneci$^1$}

\address{$^1$ Institute of Theoretical Electrical Engineering, Faculty of Electrical Engineering and Information Technology, Ruhr-University Bochum, Germany}
\address{$^2$ Experimental Physics II, Faculty of Physics and Astronomy, Ruhr-University Bochum, Germany}
\address{$^3$ Institute of Electrical Engineering and Plasma Technology, Faculty of Electrical Engineering and Information Technology, Ruhr-University Bochum, Germany}
\address{$^4$ Instituto de Plasmas e Fusão Nuclear, Instituto Superior Técnico, Universidade de Lisboa, Portugal}

\ead{efekemaneci@gmail.com}

\begin{abstract}
\noindent A zero-dimensional (volume-averaged) and a pseudo-one-dimensional (plug-flow) model are developed to investigate atmospheric-pressure plasma jet devices operated with He, He/$\mathrm{O_2}$, He/$\mathrm{N_2}$ and He/$\mathrm{N_2}$/$\mathrm{O_2}$ mixtures. The models are coupled with the Boltzmann equation under the two-term approximation to self-consistently calculate the electron energy distribution function (EEDF). The simulation results are verified against spatially resolved model calculations and validated against a wide variety of measurement data. The nitric oxide ($\mathrm{NO}$) concentration is thoroughly characterized for a variation of the gas mixture ratio, helium flow rate and absorbed power. The concentration measurements at low power are better captured by the simulation with a larger hypothetical $``$effective$"$ rate coefficient value for the reactive quenching $ \mathrm{N_2(A^3 \Sigma,B ^3\Pi)} + \mathrm{O(^3 P)} \rightarrow \mathrm{NO} + \mathrm{N(^2D)} $. This suggests that the $\mathrm{NO}$ production at low power is also covered by the species $ \mathrm{N_2(A^3 \Sigma,B ^3\Pi;}\:v>0) $ and multiple higher $ \mathrm{N_2} $ electronically excited states instead of only $ \mathrm{N_2(A^3 \Sigma,B ^3\Pi;}\:v=0) $ in this quenching. Furthermore, the $\mathrm{O(^3 P)}$ density measurements under the same operation conditions are also better predicted by the simulations with a consideration of the aforementioned hypothetical rate coefficient value. It is found that the contribution of the vibrationally excited nitrogen molecules $\mathrm{N_2}(v \geqslant 13)$ to the net $\mathrm{NO}$ formation rate gains more significance at higher power. The vibrational distribution functions (VDFs) of molecular oxygen $\mathrm{O_2}(v<41)$ and nitrogen $\mathrm{N_2}(v<58)$ are investigated regarding their formation mechanisms and their responses to the variation of operation parameters. It is observed that the $\mathrm{N_2}$ VDF shows a stronger response than the $\mathrm{O_2}$ VDF. The sensitivity of the zero-dimensional model with respect to a variation of the VDF resolutions, wall reaction probabilities and synthetic air impurity levels is presented. The simulated plasma properties are sensitive to the variation especially for a feeding gas mixture containing nitrogen.

\end{abstract}

\maketitle
%\ioptwocol
% ftp://ftp.iop.org/pub/journals/IOPLaTeXGuidelines.pdf

\section{Introduction}
\label{sec:int}

Atmospheric-pressure plasma jet devices attract growing interest over the last decades
owing to their efficiency in converting ordinary gas contents into diverse reactive species for the purposes of biomedical applications \cite{Lar18}, catalytic greenhouse gas conversion \cite{Urb18,Ste20}, vacuum-free etching \cite{Jeo98} and deposition \cite{Bab98,Reu11_apl,Sig16}. The small electrode gap of micro-scaled atmospheric-pressure plasma jet ($\rm \mu$APPJ) ensures the uniformity at low voltage and sustains the characteristics of a glow discharge. The non-equilibrium character permits an efficient electron heating while keeping the neutral species near ambient temperature suitable for the treatment of heat-sensitive materials. An accurate control of the reactive species concentrations is of fundamental importance to achieve an optimized treatment performance.

The atomic structure of the noble gas conduces to ignite and sustain an atmospheric-pressure discharge at a relatively low absorbed power, and helium is being of the preference over argon due to the lower breakdown voltage \cite{Mor04} and higher discharge stability \cite{Gol18_thesis}. A wide variety of reactive species are usually generated by an addition of reactive admixtures (e.g., O$_2$, N$_2$ typically in the order of 0-2\:\%) to the feeding noble gas. A challenge still to be addressed is a full picture of the convoluted plasma chemistry defining the reactive species concentrations. The chemical complexity is attributed to the multitude of interactions between neutral and ion, as well as atomic and molecular species. It is further increased through the introduction of several gases. Therefore, a deep understanding of the rich chemical kinetics is of the essence for a characterization and optimization of the device. In view of this purpose, zero-dimensional modelling approaches \cite{Laz12,Liu10,Gae13,Sun19} are competent to report a detailed chemical description, avoid the resulting drastic computational load and probe the underlying reaction mechanisms efficiently.

Zero-dimensional modelling approaches are implemented either in volume-averaged formalism \cite{Laz12,Liu10,Sun19} or in plug-flow formalism \cite{Gae13}. The former gives the volume-averaged plasma properties of the whole chamber, whereas the latter provides those of a small plug-volume by mapping its time-evolution to the spatial-resolution in gas flow direction \cite{Sta04}. The models are used in combination with experiments to investigate the concentrations and chemical kinetics of ozone \cite{Gae13}, atomic oxygen and nitric oxide \cite{Gae14_2}, atomic oxygen and hydroxyl radical \cite{Sch18_PCCP}, helium metastable and reactive oxygen species \cite{Mur12}, as well as argon metastable, ozone and nitrogen oxide species \cite{Sch15}. However, an effort on the benchmark against a broad range of experimental data (in view of diverse species, operation parameters, measurement techniques) is still missing for an atmospheric-pressure discharge. Such a validation is of crucial significance to further ensure predictive capability of the model calculations.

Nitric oxide (NO) is one of the key species for biomedical applications \cite{Gra12} (e.g., it plays an essential role in wound healing). It has been intensively studied, both experimentally \cite{Pip08,Pip12,Dou16,Ges13,Ges13_2,Pre20} and computationally \cite{Sun19,Gae13,Sch15,Sch14_Schmidt,Gae14}. The measurements of production rate \cite{Pip08,Pip12} and absolute density \cite{Dou16} are carried out in a Herriott cell. Spatially resolved densities in the effluent are reported \cite{Ges13,Ges13_2}. Similar experiments for a variation of the gas mixture ratio, helium flow rate and absorbed power are recently conducted on the COST Reference Microplasma Jet (COST-Jet) \cite{Pre20}. In this device, the non-reproducibility due to gas contamination and power measurement is minimized by the refinement of the mechanical and electrical design \cite{Gol16}. This is extremely suitable for the benchmark of model calculations. The dominant reaction mechanisms are discussed in simulation studies \cite{Sun19,Sch15,Sch14_Schmidt,Gae14,Gae14_2}, and the influence of operation parameters on the species concentrations is presented \cite{Gae13}. The NO concentration is characterized by the simulation results and measurement data of a high-pressure discharge \cite{Shk14} and an atmospheric-pressure plasma \cite{Gae14_2}. Nevertheless, there is still a lack on the %sufficient
validation for a variation of operation parameters.

A detailed resolution of the vibrational distribution function (VDF) has gained much attention in low-pressure plasma \cite{Gue19_rev,Ann18,Cap13_2,Kem16,Gue97,Gue95,Lou86,Cap17a} (e.g., $\mathrm{N_2}(v<46)$ \cite{Gue19_rev}, $\mathrm{O_2}(v<42)$ \cite{Ann18,Kem16}). Its influence on the electron energy distribution function (EEDF) \cite{Gue19_rev,Cap13_2,Cap17a,Lou86,Col15} and the plasma properties \cite{Ann18,Kem16} is presented. An important role of the nitrogen VDF in the production of nitric oxide and nitrogen atom is reported \cite{Gue95,Kut16}, and the VDF is intensively investigated \cite{Cap17a}. However, the influence of the detailed VDF on an atmospheric-pressure discharge is not yet fully understood. The vibrationally excited states in the zero-dimensional model are neglected \cite{Sun19,Sch15,Sch14_Schmidt} or included with a few vibrational levels \cite{Gae13,Mur12,Mur13,Mur14,Tur15,Gae14}. On the other hand, their high excited states are suggested to play a role in the ozone formation \cite{Ell12}, and this in turn affects the chemical kinetics of other species. A further study about the detailed VDF and the corresponding vibrational kinetics contributes to a better accuracy of the simulation results.

Wall reactions \cite{Liu10,Sun19,Mur12,Mur13,Mur14,Sch18_PCCP,Liu17,Sch18} and gas impurities \cite{Liu10,Sun19,Gae13,Mur12,Mur13,Mur14,Sch18_PCCP,Sch14_Schmidt,Gae14,Gae14_2} are usually taken into account in zero-dimensional models due to their underlying effects on the plasma properties. The role of wall reactions is presented such as for a feeding gas mixture of Ar/$\mathrm{H_2O}$ \cite{Liu17} or He/$\mathrm{H_2O}$ \cite{Sch18}. The influence of gas impurities is shown, e.g., on a He \cite{Liu10,Sun19,Sch18_PCCP} or He/$\mathrm{O_2}$ \cite{Mur12,Mur13,Mur14} discharge. In parallel with these studies, the highlight of this work is to make a systematic analysis gradually for several gas mixtures: He, He/$\mathrm{O_2}$, He/$\mathrm{N_2}$ and He/$\mathrm{N_2}$/$\mathrm{O_2}$.

The main goals of this paper are as follows:
\begin{itemize}
  \item For a confirmed prediction capability, the developed zero-dimensional and pseudo-one-dimensional models are well benchmarked against spatially resolved simulation results and numerous measurements data for distinct species densities under different operation conditions from various literature, as shown in section \ref{sec:benchmarking}.
  \item The NO concentration in the COST-Jet is intensively characterized by the simulations and experiments for a variation of the gas mixture ratio, helium flow rate and absorbed power (see section \ref{sec:NOproduction}).
  \item The role of the detailed resolution of $\mathrm{O_2}(v<41)$ and $\mathrm{N_2}(v<58)$ VDFs in an atmospheric-pressure discharge is presented in section \ref{sec:Sensitivity}. 
\end{itemize}
The description of the atmospheric-pressure plasma jets simulated in this work is provided in section \ref{sec:set}. The developed models are described in section \ref{sec:mod} and the considered chemical kinetics are discussed in section \ref{sec:Chem_kin}. The spatial evolution of NO concentration in gas flow direction, and the main gain/loss channels are exhibited in section \ref{sec:NOproduction}. The variation of the VDFs and their spatial evolution to the operation parameters is demonstrated in section \ref{sec:VDF}, The sensitivity of the plasma properties to the wall reaction probabilities and the synthetic air ($\mathrm{N_2}$:$\mathrm{O_2}$=8:2) impurity levels is addressed in section \ref{sec:Sensitivity}. The main results are summarized in the conclusion given in section \ref{sec:con}.

\section{Setup}
\label{sec:set}

Atmospheric-pressure plasma jets of planar electrode configuration in five different sizes \cite{Gol20,Nie10,Nay19,Was10,Bib11,Kor20,Sch14} are simulated for the purpose of benchmark. The specific operation conditions are presented in section \ref{sec:benchmarking}. A plasma jet \cite{Nay19} is investigated as a preliminary test of the model predictive capability to a pulsed-modulated discharge.
The old versions of the radio-frequency driven COST-Jet (i.e., a $\rm \mu$APPJ) \cite{Nie10,Was10,Bib11,Sch14} might be insufficiently sealed. The resulting unknown high impurity levels in the experiments \cite{Nie10} may impact the results and need to be considered in the simulations.
Furthermore, a power transfer efficiency around 5\:\% is usually assumed in the modelling of these old versions for converting the provided generator input power to the absorbed power in the plasma. The irreproducible experimental results of a $\rm \mu$APPJ are mainly ascribed to the gas impurity and the power uncertainty. Hence, they are minimized by the COST-Jet \cite{Gol20,Kor20} with a large amount of effort \cite{Gol16} (e.g., the sealing improvement and the absorbed power measurements with integrated probes).

The COST-Jet is investigated with a focus in this paper owing to the above-mentioned advantages. In section \ref{sec:NOproduction}, the operation conditions are rightfully addressed based on the measurements of Preissing $et$ $al.$ \cite{Pre20}. In sections \ref{sec:VDF} and \ref{sec:Sensitivity}, the jet is simulated with a typical operation condition provided in \cite{Gol16}: the plasma volume of $1\times 1\times 30$ mm$^3$ sustained by an absorbed power of 0.6 W at a pressure of 101325 Pa and a gas temperature of 345 K, is fed with 1400 sccm He, 1400 sccm He + 0.5\% $\mathrm{O_2}$, 1400 sccm He + 0.5\% $\mathrm{N_2}$ or 1400 sccm He + 0.5\% $\mathrm{N_2}$ + 0.5\% $\mathrm{O_2}$, unless stated otherwise. The computational method for a prediction of the plasma properties in the aforementioned setups is explained by the physical and chemical model in section \ref{sec:mod} and \ref{sec:Chem_kin}, respectively.

\section{Model}
\label{sec:mod}

The species particle balance equations and an electron energy balance equation are incorporated in the model to calculate the plasma properties: the species concentrations and an effective electron temperature ${\rm T}_{\rm e}$ (eV). The electron density is calculated by imposing a constraint on its wall flux based on the quasi-neutrality implementation \cite{Kem14_2}, and the gas temperature value ${\rm T}_{\rm g}$ (K) is fixed by the measurements. Both the zero-dimensional volume-averaged formalism \cite{Laz12,Liu10} and the pseudo-one-dimensional plug-flow formalism \cite{Sta04,Gae13,Sch18_PCCP} are implemented. The balance equations are averaged over the whole plasma domain in the zero-dimensional formalism, and steady or transient volume-averaged plasma quantities are calculated \cite{Kem14,Kem17}. In the pseudo-one-dimensional formalism, the equations are averaged over a much smaller plug-volume, that is infinitesimally small along the unidirectional laminar net mass flow field. The time-evolution in this volume is mapped to the spatial position in the discharge channel since the plug co-moves with the net mass flow.

The EEDF is calculated based on the steady-state solution of the Boltzmann equation under the two-term approximation. The open-source simulation tool {\em LisbOn KInetics Boltzmann} (LoKI-B) published by Tejero-del-Caz $et$ $al.$ \cite{Tej19} is adapted for a self-consistent calculation. Super-elastic collisions are incorporated in the calculation for a better resolution of the electron kinetics \cite{Cap00,Cap16,Gro16}. The time-variation of the plasma composition in the transient simulations (e.g., pulse-modulated zero-dimensional and pseudo-one-dimensional models) necessitates multiple subsequent LoKI-B calculations to involve the corresponding evolution of the electron kinetics. However, the electron kinetics is mainly controlled by the virtually invariant dominant background helium gas density. This is furthermore confirmed by the insensitivity of the simulation results to the number of calling LoKI-B (multiple during the pulse/plug-flow or single in-advance). In order to reduce the computationally-demanding transient simulation duration, LoKI-B is deployed in advance by a single call to calculate the EEDF for a corresponding steady-state plasma composition.

A time-dependent solution of the Boltzmann equation would better represent %resolve 
the transient phenomena \cite{Cap17a}. However, at atmospheric pressure, the time necessary to establish a steady-state EEDF is in the order of nanoseconds, approximated by the reduced momentum-transfer collision 
frequency. The range of the value is also numerically verified (e.g., a few nanoseconds in the kinetic models \cite{Cap00}). Therefore, the stationary solution of the Boltzmann equation is considered in the transient simulations (microsecond-pulsed modulation) of this work, 
in parallel with the available implementations of time-dependent (nanosecond range) models \cite{Hof20,Lie18,Luo18}. 

The generic form of the volume-averaged particle balance equation is
\begin{equation}
\frac{d n_i}{d t} = \left. \sum_{j}  {\cal W}_{ij}R^{j}_{i} \right|_{V} 
+ \left. \sum_{j}  {\cal W}_{ij}R^{j}_{i} \right|_{W},
\end{equation}
where $n_i$ is the volume-averaged density of the species denoted by $i$, ${\cal W}_{ij}$ is the net stoichiometric coefficient for a gain/loss mechanism denoted by $j$, and $R^j_i$ is the reaction rate.
The subscript ``$_V$'' denotes the chemical reactions inside the
plasma volume and ``$_W$'' the flux (e.g., convective and diffusive) source
term including the heterogeneous reactions on the solid walls.

The rate of a plasma volume chemical reaction $j$ is
\begin{equation} 
\left. R^{j}_{i} \right|_{V} = k^j 
\prod_l n_l^{\nu_{lj}},
\end{equation}
where $k^j$ is the rate coefficient, $l$ the reactant and %$\nu_{jl}$ 
$\nu_{lj}$ is the forward stoichiometric coefficient. The rate coefficients are tabulated in \ref{sec:chemkin}. The electronic rate coefficients are mostly computed from the enlisted set of cross-sections available in the literature together with the self-consistently calculated EEDF.

The flow-in and flow-out rates are excluded in the pseudo-one-dimensional formalism due to the motion of the plug with the net mass flow \cite{Sta04}. The flow-in rate in the zero-dimensional formalism is given by the relation \cite{Kem19}
\begin{equation}
\left. R_{{\rm He,O}_2{\rm ,N}_2}^{{\rm F-in}} \right|_{W} =  
c \frac{Q_{{\rm He,O}_2{\rm ,N}_2}P_{{\rm atm}}}{V {\rm k}_{B} T_{{\rm in}}}, 
\end{equation}
where $Q_{{\rm He,O}_2{\rm ,N}_2}$ (sccm) is the partial mass flow rate of
each species fed into the plasma domain, $P_{{\rm atm}}$ the atmospheric 
pressure, $V$ the plasma volume, $T_{{\rm in}}$ the feeding gas temperature and 
$c$ is the unit conversion factor.
The flow-out rate of a species $i$ is \cite{Kem19}
\begin{equation} \label{eqn:flow-out}
	\left. R_i^{{\rm F-out}} \right|_{W} = c  Q \frac{P_{{\rm atm}}{\rm T}_{\rm g}}{V P  T_{{\rm in}}} n_i,
\end{equation}
where $Q$ denotes the total mass flow rate and $P$ the plasma pressure.

The ion and neutral wall flux at the surfaces perpendicular to the mass flow field is ignored
in the zero-dimensional formalism due to the negligibly small area (mostly forming below 2\:\% of the total plasma surface area). This term is also neglected in the pseudo-one-dimensional formalism since the convective mass flow mechanism prevails the flux vector \cite{Sta04}. The wall loss of an ion $i$ at the surfaces parallel to the mass flow field is defined as \cite{Kem17,LieBook2005,Kim06}
\begin{equation} 
 R^+_i \left. \right|_W  =  \left( u_{iB} \frac{h^+_{iX} 2 A^f_X}{V^f} + u_{iB} \frac{h^+_{iY} 2 A^f_{Y}}{V^f} \right) n_{i}, 
\label{eqn:ion_wall}
\end{equation} 
where $u_{iB}$ denotes the Bohm velocity, $h^+_{iX}$ and $h^+_{iY}$ are the ion
{\em edge-to-center ratios}, the superscript $^f$ specifies either the zero-dimensional 
or the pseudo-one-dimensional implementation, $V^f$ is the corresponding volume, 
$A^f_X$ and $A^f_Y$ are their surface areas. Namely, the values of $V^f$,$A^f_X$ and $A^f_Y$ 
are either the plasma volume and surface in the zero-dimensional formalism, or the plug volume and surface in the pseudo-one-dimensional formalism.

The {\em edge-to-center ratio} definition is adapted in the ion wall flux, following conventional global model studies \cite{Lee95,Cha16}. It should be noted that the {\em edge-to-volume-averaged ratio} \cite{Kem15} physically represents a more realistic description at intermediate- or atmospheric-pressure (so-called collisional regime \cite{LieBook2005}), however the resulting variations in the calculations are numerically negligible \cite{Kem17}. The ion velocity at the plasma-sheath edge is expected to deviate from the Bohm velocity at high pressure values due to the larger collisionality of the sheath region compared to the low-pressure plasma \cite{Fra00}. Nevertheless, the deviation is negligible at atmospheric-pressure since the ratio of the ion-neutral collision to the direct ionization frequencies of the helium atomic ion is less than the critical value of $10$ (larger values represent the transition into the collisional sheath regime) based on the analyses by Franklin \cite{Fra00} and Valentini $et$ $al.$ \cite{Val96,Val15}. Additionally, assuming an effective ionization frequency to include the space-charge dominating excimer ions as well as the prevailing mechanisms of Penning and stepwise ionization schemes lead to a much smaller ratio compared to the critical value. In other words, the validity range of the Bohm criterion is not breached at atmospheric-pressure. Furthermore, our model calculations are not sensitive to the ion wall loss, hence any deviation from the Bohm velocity is not influential on the simulation results.

The ion {\em edge-to-center ratios} in the collisional regime are defined as \cite{LieBook2005,Kem17}
\begin{equation}
\begin{array}{l l}
h^+_{iX} & = \frac{1}{1+\alpha} (u_{iBE}/u_{iB}) \left(  1 + \left( \frac{X u_{iBE}}{\pi D_{i+}}\right)^2 \right)^{-1/2}, \\
\\[\dimexpr-\normalbaselineskip+8pt]
h^+_{iY} & =  \frac{1}{1+\alpha} (u_{iBE}/u_{iB}) \left(  1 + \left( \frac{Y u_{iBE}}{\pi D_{i+} }\right)^2 \right)^{-1/2}, 
\label{eqn:ha_coa}
\end{array}
\end{equation}
where $\alpha$ represents the degree of electronegativity, $u_{iBE}$ the electronegative Bohm velocity \cite{Tho09_2,Bra88}, $u_{iB}$  the electropositive Bohm velocity, $X$ and $Y$ the plasma dimensions, and $D_{i+}$ denotes the multiple-ion ambipolar diffusion coefficient \cite{Kem14}.

The reaction rate due to diffusion of a reactive neutral $i$ to the wall and its subsequent heterogeneous reaction in the zero-dimensional and pseudo-one-dimensional formalisms is given as \cite{Boo91,Cha87} 
\begin{equation}
	\left. R^{N}_i \right|_W = n_i \left( \frac{\Lambda_0^2}{D_i}  + \frac{2V^f (2 - \gamma_i)}{2(A^f_X+A^f_Y) v_i \gamma_i}  \right)^{-1},
\label{eqn:wal_rat_neutral}
\end{equation}
where $\Lambda_0$ is the effective diffusion length, $D_i$ the diffusion coefficient, $V^f$ the volume, $2(A^f_X+A^f_Y)$ the net surface area, $v_i$ the mean neutral velocity and $\gamma_{i}$ is the wall reaction probability. The superscript $^f$ denotes the corresponding parameters in the considered formalism: either those of the plasma in the zero-dimensional or those of the plug in the pseudo-one-dimensional formalism.

The volume-averaged electron energy balance equation is given in the form
\begin{equation}
	\frac{d}{d t} \left( \frac{3}{2} n_e \mathrm{T_e} \right) = Q_{{\rm abs}} - (Q_{{\rm Che}} + Q_{{\rm Ela}} + Q_{\rm W}),
\end{equation}
where $n_e$ represents the electron density, $\mathrm{T_e}$ the effective electron temperature, $Q_{{\rm abs}}$ the electrical power absorbed by the plasma as well as the energy loss due to chemical reactions $Q_{{\rm Che}}$, elastic collisions $Q_{{\rm Ela}}$ and wall flux $Q_{\rm W}$.

The power absorbed by the plasma, assuming a homogeneous distribution, is defined by the relation
\begin{equation}
	Q_{{\rm abs}}= \beta {\mathcal P}/V,
\end{equation}
where ${\mathcal P}$ is the input power, $V$ the plasma volume and $\beta$ is the power transfer efficiency. The electron energy loss in the chemical reactions is given in the following form 
\begin{equation}
	Q_{{\rm Che}}= \left. \sum_j {\cal E}_j R^j_e \right|_V,	
\end{equation}
where $R^j_e$ denotes the rate of an electronic reaction $j$ and ${\cal E}_j$ the net energy.
The principle of {\em detailed balancing} is used in calculating the backward rate coefficients of the electron-impact excitation reactions \cite{LieBook2005}. The energy loss due to elastic collisions
$Q_{{\rm Ela}}$ is calculated by the individual elastic rate coefficients
\cite{Kem14,Kem14_2}. The considered elastic collisions are tabulated in \ref{tab:elaHeN2O2} with a reference to the considered cross-section. The energy loss due to the wall flux is given in the form \cite{Kem17,LieBook2005}
\begin{equation}
Q_W = 
\left. \sum_{i \in Ions} \left( {\cal E}_P 
+ {\cal E}_e + {\cal E}_s  \right) R^+_{i} \right|_{W},
\end{equation}
where ${\cal E}_P$ is the plasma potential, ${\cal E}_e$ is the mean energy loss per electron lost and ${\cal E}_s$ represents the sheath potential. The electron energy balance equation is identically implemented in the zero-dimensional and the pseudo-one-dimensional formalism. The only distinction is the aforementioned calculation of the ion wall loss rate due to the differing volume and surface area.

\section{Chemical kinetics}
\label{sec:Chem_kin}

The gas mixtures He, He/O$_2$, He/N$_2$ and He/N$_2$/O$_2$ are considered in this study. The species, reactions and elastic collision sets are enlisted in \ref{sec:chemkin}: tables
\ref{tab:SpeciesHeN2O2},\ref{tab:ReactionListHe},\ref{tab:ReactionListHeO2},\ref{tab:ReactionListHeN2},\ref{tab:ReactionListHeNxOy},\ref{tab:ReactionListHeO2Vib},\ref{tab:ReactionListHeO2vprime},\ref{tab:ReactionListHeN2Vib},\ref{tab:ReactionListHeN2vprime},\ref{tab:ReactionListN2O2Vib},\ref{tab:wrHeN2O2},\ref{tab:iwrHeN2O2} and \ref{tab:elaHeN2O2}. The reactions are separated into several tables for the sake of clarity and tabulated according to the mixture composition and reaction type. The bulk of the He and He/O$_2$ chemical sets is adopted from a study of Waskoenig $et$ $al.$ \cite{Was10}, except the vibrational kinetics. This maximizes the consistency in model-to-model benchmark of our zero-dimensional simulation results to that of the one-dimensional model provided by Waskoenig $et$ $al.$ (see section \ref{sec:benchmarking}). Comparatively, the chemical sets here are supplemented with the reaction mechanisms of $ \mathrm{O_4^+} $ and $ \mathrm{O_4^-}$ for a better agreement of the calculated $ \mathrm{O(^3 P)}$ density with the measurement (especially, due to reactions 49, 73 and 74 in table \ref{tab:ReactionListHeO2}). Additionally, more detailed helium excimer reaction mechanisms are included in this work. The chemical sets of He/N$_2$ and He/N$_2$/O$_2$ are established based on a wide variety of studies available in literature (explicitly given in \ref{sec:chemkin}). Both available databases \cite{LXCat,nist15} and individual studies (e.g., \cite{Tho08,Nor15,Kim15,Kos92,CapBook,Sak12}) are referred in the data collection. A recent set of electron-impact cross-sections are adopted for N$_2$O and NO$_2$ \cite{Son19}.

The vibrationally excited molecules $ \mathrm{O_2}(v<41) $ and $ \mathrm{N_2}(v<58) $ are incorporated in the species set. The detailed vibrational kinetics in tables \ref{tab:ReactionListHeO2Vib},\ref{tab:ReactionListHeO2vprime},\ref{tab:ReactionListHeN2Vib},\ref{tab:ReactionListHeN2vprime} and \ref{tab:ReactionListN2O2Vib} are divided into  electron-Vibrational (e-V), Vibrational-Translational (V-T) and Vibrational-Vibrational (V-V) energy transfer mechanisms, mainly based on databases \cite{IST-Lisbon-LXCat,Phys4Entry} and studies \cite{Lap13,Lap14,Gue95,CapBook,Kem16,Gue19_rev}. Moreover, the Penning ionization and charge exchange reactions between helium and vibrationally excited molecules are integrated into the kinetics.
The resonant vibrational excitation cross-sections \cite{Lap13,Lap14} are favoured in the e-V transfer due to their efficiency. Only the cross-sections of the first six vibrational levels are used in the solution to the Boltzmann equation, while those of the higher levels are directly evaluated to the corresponding rate coefficients according to the established EEDF. The reverse rate coefficients of all the e-V excitation are calculated via the principle of {\em detailed balancing} \cite{LieBook2005}. The V-T transitions are substantially dominated by the strong He background gas, however the transitions due to $\mathrm{O(^3 P)}$, $\mathrm{N(^4 S)}$, O$_2$ and N$_2$ are considered as well. The rate coefficients of N$_2(v>0)$-O$_2$ and O$_2(v>0)$-N$_2$ V-T reactions are estimated by those of N$_2(v>0)$-N$_2$ and O$_2(v>0)$-O$_2$, respectively due to lack of data \cite{Gue95,Gue19_rev}. Only the N$_2(v>0)$-N channel is taken into account among the nitrogen atom induced V-T mechanisms, whereas the inefficient O$_2(v>0)$-N channel is neglected \cite{Gue19_rev}. Furthermore, the dissociation mechanism of $\mathrm{O(^3 P)}$ + O$_2(v>0)$ \cite{Esp02} and the NO formation reaction of $\mathrm{O(^3 P)}$ + N$_2(v \geqslant 13)$ \cite{Gue19_rev} are involved in the calculations. The ozone production channel of O$_2$ + O$_2(v>0)$ \cite{Ell12} is ignored due to the insensitivity of the simulation results to its presence. The V-V mechanisms are thoroughly considered, however $ {\rm N}_2(v \geqslant 0) + {\rm O}_2(w>1) \rightarrow {\rm N}_2(v+1) + {\rm O}_2(w-1) $ is excluded on account of its inefficient contribution to the quenching of O$_2(w>0)$ \cite{Bil94}. Both ${\rm O}_2(v'=41)$ and ${\rm N}_2(v'=58)$, referred as pseudo levels, are assumed to instantaneously dissociate due to the proximity of the vibrational energy to the dissociation energy \cite{Cac78}. In total, the He/$ \mathrm{N_2} $ set includes $72$ distinct species and $5831$ reactions, the He/$ \mathrm{O_2} $ set $58$ species and $3216$ reactions and the He/$ \mathrm{N_2} $/$ \mathrm{O_2} $ set 138 species and $11799$ reactions.

The considered neutral and ion reactions at the wall are tabulated in \ref{tab:wrHeN2O2} and \ref{tab:iwrHeN2O2}, respectively. The quenching of the excited states and the recombination of nitrogen atom on the chamber surface are taken into account. The oxygen atom wall recombination is ignored since volume processes are expected to dominate over surface reactions under the considered operation conditions \cite{Was10}. Nevertheless, the influence of the recombination probability value on the simulation results is discussed in section \ref{sec:benchmarking}. An ion impinging on the surface is assumed to capture an electron via the Auger effect or resonant electron transfer and then returns back to the plasma region.

By modifications of the underlying production channels based on the above-mentioned reference chemical kinetics, four sets of additional analyses are conducted on the NO density in He/N$_2$/O$_2$ mixtures (see section \ref{sec:NOproduction}): 
\begin{enumerate}
\item[(1)] A rate coefficient value of $1 \times 10^{-19}$ $ \mathrm{m^3/s}$ is used for the reaction $\mathrm{O(^3 P)} + \mathrm{N_2}(v \geqslant 13) \rightarrow \mathrm{NO} + \mathrm{N (^4S)}$ in this study, suggested by Guerra $et$ $al.$ \cite{Gue95,Gue97,Gue19_rev} and Pintassilgo $et$ $al.$ \cite{Pin09}.
A distinct value of $1 \times 10^{-17}$ $\mathrm{m^3/s}$ is used by Gordiets $et$ $al.$ \cite{Gor95} and the influence of this value on the NO density is quantified. 
\item[(2)] A NO wall formation by the adsorbed $\rm O(^3 P)$ and $\rm N(^4 S)$ on the surface is ignored due to negligible probability measured at a lower pressure regime \cite{Mar10}. However, the probability is not confirmed for atmospheric-pressure plasma jets. 
The presence of a NO wall formation $\rm N (^4 S) + wall \rightarrow NO$ (i.e., an effective form of the surface reactions given by a set of adsorbed layer reaction mechanisms \cite{Gor96}) is further considered with a maximum probability of 1 based on set (1), and its role in the NO density is shown. This wall reaction is constructed under the presumption that $\rm O(^3 P)$ is sufficiently adsorbed on the surface due to its much larger concentration compared to that of $\rm N(^4 S)$ under the considered operation conditions.  
\item[(3)] A well accepted value of $7 \times 10^{-18}$ $ \mathrm{m^3/s}$ \cite{CapBook, Gue19_rev, Kos92} is used for the reactive quenching $\rm N_2(A^3 \Sigma) + O(^3 P) \rightarrow NO + N(^2D) $ in this work. 
The value is confirmed for the vibrational ground state $\mathrm {N_2(A^3 \Sigma},v=0)$. %\cite{Pip81,Pip82,Tho96,Yon19,Ben97,Dil99}, 
Note that a much larger value is observed for the higher vibrational quantum numbers $\mathrm {N_2(A^3 \Sigma},v>0)$ \cite{Pip81,Pip82,Tho96,Yon19,Ben97,Dil99}. Furthermore, the virtually identical NO density is obtained between the measurements and the simulations considering the reactive quenching of multiple $ \mathrm{N_2} $ electronically excited states by $ \mathrm{O(^3 P)} $ \cite{Shk14}. In order to estimate the role of the vibrationally excited molecules $\mathrm{N_2(A^3 \Sigma},v>0)$ and of multiple higher $ \mathrm{N_2} $ electronically excited states, a larger hypothetical rate coefficient value of $7 \times 10^{-15}$ $ \mathrm{m^3/s}$ is considered for the aforementioned quenching reaction and the resulting NO density is revealed. 
\item[(4)] A reactive quenching of $\mathrm{N_2(B ^3\Pi)}$ by $\mathrm{O(^3 P)}$ is neglected in many chemical kinetics studies \cite{Sak12, CapBook, Gue19_rev, Kos92, Sun19} except a handful of publications \cite{Ada15,Ver20_2,Shk14} assigning a rate coefficient value of $3 \times 10^{-16}$ $ \mathrm{m^3/s}$. On the other hand, larger quenching rates of $\mathrm{N_2(B ^3\Pi},v>0)$ are observed at its higher vibrational levels \cite{Pip92}. Thus, the same hypothetical rate coefficient value of $7 \times 10^{-15}$ $ \mathrm{m^3/s}$ for the reactive quenching $\rm N_2(B ^3\Pi) + O(^3 P) \rightarrow NO + N(^2D) $ is additionally included to estimate the role of $\mathrm{N_2(B ^3\Pi},v\geqslant0)$ and of multiple higher $ \mathrm{N_2} $ electronically excited states based on set (3). Their influence on the NO concentration is presented. 
\end{enumerate}
It should be noted that the state-by-state resolution of the vibrational distribution function $\mathrm{N_2(A^3 \Sigma,B ^3\Pi;}\:v>0)$ is not preferred in the current study due to lack of the necessary data and the excessive computational load. Moreover, the kinetics of the above-mentioned multiple $ \mathrm{N_2} $ electronically excited states are intensively discussed in a nanosecond-pulsed air discharge at a pressure of 13332 Pa \cite{Shk14}. Compared to the measured peak $ \mathrm{NO} $ density, a factor of $\sim$5 is under-estimated by the simulations considering the reactive quenching of $\mathrm{N_2(A^3 \Sigma, B ^3\Pi, C ^3\Pi)}$ by $ \mathrm{O(^3 P)} $. The under-estimation is minimized by an introduction of the triplet states $\mathrm{N_2(W ^3\Delta_u, B^{\prime}{}^{3}\Sigma^-_u, E ^3\Sigma^+_g)}$ and the singlet states $\mathrm{N_2(a^{\prime}{}^{1}\Sigma^-_u, a ^1\Pi_g, w ^1\Delta_u, a^{\prime\prime}{}^{1}\Sigma^+_g)}$ in this quenching with a gas kinetic rate coefficient value of $3 \times 10^{-16}$ $ \mathrm{m^3/s}$ (except that of $\mathrm{N_2(A^3 \Sigma)}$ with $\sim$10$^{-18}$ $ \mathrm{m^3/s}$). The role of these multiple higher electronically excited states is partially estimated in our study by the hypothetically large rate coefficient value of $7 \times 10^{-15}$ $ \mathrm{m^3/s}$ in sets (3) and (4). An $``$effective$"$ value of $1.5 \times 10^{-15}$ $ \mathrm{m^3/s}$ for set (4) can be suggested by the aforementioned factor of $\sim$5 (observed in \cite{Shk14}). However, the NO density is still under-estimated by our simulations using this value relative to the measurements at low power (e.g., by a factor of around 2 at 0.6 W). Although $7 \times 10^{-15}$ $ \mathrm{m^3/s}$ is evidently not regarded as a realistic rate coefficient value for the reactive quenching channels of $\mathrm{N_2(A^3 \Sigma, B ^3\Pi)}$ in set (4), a strong signal is imposed that their vibrationally excited levels and the multiple higher $ \mathrm{N_2} $ electronically excited states play an underlying role in the NO formation at low power. A development of the chemical kinetics involving these levels and states is desirable in the future.

The sensitivity of the simulation results with respect to the considered wall reaction mechanisms is investigated. A small oxygen atom wall recombination probability value is observed by Booth $et$ $al.$ \cite{Boo19} at relatively low pressure values; however, the measurements in the considered pressure regime are still absent. Furthermore, a much larger value is assigned by a study of an atmospheric-pressure plasma jet by Schr{\"o}ter $et$ $al.$ \cite{Sch18_PCCP}, in contradiction with the negligible value suggested by Waskoenig $et$ $al.$ \cite{Was10}. Likewise, based on the estimated values at low-pressure plasma sources \cite{Jan10,Mar13,Mar13_2}, it is assumed that the ozone wall formation is negligible under the considered operation conditions. The influence of the oxygen and nitrogen atom wall recombination as well as the first-order ozone wall formation probabilities are analysed in section \ref{sec:Sensitivity}.

\section{Results}
\label{sec:res}

The developed models in section \ref{sec:mod} and \ref{sec:Chem_kin} are firstly benchmarked against the spatially resolved simulation results and a wide variety of measurements available from literature \cite{Gol20,Nie10,Nay19,Was10,Bib11,Kor20,Sch14}. The NO concentration is then characterized with a focus on the comparison between the model calculations and the recently published measurements \cite{Pre20}. Lastly, the VDFs, the wall reaction probabilities and the synthetic air impurity levels are computationally investigated based on the validity of the models. In this section, a pressure of 101325 Pa and a gas temperature of 345 K is used in the simulations, unless stated otherwise.

\subsection{Benchmark against measurements and simulation results}
\label{sec:benchmarking}

\begin{figure}[t!]
	\begin{minipage}[b]{0.5\textwidth}
		\includegraphics[width=0.95\textwidth]{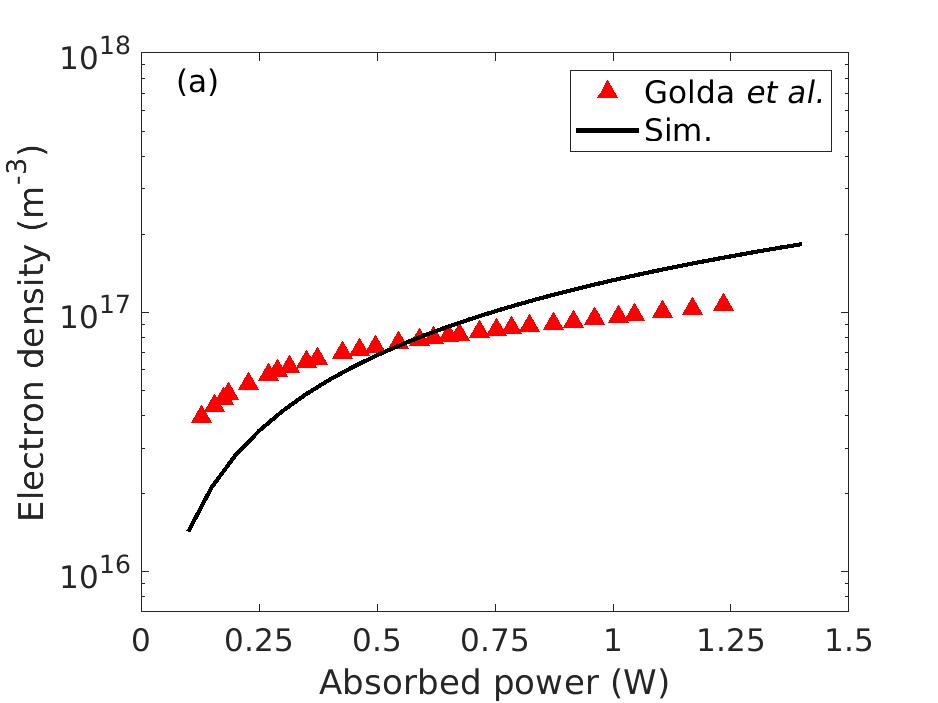} \\ 
\end{minipage}
\begin{minipage}[b]{0.5\textwidth}
	\includegraphics[width=0.95\textwidth]{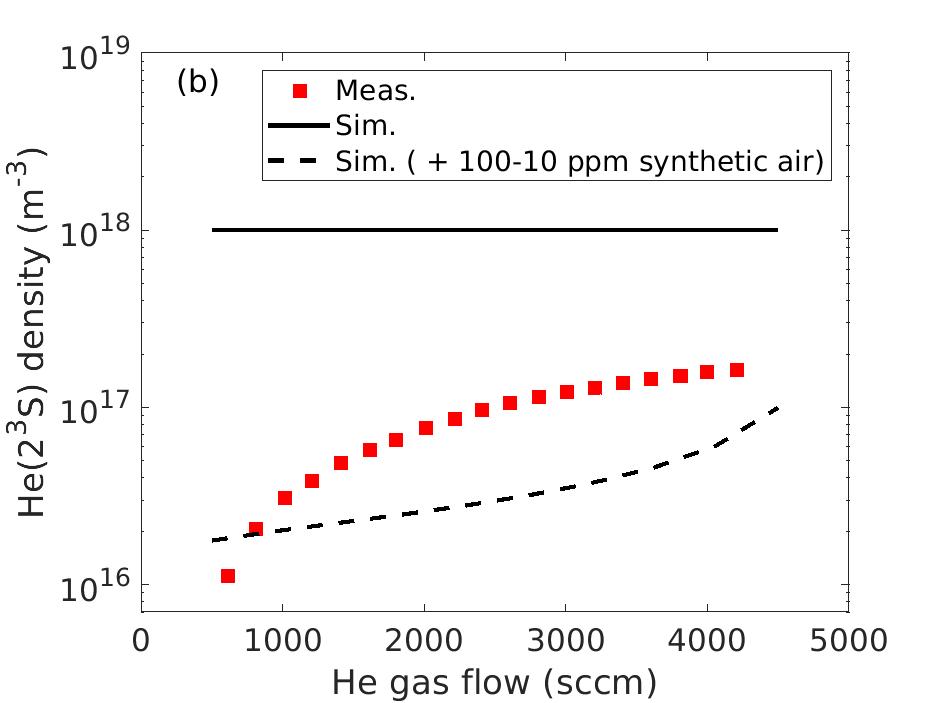} \\ 
\end{minipage}
\caption{$(a)$ The volume-averaged electron density (\textcolor{red}{$\blacktriangle$}) in a He COST-Jet reported by Golda $et$ $al.$ \cite{Gol20} and the zero-dimensional simulation results (\textcolor{black}{\full}). $(b)$ The measurements of helium metastable density close to the powered electrode (\textcolor{red}{$\blacksquare$}) in a He $\rm \mu$APPJ by Niermann $et$ $al.$ \cite{Nie10} as well as the zero-dimensional simulation results for pure helium (\textcolor{black}{\full}) and helium diluted by a linearly decreasing synthetic air (N$_2$:O$_2$=8:2) impurity between $100-10$ ppm with respect to the increasing helium flow rate from 500 sccm to 4500 sccm (\textcolor{black}{\dashed}). We refer to the text for the operation conditions.
}\label{fig:gol20}
\end{figure}

The volume-averaged electron density in a He COST-Jet is reported by Golda $et$ $al.$ \cite{Gol20}. The plasma is operated at a net mass flow rate of $1000$ sccm and a pressure of 99000 Pa. The gas temperature is fixed at about $345$ K \cite{Gol16}. Similar density values and trend are observed between the reported electron density and the zero-dimensional simulation results shown in figure \ref{fig:gol20}$(a)$. Compared with the simulations for an assumption of a Maxwellian EEDF (not shown here), the self-consistent EEDF calculation reduces the electron density (around 28\:\%) and increases the electron temperature (about from $2.0$ eV to $2.5$ eV) due to the depletion in the EEDF tail. It should be emphasized that the electron density is associated with much less susceptibility to the impurities relative to the other species (e.g., helium metastable). Furthermore, the plasma source under investigation is tightly controlled during the measurements to achieve a maximum purity.

The helium metastable density measurements taken close to the powered electrode (i.e., highest density value) in a He $\rm \mu$APPJ with dimensions $1 \times 1.8 \times 40$ mm$^3$ are provided by Niermann $et$ $al.$ \cite{Nie10}. The fed helium gas flow rate value ranges from $500$ sccm to $4500$ sccm at a pressure of $ 1 \times 10^5 $ Pa. The gas temperature is estimated as $333$ K. An input power of $23$ W is fed into the discharge and a power transfer efficiency of $5\:\%$ is assumed in the simulations. Both pure helium and helium with a linearly decreasing synthetic air impurity between $100-10$ ppm with respect to the increasing helium flow rate are simulated due to either the back diffusion from the effluent \cite{Ver93} or the air leakage into the feeding gas. The metastable density measurement data and the zero-dimensional simulation results are shown in figure \ref{fig:gol20}$(b)$. The Penning ionization is a significant loss mechanism, and an almost two orders of magnitude smaller metastable density is predicted by the simulations in the presence of the impurity compared with that of pure helium. Note that the inverse-exponentially increasing trend of the measured metastable density can be better explained by the simulation results considering an exponentially decreasing impurity (due to the unknown exponential factor, the results are not shown here). This suggests that the impurity levels are strongly correlated with the feeding gas flow rate.

\begin{figure}[t!]
\begin{minipage}[b]{0.5\textwidth}
	\includegraphics[width=0.95\textwidth]{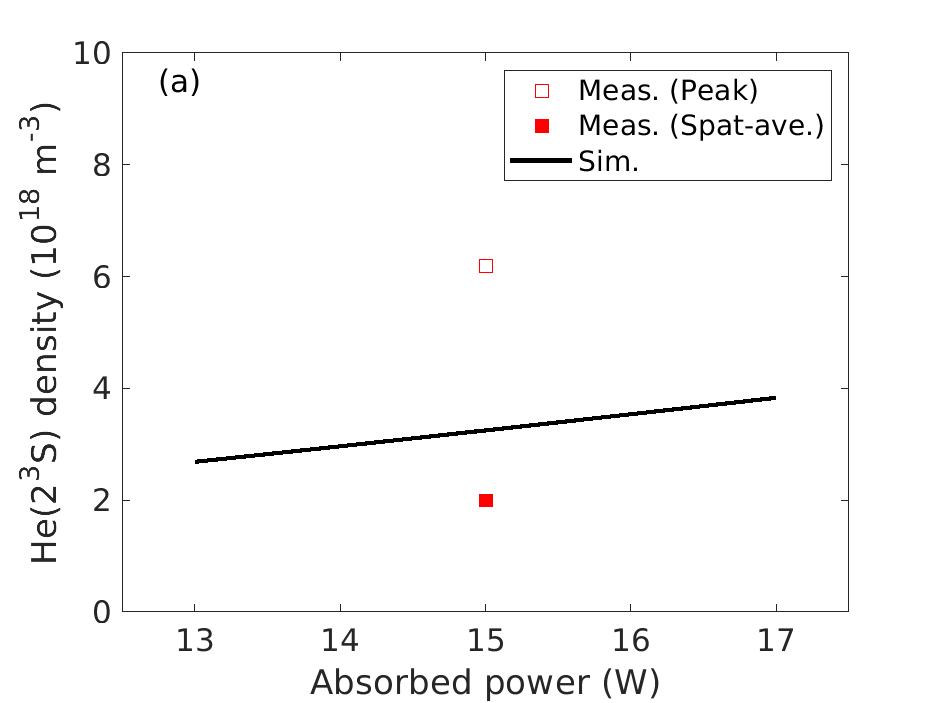} \\ 
\end{minipage}
\begin{minipage}[b]{0.5\textwidth}
	\includegraphics[width=0.95\textwidth]{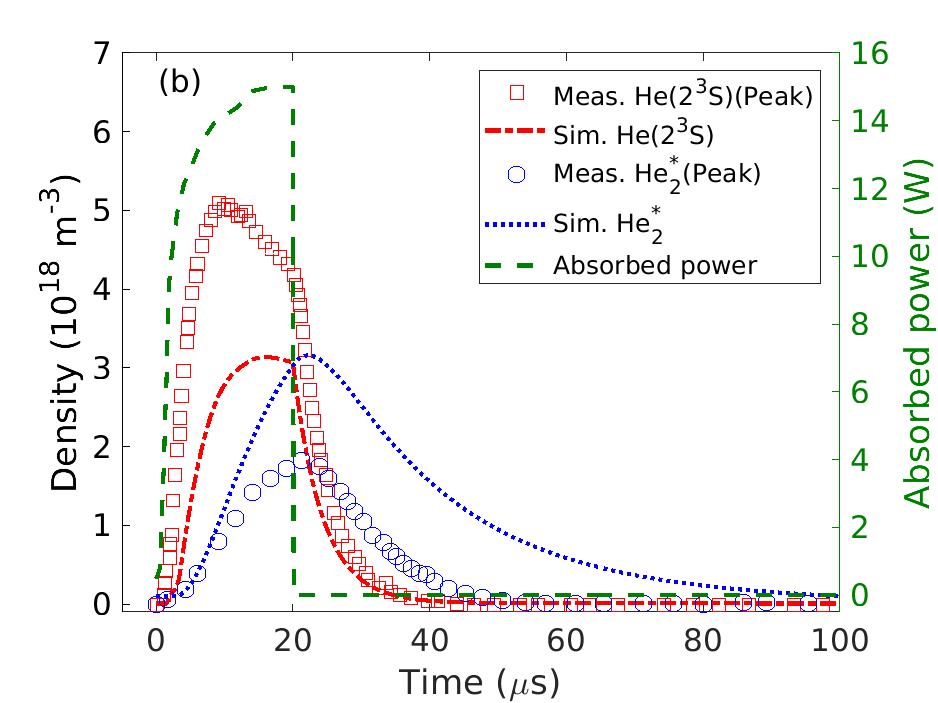} \\ 
\end{minipage}
\caption{The measurements of $ \mathrm{He(2 ^3S)}$ and $\mathrm{He_2^*}$ densities (symbols) in a He atmospheric-pressure plasma jet by Nayak $et$ $al.$ \cite{Nay19} together with the zero-dimensional simulation results. $(a)$ The peak $ \mathrm{He(2 ^3S)}$ density between the electrodes (\textcolor{red}{$\square$}) and its spatially-averaged value (\textcolor{red}{$\blacksquare$}) in a continuous power input as well as the volume-averaged simulation results (\textcolor{black}{\full}). $(b)$ The time-resolved peak $ \mathrm{He(2 ^3S)}$ (\textcolor{red}{$\square$}) and $\mathrm{He_2^*}$ (\textcolor{blue}{$\circ$}) density measurements in a pulse-modulated power input (\textcolor{OliveGreen}{\dashed}) as well as the zero-dimensional simulation results for $ \mathrm{He(2 ^3S)}$ (\textcolor{red}{\chain}) and $\mathrm{He_2^*}$ (\textcolor{blue}{\dotted}). We refer to the text for the operation conditions.
}\label{fig:nay19}
\end{figure}

The helium metastable density in a He atmospheric-pressure plasma jet for both continuous and pulse-modulated power input is measured by Nayak $et$ $al.$ \cite{Nay19}. The plasma is confined in a volume of $2 \times 9.5 \times 19.1$ mm$^3$ with a mass flow rate of $5000$ sccm He at a gas temperature of around $335$ K. An absorbed power of $15$ W is measured in the continuous discharge. The pulse-modulation is applied in a period of $100 \:\mu$s at a power-on duration of $20 \:\mu$s and a peak power of around $15$ W. 
The peak (inter-electrode) and the spatially-averaged $ \mathrm{He(2 ^3S)}$ density measurements for the continuous power input as well as the corresponding zero-dimensional simulation results for a variation of the absorbed power are shown in figure \ref{fig:nay19}$(a)$. A linearly increasing $ \mathrm{He(2 ^3S)}$ density profile is observed in the simulations with respect to the power. The simulation results represent the volume-averaged density and show a better proximity to the spatially-averaged measurement value than that of the peak. The time-resolved peak $ \mathrm{He(2 ^3S)}$ and $ \mathrm{He_2^*}$ density measurements as well as the simulation results during a modulation cycle in the pulsed operation are shown in figure \ref{fig:nay19}$(b)$. Relative to the measurements, the simulations suggest a slightly smaller He($2^3$S) density. Note that only the peak density measurements between the electrodes are provided by Nayak $et$ $al.$. The spatially-averaged measurements would show a better agreement with the simulations. 
The over-estimated $ \mathrm{He_2^*}$ density in the calculations relative to the peak measurements can be partially attributed to the role of the humid air impurities reported by Nayak $et$ $al.$. An inclusion of $10$ ppm synthetic air impurity in this work reduces the simulated $ \mathrm{He_2^*}$ density by a factor of about $0.15$ ($ \mathrm{He(2 ^3S)}$ density about 0.25). Hence, this inclusion improves the agreement between the simulations and the spatially-averaged measurements.

\begin{figure}[t!]
\begin{minipage}[b]{0.5\textwidth}
	\includegraphics[width=0.95\textwidth]{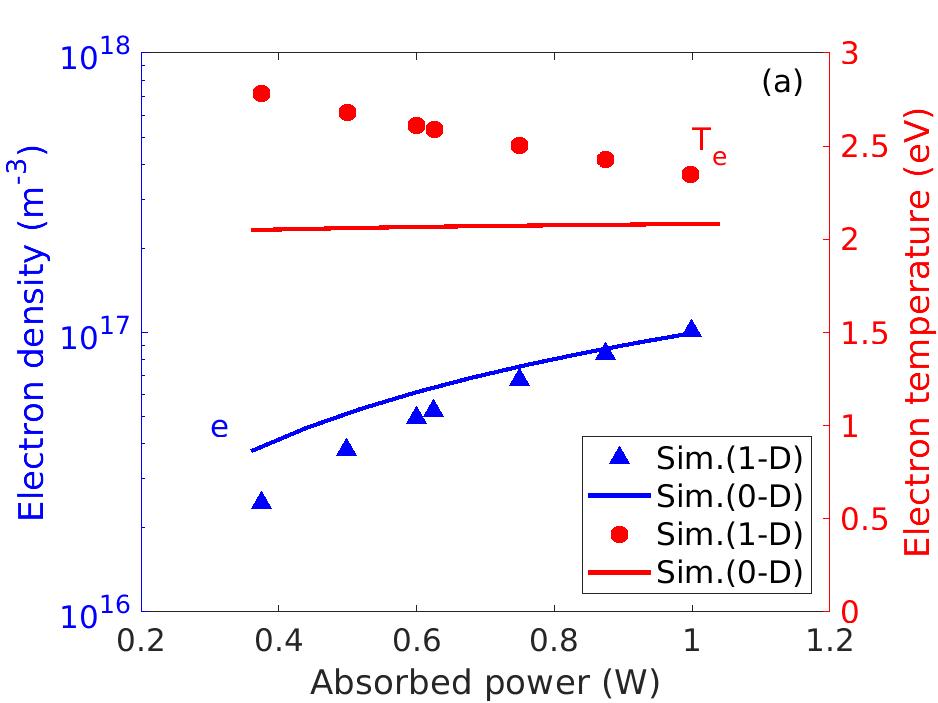} \\ 
\end{minipage}
\begin{minipage}[b]{0.5\textwidth}
	\includegraphics[width=0.95\textwidth]{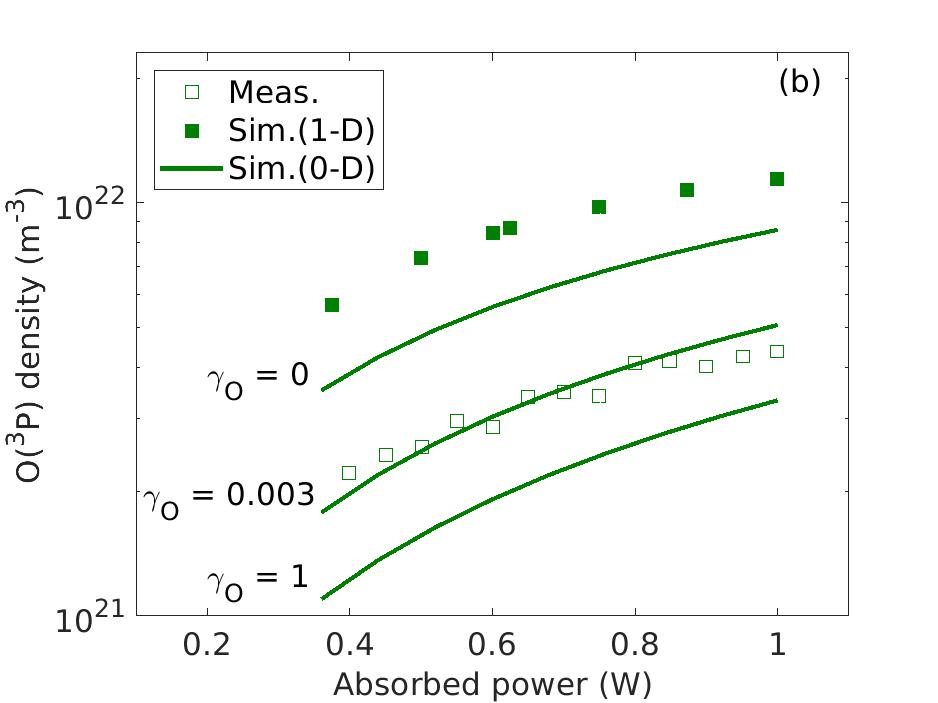} \\ 
\end{minipage}
\caption{The one-dimensional (1-D) spatially resolved simulation results (the time- and space-averaged electron density (\textcolor{blue}{$\blacktriangle$}) and temperature (\tikzcircle[red, fill=red]{2.0pt}) as well as the oxygen atom density (\textcolor{OliveGreen}{$\blacksquare$}) in the discharge center) and the corresponding oxygen atom density measurements (\textcolor{OliveGreen}{$\square$}) in a He/$\mathrm{O_2}$ $\rm \mu$APPJ by Waskoenig $et$ $al.$ \cite{Was10} together with the calculations (lines) by the zero-dimensional (0-D) model for a variation of the absorbed power. $(a)$ electron density and temperature, $(b)$ oxygen atom density. We refer to the text for the operation conditions.
}\label{fig:was10}
\end{figure}

A spatially resolved one-dimensional (1-D) model is developed and its simulation results together with a set of oxygen atom density measurements in a He/$\mathrm{O_2}$ $\rm \mu$APPJ are reported by Waskoenig $et$ $al.$ \cite{Was10}. The plasma dimensions are given as $1 \times 1 \times 40$ mm$^3$ at a fixed gas temperature value of $345$ K and a pressure of $1\times10^5$ Pa. A total mass flow rate of $1000$ sccm helium and oxygen is fed into the plasma domain with the ratio 1000:5. A power transfer efficiency of $5\:\%$ is calculated. The simulation results and the measurements are shown in figure \ref{fig:was10}$(a)$: the time- and space-averaged electron density and temperature calculated by Waskoenig's one-dimensional simulation as well as our zero-dimensional model, \ref{fig:was10}$(b)$: the oxygen atom density measurements in the discharge center as well as the one- and zero-dimensional model calculations. The zero-dimensional simulation results for an assumed and maximum oxygen atom wall recombination probability (see reaction 9 in table \ref{tab:wrHeN2O2}) are also provided in figure \ref{fig:was10}$(b)$. It should be emphasized that the deviations between the simulation results of both models are ascribed to the supplement of $ \mathrm{O_4^+} $ and $ \mathrm{O_4^-}$ into the zero-dimensional calculations for an improved chemical description. A much better agreement of electron temperature and oxygen atom density is achieved without this supplement. Furthermore, the oxygen atom concentration is over-estimated by both modeling approaches relative to the measurements. This over-estimation is minimized with the assumed wall recombination probability of 0.003. The electron density and temperature calculations are negligibly altered for a variation of the wall recombination probability between a maximum and a vanishing value, below 2\:\%. Furthermore, the simulation results are not highly sensitive to the exclusion of the vibrationally excited levels. The absence of these levels (not shown here) raises the oxygen atom density by 5.74\:\%, the electron density by 5.23\:\%, whereas it reduces the electron temperature by 0.39\:\%.

\begin{figure}[tp]
        \centering
	\subfigure{\includegraphics[scale=0.25,clip=true,trim=0cm 0cm 0cm 0cm,]{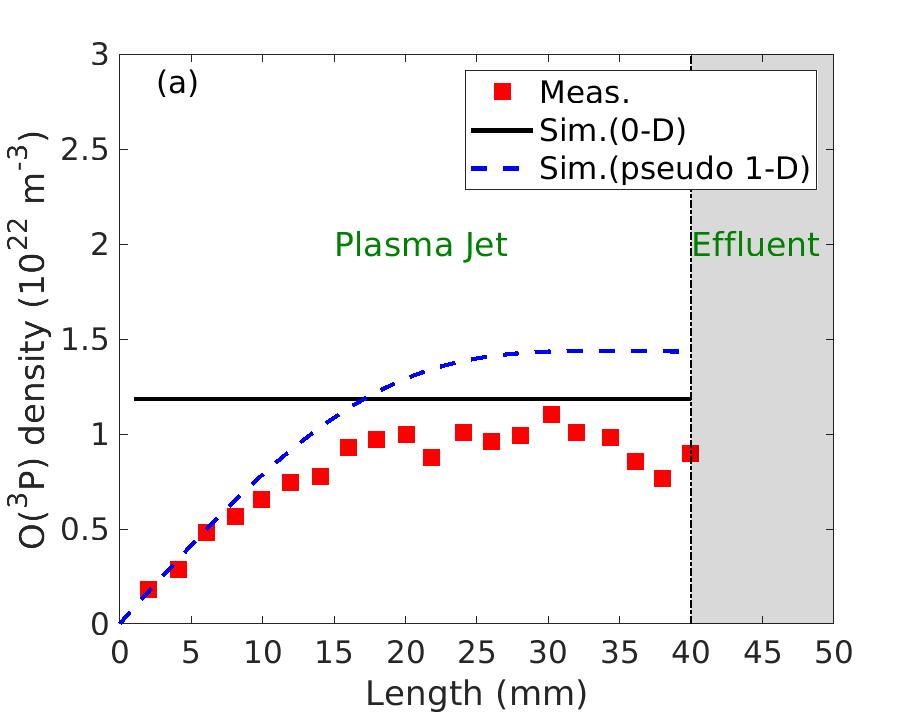}} \\
	\subfigure{\includegraphics[scale=0.25,clip=true,trim=0cm 0cm 0cm 0cm,]{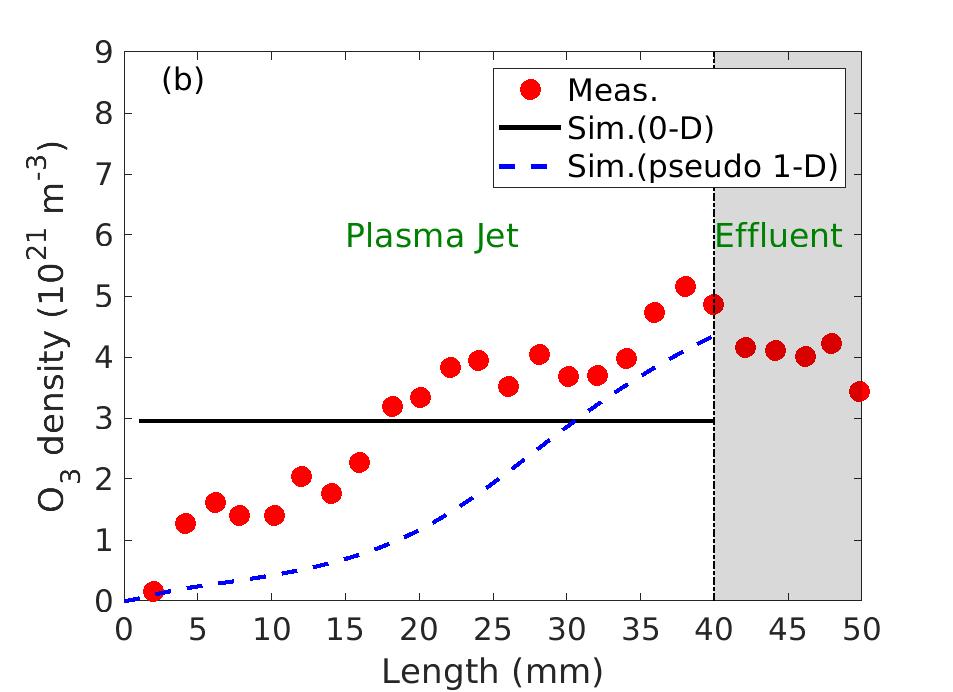}} \\
	\subfigure{\includegraphics[scale=0.25,clip=true,trim=0cm 0cm 0cm 0cm,]{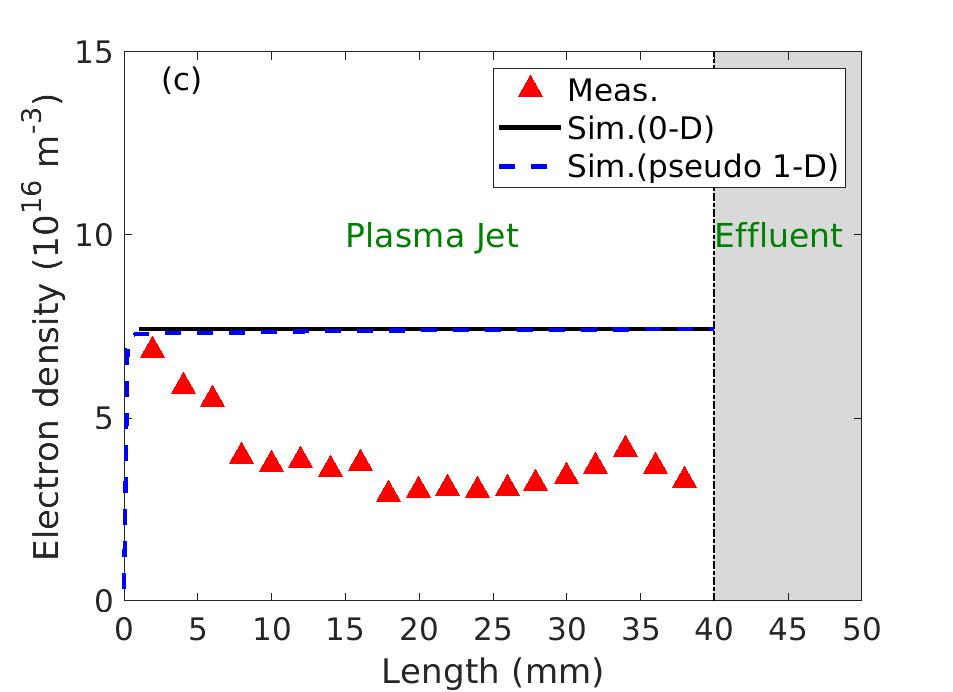}}
	\caption{The spatially resolved measurements in gas flow direction of a He/$\mathrm{O_2}$ $\rm \mu$APPJ by Bibinov $et$ $al.$ \cite{Bib11} as well as both the zero-dimensional (0-D) and pseudo-one-dimensional (pseudo 1-D) simulation results. $(a)$ oxygen atom, $(b)$ ozone and $(c)$ electron densities. We refer to the text for the operation conditions.
}\label{fig:bib11}
\end{figure}

The spatial variation of the oxygen atom, ozone and electron densities in gas flow direction is measured inside the chamber of a He/$\mathrm{O_2}$ $\rm \mu$APPJ by Bibinov $et$ $al.$ \cite{Bib11}. The plasma is formed within a volume of $1 \times 1.3 \times 40$ mm$^3$ at a gas temperature of $370$ K. The helium and oxygen mass flow rates are fixed at $1500$ sccm and $22.5$ sccm, respectively. An input power of $30$ W is fed into the plasma and a power transfer efficiency of $5$\:\% is assumed. The measurements together with the simulation results for both the pseudo-one-dimensional plug-flow and the zero-dimensional modelling formalisms are displayed in figure \ref{fig:bib11}. The calculated pseudo-one-dimensional densities are in agreement with the measurements. A close proximity is obtained between the simulation results of both models, e.g., the spatial averages of the pseudo-one-dimensional results are virtually identical with the zero-dimensional results.

\begin{figure}[t!]
	\begin{minipage}[b]{0.5\textwidth}
	    \includegraphics[width=0.95\textwidth]{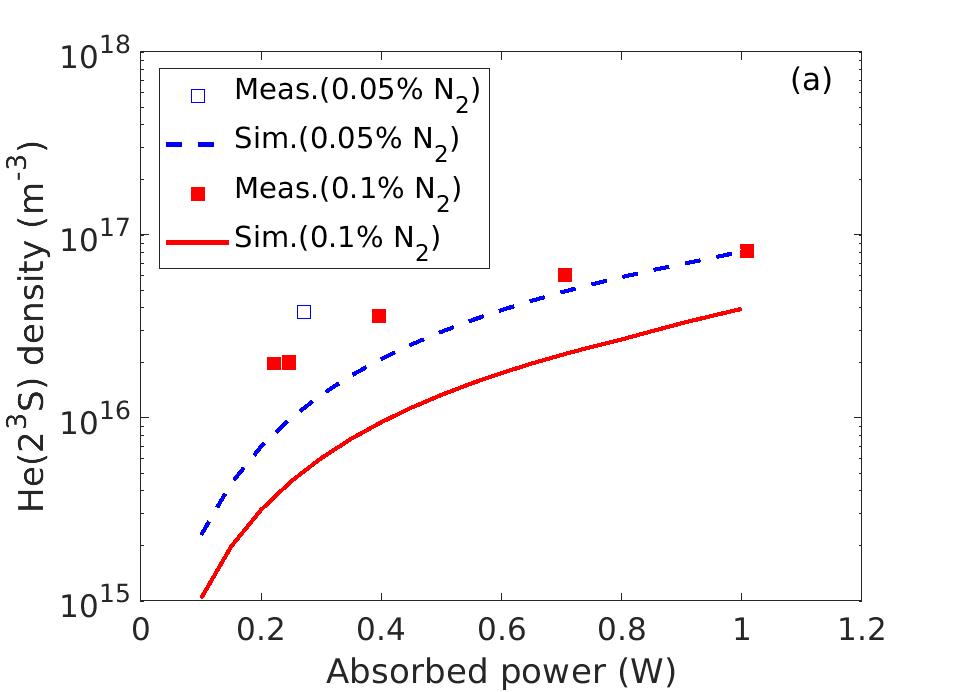} \\ 
\end{minipage}
\begin{minipage}[b]{0.5\textwidth}
    \includegraphics[width=0.95\textwidth]{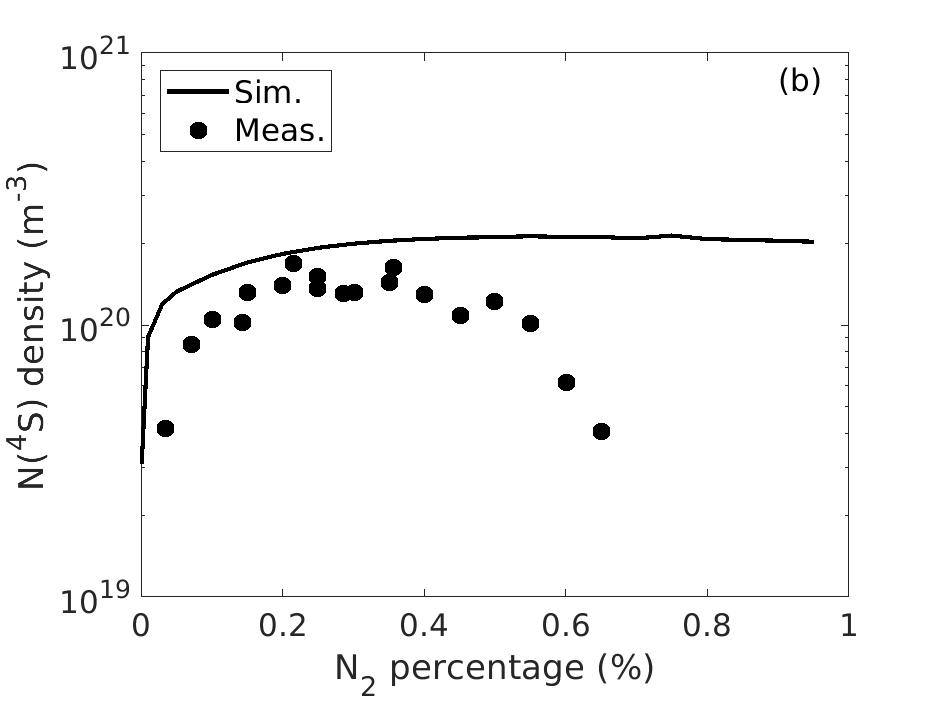} \\ 
\end{minipage}
\caption{The measurements (symbols) and the zero-dimensional simulation results (lines) in He/$\mathrm{N_2}$ $\rm \mu$APPJs. $(a)$ The space- and time-averaged helium metastable density measurements (\textcolor{blue}{$\square$},\textcolor{red}{$\blacksquare$}) by Korolov $et$ $al.$ \cite{Kor20,Kor21}. $(b)$ The nitrogen atom density measurements (\tikzcircle[black, fill=black]{2.0pt}) in the effluent (at 4 mm away from the jet nozzle) by Schneider $et$ $al.$ \cite{Sch14}. We refer to the text for the operation conditions.
}\label{fig:a4_Korolov}
\end{figure}

The space- and time-averaged helium metastable density measurements in a He/$\mathrm{N_2}$ COST-Jet are reported by Korolov $et$ $al.$ \cite{Kor20,Kor21}. The plasma jet is operated at a pressure of about $ 1 \times 10^5 $ Pa with an estimated gas temperature of $345$ K. The helium mass flow rate is fixed at $1000$ sccm with an additional nitrogen mass flow rate of $0.5$ sccm or $1$ sccm. The absorbed power values for the measurements are calculated from the particle-in-cell model described in \cite{Kor20}. The measurement data and the zero-dimensional simulation results are shown in figure \ref{fig:a4_Korolov}$(a)$. An inverse proportionality between the metastable density and nitrogen content is observed in both the measurement data and the simulation results. However, a slightly smaller density value is predicted by the simulations relative to the measurements. The difference is attributed to the termolecular Penning ionization (reaction 51 in table \ref{tab:ReactionListHeN2}) and the metastable quenching via vibrationally excited nitrogen molecules (reaction 11 in table \ref{tab:ReactionListHeN2Vib}). The termolecular Penning ionization is the most dominant metastable loss mechanism, forming about $40$\:\% of the net loss rate. The rate coefficient is determined by Pouvesle $et$ $al.$ \cite{Pou82} and later reported with a much smaller value \cite{Pou88}. The difference between the simulations and measurements is lowered by the latter value, however it is not explicitly preferred in the simulations due to lack of evaluation on the accuracy. The metastable quenching rate coefficient via vibrationally excited molecular nitrogen contributes about $15$\:\% of the net metastable loss rate. In parallel with Sommerer $et$ $al.$ \cite{Som92_2}, we assume that the rate coefficient is identical to that of the quenching via ground state molecular nitrogen estimated by Pouvesle $et$ $al.$ \cite{Pou88}. Note that this estimation does not completely exclude the quenching via vibrational levels since it relies on the optical absorption of the $2^3$S - $3^3$P helium transition. A thorough analysis of the aforementioned rate coefficients is of the essence for a better resolution on the subject.

The nitrogen atom density in a He/$\mathrm{N_2}$ $\rm \mu$APPJ effluent (e.g., at $4$ mm away from the jet nozzle) is measured by Schneider $et$ $al.$ \cite{Sch14}. A plasma volume of $1 \times 1 \times 30$ mm$^3$ is fed with a fixed helium mass flow rate of $1400$ sccm and a variable nitrogen mass flow rate. An absorbed power of $0.8$ W is assumed in the simulations based on the estimated relation between the applied electrode voltage and absorbed power \cite{Sch14}. Additionally, a gas temperature of $345$ K is estimated following the similarity of the operation conditions to those of Korolov $et$ $al.$ \cite{Kor20}. The measurement data and the zero-dimensional simulation results are given in figure \ref{fig:a4_Korolov}$(b)$ for a variation of the nitrogen percentage in the helium background gas. The measured density profile of below $0.5$\:\% $\mathrm{N_2}$ is captured by the model, however the experimentally predicted decay at larger percentage values is not observed in the simulations. A likely reason for this discrepancy is the presence of a solid wall in the plasma chamber, that is absent in the effluent region. The disappearance of the wall induces deviations in the presumed transport (e.g., diffusion and flow) properties, which are more definitive on the effluent concentration due to the vanishing electron impact reactions. Additionally, the quenching of $\mathrm{N(^2 D)}$ at the wall plays a relatively more important role in the $\mathrm{N(^4 S)}$ production with an increasing nitrogen content in the background gas. The difference of transport properties and the lack of the wall quenching in the effluent might lead to the decay of the measured nitrogen atom density for the plasma at a large nitrogen percentage.

\subsection{Characterization of nitric oxide} 
\label{sec:NOproduction}

\begin{figure}[tp]
        \centering
	\subfigure{\includegraphics[scale=0.25,clip=true,trim=0cm 0cm 0cm 0cm,]{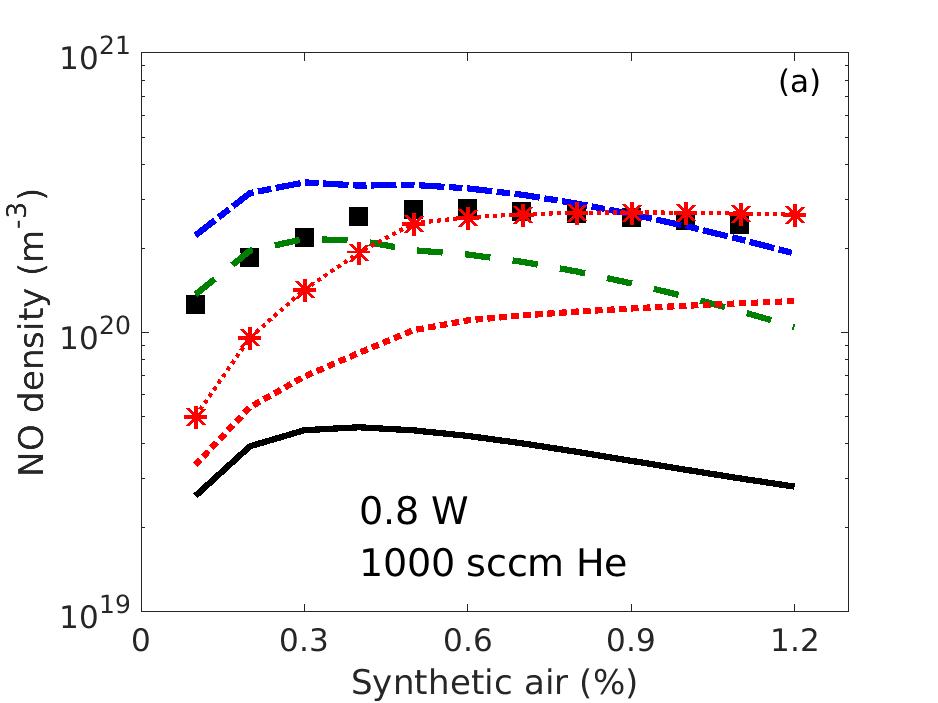}} \\
	\subfigure{\includegraphics[scale=0.25,clip=true,trim=0cm 0cm 0cm 0cm,]{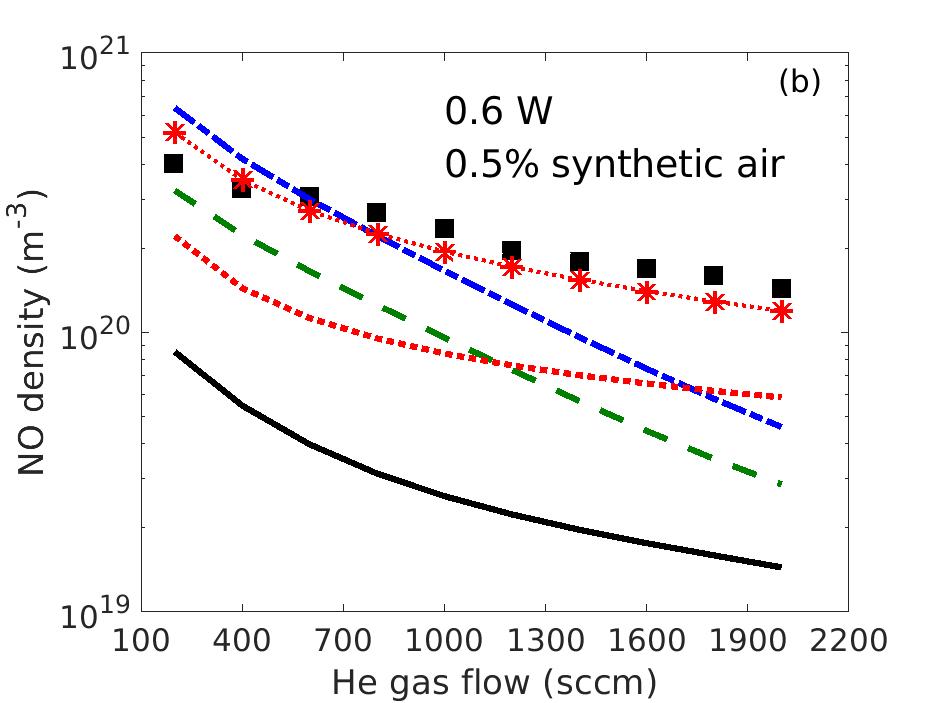}} \\
	\subfigure{\includegraphics[scale=0.25,clip=true,trim=0cm 0cm 0cm 0cm,]{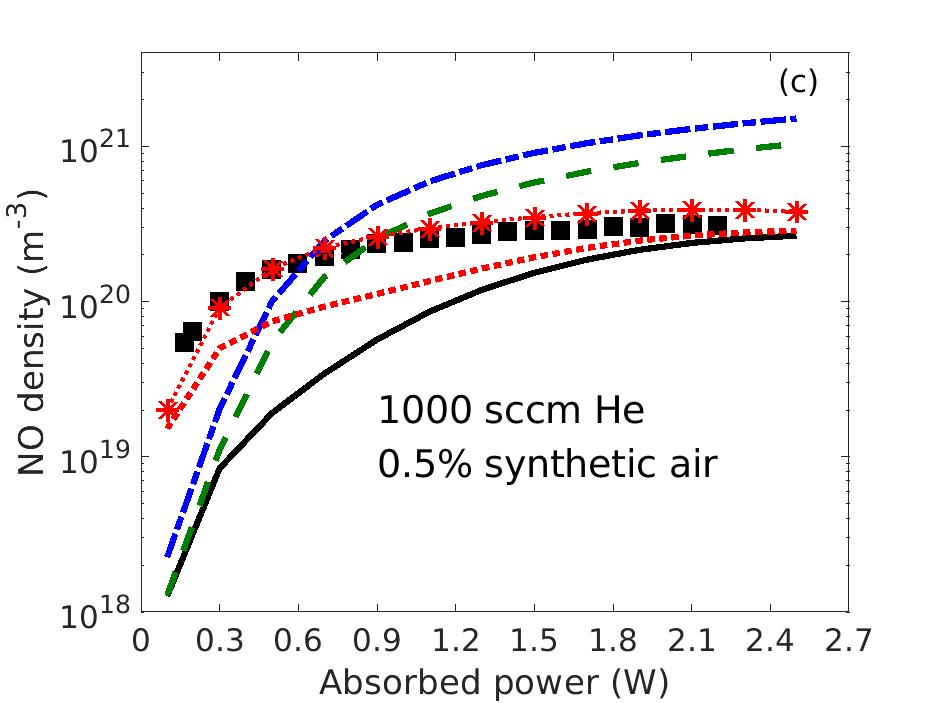}}
	\caption{The absolute $\mathrm{NO}$ density measurements (\textcolor{black}{$\blacksquare$}) at the COST-Jet nozzle by Preissing $et$ $al.$ \cite{Pre20} together with the zero-dimensional simulation results with respect to $(a)$ the variation of synthetic air mixture ratio ($0.8$ W and 1000 sccm He), $(b)$ the variation of He gas flow ($0.6$ W and 0.5\:\% synthetic air mixture) and $(c)$ the variation of absorbed power (1000 sccm He and $0.5$\:\% synthetic air). The solid lines (\textcolor{black}{\full}) represent the simulation results with the reference chemical kinetics. The dashed lines (\textcolor{OliveGreen}{\dashed}) denote the simulation results with a rate coefficient of $1 \times 10^{-17}$ $ \mathrm{m^3/s}$ for reaction (\ref{eqn:N2v13O3p_NON4s}) instead of the reference value $1 \times 10^{-19}$ $ \mathrm{m^3/s}$ in set (1). The dashed-dotted lines (\textcolor{blue}{\chain}) denote those with the same rate coefficient $1 \times 10^{-17}$ $ \mathrm{m^3/s}$ as well as an additional wall reaction $\rm N(^4S) + wall \rightarrow NO$ with a probability of $1$ in set (2). The dotted lines (\textcolor{red}{\dotted}) show the simulation results with a rate coefficient value of $7 \times 10^{-15}$ $ \mathrm{m^3/s}$ for reaction (\ref{eqn:N2aO3p_NON2d}) instead of the reference value $7 \times 10^{-18}$ $ \mathrm{m^3/s}$ in set (3). The star lines (\textcolor{red}{$\ast\!\cdot\!\cdot\!\ast\!\cdot\!\cdot\!\ast$}) show those with the same rate coefficient $7 \times 10^{-15}$ $ \mathrm{m^3/s}$ as well as an additional reaction (\ref{eqn:N2bO3p_NON2d}) with a rate coefficient value of $7 \times 10^{-15}$ $ \mathrm{m^3/s}$ in set (4). 
}\label{fig:B2_air_flow_power}
\end{figure}

The absolute $\mathrm{NO}$ concentration produced by a He/$\mathrm{N_2}$/$\mathrm{O_2}$ COST-Jet is recently measured by Preissing $et$ $al.$ \cite{Pre20}. A plasma volume of $1 \times 1 \times 30$ mm$^3$ is fed with a synthetic air mixture in a helium background at a pressure of $98000$ Pa for an estimated gas temperature of $345$ K. The measurement data at the beginning of the effluent (at 0.5 mm away from the jet nozzle) and the zero-dimensional simulation results are shown in figure \ref{fig:B2_air_flow_power}: $(a)$ for a variation of the synthetic air mixture ratio at a helium mass flow rate of $1000$ sccm and at an absorbed power of $0.8$ W, $(b)$ for a variation of the helium gas flow diluted with 0.5\:\% synthetic air mixture at an absorbed power of $0.6$ W and $(c)$ for a variation of the absorbed power at $1000$ sccm helium mixed with $0.5$\:\% synthetic air. The simulations are presented for four more distinct sets of NO formation mechanisms in addition to the reference chemical kinetics (see section \ref{sec:Chem_kin}):
\begin{enumerate}
    \item[(1)] A rate coefficient value of $1 \times 10^{-17}$ $ \mathrm{m^3/s}$ \cite{Gor95} is used for the reaction 
    \begin{equation}
    \label{eqn:N2v13O3p_NON4s}
    {\rm N}_2(v \geqslant 13) + {\rm O(^3P)} \rightarrow {\rm NO} + {\rm N(^4S)}
    \end{equation}
    instead of $1 \times 10^{-19}$ $ \mathrm{m^3/s}$ \cite{Gue95, Gue97, Pin09, Gue19_rev}.
    \item[(2)] A rate coefficient value of $1 \times 10^{-17}$ $ \mathrm{m^3/s}$ is used for reaction (\ref{eqn:N2v13O3p_NON4s}) with an additional NO wall formation $\rm N(^4S) + wall \rightarrow NO$ \cite{Mar10,Gor96} defined by a maximum probability of 1.
    \item[(3)] A hypothetical rate coefficient value of $7 \times 10^{-15}$ $ \mathrm{m^3/s}$ is assigned for the reactive quenching  
    \begin{equation}
    \label{eqn:N2aO3p_NON2d}
    \mathrm{N_2(A^3 \Sigma)} + \mathrm{O(^3 P)} \rightarrow \mathrm{NO} + \mathrm{N(^2D)}
    \end{equation}
    instead of $7 \times 10^{-18}$  $ \mathrm{m^3/s}$ \cite{CapBook, Gue19_rev, Kos92}, in order to imitate the role of $\mathrm{N_2(A^3 \Sigma}, v>0)$ \cite{Pip81,Pip82,Tho96,Yon19,Ben97,Dil99} and of multiple higher $ \mathrm{N_2} $ electronically excited states \cite{Shk14}.
    \item[(4)] A hypothetical rate coefficient value of $7 \times 10^{-15}$ $ \mathrm{m^3/s}$ is assigned for reaction (\ref{eqn:N2aO3p_NON2d}) as well as an additional reactive quenching
    \begin{equation}
    \label{eqn:N2bO3p_NON2d}
    \mathrm{N_2(B ^3\Pi)} + \mathrm{O(^3 P)} \rightarrow \mathrm{NO} + \mathrm{N(^2D)}
    \end{equation}
    with an identical value of $7 \times 10^{-15}$ $ \mathrm{m^3/s}$ to estimate the influence of $\mathrm{N_2(B ^3\Pi}, v \geqslant 0)$ \cite{Ada15,Ver20_2,Pip92} and of multiple higher $ \mathrm{N_2} $ electronically excited states \cite{Shk14}.
\end{enumerate}

Considerably similar NO density trends are obtained by the measurements and simulations in figure \ref{fig:B2_air_flow_power} for a variation of $(a)$ the synthetic air mixture ratio, $(b)$ the helium gas flow and $(c)$ the absorbed power. Accordingly, a maximal net NO production is achieved by 0.5\:\% synthetic air mixture, a low helium gas flow rate and a high absorbed power, while the NO concentration saturates at around $2$ W for a variation of the power. However, the concentration values are under-estimated by the simulations with the reference chemical kinetics (solid lines) relative to the experimental data (similarly observed in \cite{Shk14}, see section \ref{sec:Chem_kin}). The amount of under-estimation is mainly correlated with the absorbed power (i.e., the amount is enhanced with decreasing power). The introduction of a rate coefficient value of $1 \times 10^{-17}$ $ \mathrm{m^3/s}$ for reaction (\ref{eqn:N2v13O3p_NON4s}) in set (1) (dashed lines) reduces the difference between the measurements and the simulations, whereas an additional consideration of the NO wall formation in set (2) (dashed-dotted lines) further improves the agreement between them. A much better agreement is achieved by the use of a rate coefficient value of $7 \times 10^{-15}$ $ \mathrm{m^3/s}$ for reaction (\ref{eqn:N2aO3p_NON2d}) in set (3) (dotted lines), while virtually identical NO density calculations to the measurements are obtained by an additional consideration of reaction (\ref{eqn:N2bO3p_NON2d}) in set (4) (star lines). The inclusion of state-by-state vibrational levels $\mathrm{N_2(A^3 \Sigma, B ^3\Pi};\:v>0)$ and of multiple $\mathrm{N_2}$ electronically excited states would better address the resolution, however it is out of the scope in the current study due to lack of data. A thorough analysis and quantification of all these rate coefficients are the prerequisites for a more accurate NO resolution in the simulations.

\begin{figure}[tp]
        \centering
	\subfigure{\includegraphics[scale=0.25,clip=true,trim=0cm 0cm 0cm 0cm,]{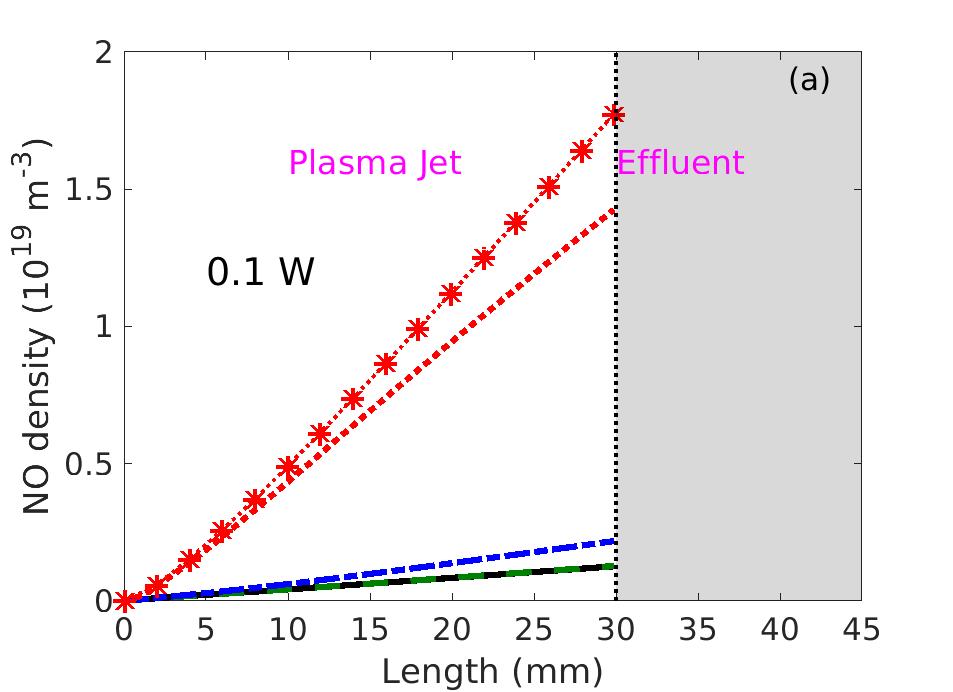}} \\
	\subfigure{\includegraphics[scale=0.25,clip=true,trim=0cm 0cm 0cm 0cm,]{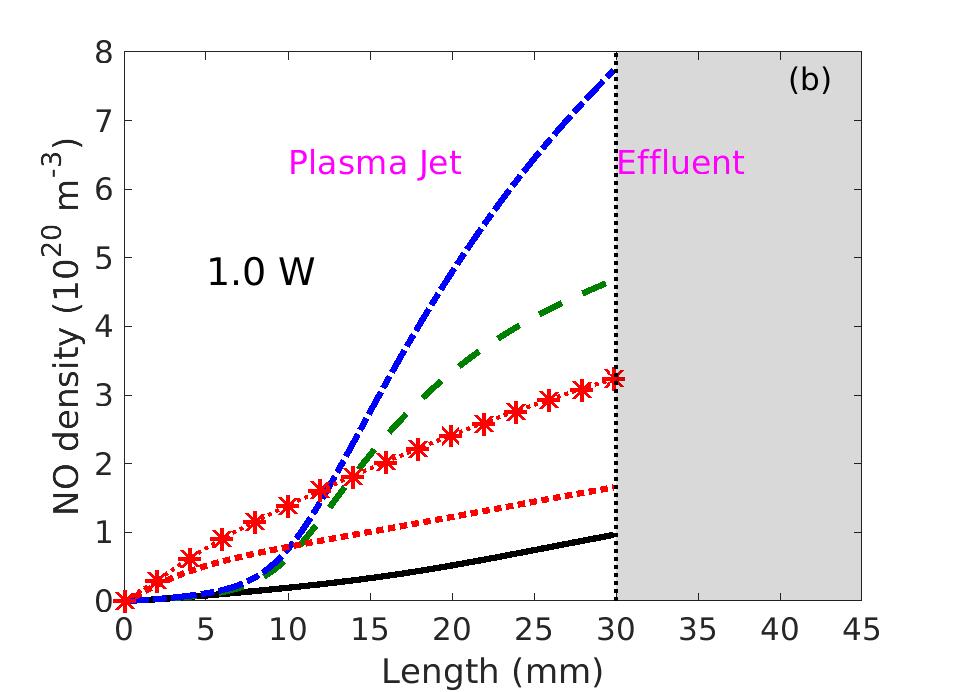}} \\
	\subfigure{\includegraphics[scale=0.25,clip=true,trim=0cm 0cm 0cm 0cm,]{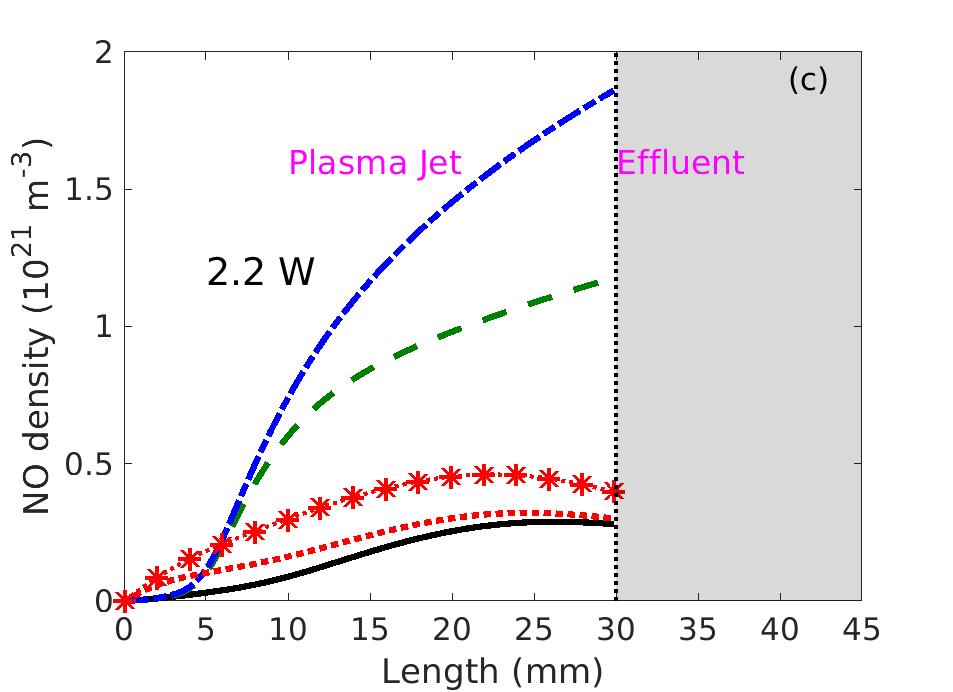}}
	\caption{The pseudo-one-dimensional simulation results of $\mathrm{NO}$ density in the COST-Jet fed with 1000 sccm He + $0.5$\:\% synthetic air mixture at absorbed power values of $(a)$ 0.1 W, $(b)$ 1.0 W and  $(c)$ 2.2 W. The solid lines (\textcolor{black}{\full}) represent the simulation results with the reference chemical kinetics. The dashed lines (\textcolor{OliveGreen}{\dashed}) denote the simulation results with a rate coefficient of $1 \times 10^{-17}$ $ \mathrm{m^3/s}$ for reaction (\ref{eqn:N2v13O3p_NON4s}) instead of the reference value $1 \times 10^{-19}$ $ \mathrm{m^3/s}$ in set (1). The dashed-dotted lines (\textcolor{blue}{\chain}) denote those with the same rate coefficient $1 \times 10^{-17}$ $ \mathrm{m^3/s}$ as well as an additional wall reaction $\rm N(^4S) + wall \rightarrow NO$ with a probability of $1$ in set (2). The dotted lines (\textcolor{red}{\dotted}) show the simulation results with a rate coefficient value of $7 \times 10^{-15}$ $ \mathrm{m^3/s}$ for reaction (\ref{eqn:N2aO3p_NON2d}) instead of the reference value $7 \times 10^{-18}$ $ \mathrm{m^3/s}$ in set (3). The star lines (\textcolor{red}{$\ast\!\cdot\!\cdot\!\ast\!\cdot\!\cdot\!\ast$}) show those with the same rate coefficient $7 \times 10^{-15}$ $ \mathrm{m^3/s}$ as well as an additional reaction (\ref{eqn:N2bO3p_NON2d}) with a rate coefficient value of $7 \times 10^{-15}$ $ \mathrm{m^3/s}$ in set (4). 
}\label{fig:B2_power_plug}
\end{figure}

The pseudo-one-dimensional simulation results corresponding to the COST-Jet measurements by Presissing $et$ $al.$ \cite{Pre20}  are shown in figure \ref{fig:B2_power_plug} for a helium mass flow rate of $1000$ sccm with $0.5$\:\% synthetic air mixture at absorbed power values of $(a)$ 0.1 W, $(b)$ 1.0 W and $(c)$ 2.2 W. The density profiles for the above-mentioned distinct sets of NO formation mechanisms are individually shown. For each set of formation mechanisms, a close similarity is observed between the $\mathrm{NO}$ density at the jet nozzle and the corresponding volume-averaged value (not shown here) calculated by the zero-dimensional simulations. This confirms that the difference between the measurement data and the zero-dimensional simulation results in figure \ref{fig:B2_air_flow_power} is not generated by the lack of spatial resolution. The role of the power in the NO density is additionally verified by the pseudo-one-dimensional simulation results. An approximately linear density growth in gas flow direction is obtained at an absorbed power of $0.1$ W in figure \ref{fig:B2_power_plug}$(a)$. The steepest growths are produced by set (3) (dotted lines) and set (4) (star lines),  while much flatter growth rates are observed among the remaining chemical sets at this low power operation. The influence of both set (1) and (2) increases with the increasing power, leading to a parabolic NO growth in figures \ref{fig:B2_power_plug}$(b)$ and \ref{fig:B2_power_plug}$(c)$. On the other hand, the reference chemical kinetics (solid lines), set (3) (dotted lines) and set (4) (star lines) induce relatively lower density values at larger power. They impose a saturation at an absorbed power of $2.2$ W in figure \ref{fig:B2_power_plug}$(c)$.

\begin{figure}[tp]
        \centering
	\subfigure{\includegraphics[scale=0.25,clip=true,trim=0cm 0cm 0cm 0cm,]{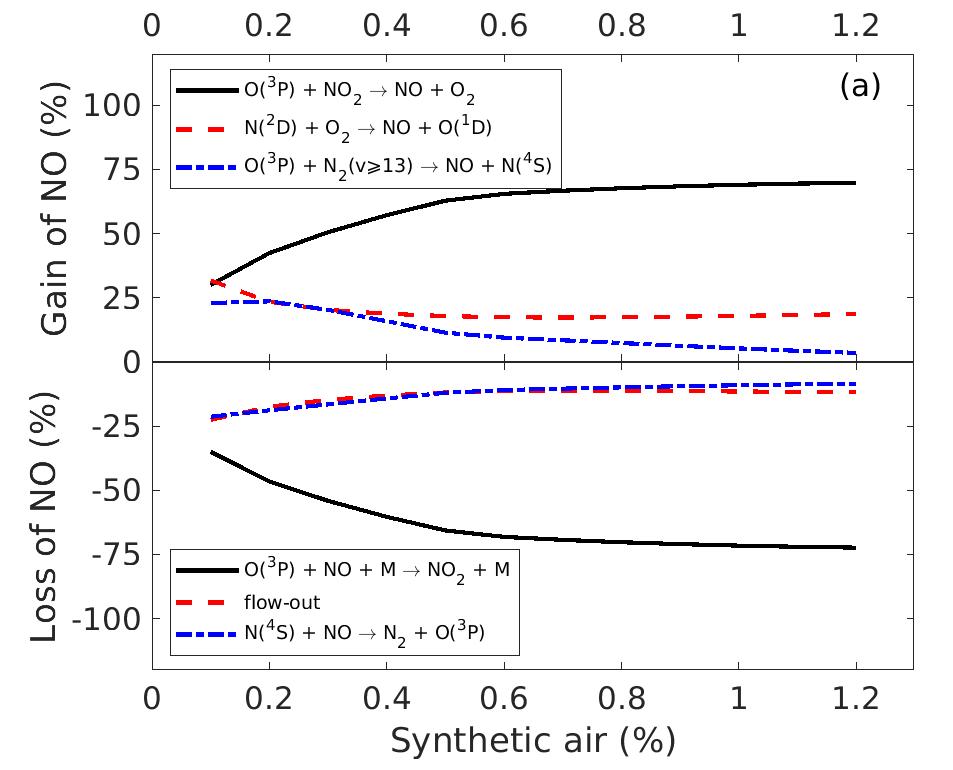}} \\
	\subfigure{\includegraphics[scale=0.25,clip=true,trim=0cm 0cm 0cm 0cm,]{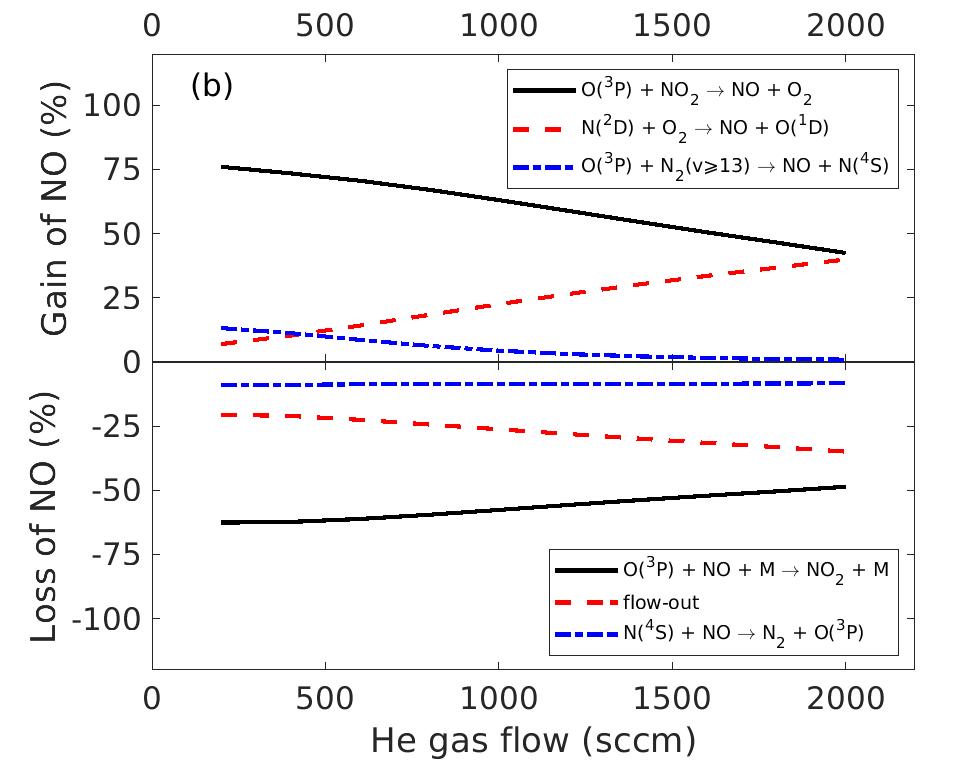}} \\
	\subfigure{\includegraphics[scale=0.25,clip=true,trim=0cm 0cm 0cm 0cm,]{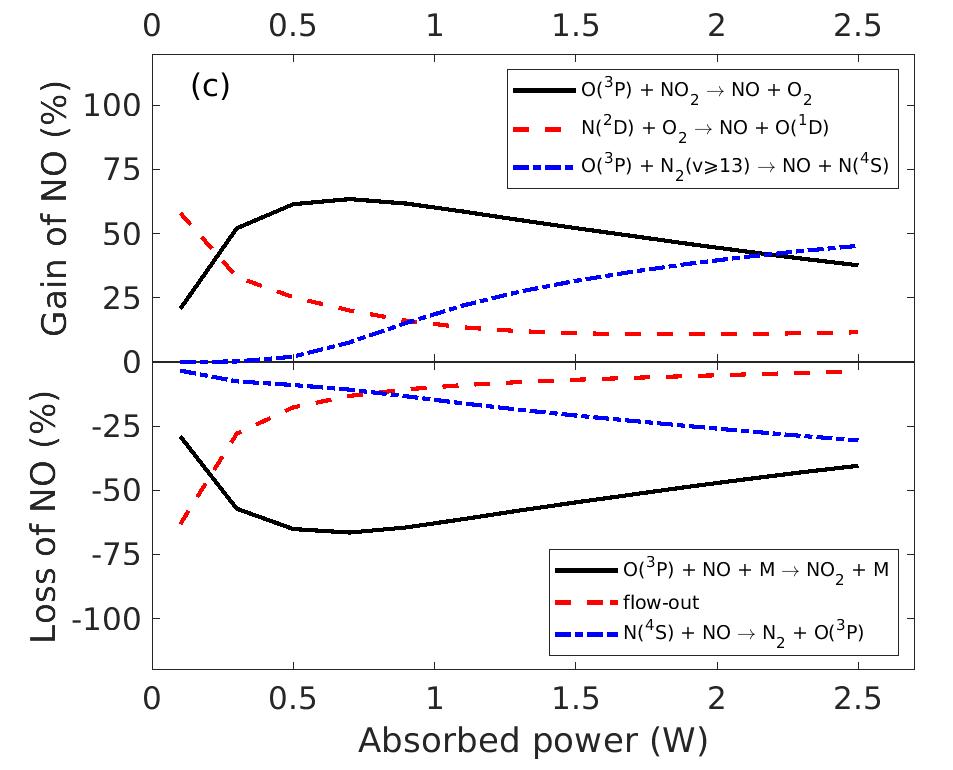}}
	\caption{The contributions of the dominant $\mathrm{NO}$ gain and loss channels in the COST-Jet for a variation of $(a)$ the synthetic air mixture ($0.8$ W and 1000 sccm He), $(b)$ the He gas flow ($0.6$ W and 0.5\:\% synthetic air mixture) and $(c)$ the absorbed power (1000 sccm He and $0.5$\:\% synthetic air mixture). These values are calculated for the zero-dimensional simulation results using the reference chemical kinetics in figure \ref{fig:B2_air_flow_power}. 
}\label{fig:B2_gainLoss}
\end{figure}

The contributions of the dominant NO gain and loss channels associated with the
aforementioned COST-Jet measurements \cite{Pre20} are provided in figure \ref{fig:B2_gainLoss} for a variation of $(a)$ the synthetic air mixture ratio, $(b)$ the helium gas flow and $(c)$ the absorbed power. The reaction rates are calculated by the zero-dimensional model using the reference chemical kinetics. The NO production rate is mostly manipulated by the reactions
\begin{equation}
\label{eqn:O3pNO2_NOO2}
\mathrm{O(^3 P)} + \mathrm{NO_2} \rightarrow \mathrm{NO} + \mathrm{O_2},
\end{equation}
\begin{equation}
\label{eqn:N2dO2_NOO1d}
\mathrm{N(^2 D)} + \mathrm{O_2} \rightarrow \mathrm{NO} + \mathrm{O(^1 D)},
\end{equation}
both in agreement with earlier studies \cite{Sun19,Mur14,Mal16,Gae14_2}. The former reaction generally plays a relatively more crucial role, while the latter gains significance at high helium gas flow and low absorbed power. In comparison with these, the contribution of reaction (\ref{eqn:N2v13O3p_NON4s}) is likewise substantial at low synthetic air percentage and low helium gas flow. Furthermore, it solely dominates the production rate at high absorbed power. Although the source channel via reaction (\ref{eqn:N2aO3p_NON2d}) contributes less than $10$\:\%, it forms a significant portion of the production within set (3) (see figure \ref{fig:B2_air_flow_power}). Additionally, this channel indirectly enhances the influence of reaction (\ref{eqn:N2dO2_NOO1d}) due to its essence on the first stage $\mathrm{N(^2 D)}$ generation \cite{Ion06}. The NO loss channels are mainly governed by its flow-out rate and the reactions
\begin{equation}
\label{eqn:O3PNOM_NO2M}
\mathrm{O(^3 P)} + \mathrm{NO} + \mathrm{M} \rightarrow \mathrm{NO_2} + \mathrm{M},
\end{equation}
\begin{equation}
\label{eqn:N4SNO_N2O3P}
\mathrm{N(^4 S)} + \mathrm{NO} \rightarrow \mathrm{N_2} + \mathrm{O(^3 P)}, 
\end{equation}
both confirmed via observations \cite{Sun19,Mur14,Mal16,Gae14,Gae14_2}. The loss percentages, except that of reaction (\ref{eqn:O3PNOM_NO2M}), are virtually invariant for a variation of the synthetic air mixture and the helium gas flow. The strong influence of reaction (\ref{eqn:O3PNOM_NO2M}) is minimized by the other two dominant loss channels at either low or high absorbed power.

\begin{figure}[t!]
	\begin{minipage}[b]{0.5\textwidth}
	\includegraphics[width=0.95\textwidth]{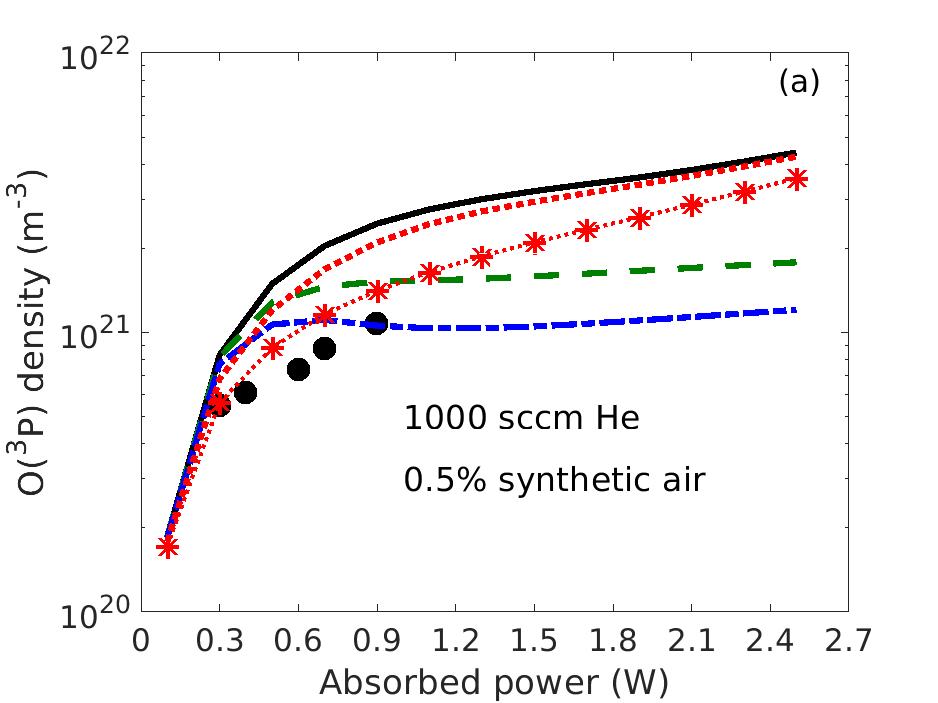} \\ 
\end{minipage}
\begin{minipage}[b]{0.5\textwidth}
    \includegraphics[width=0.95\textwidth]{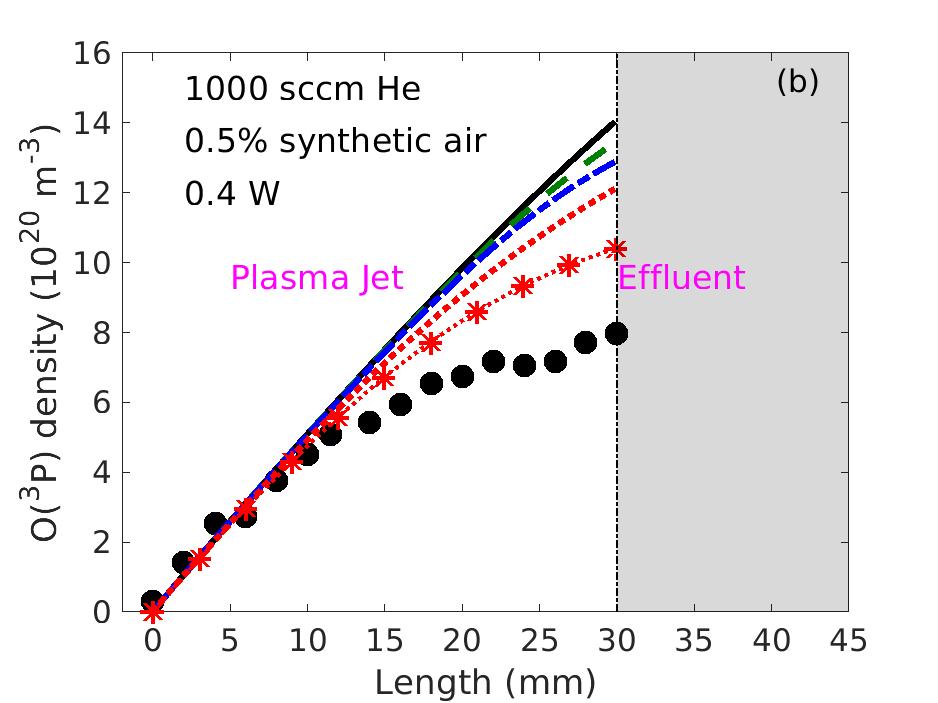} \\ 
\end{minipage}
\caption{$(a)$ The $\mathrm{O(^3P)}$ density measurements (\tikzcircle[black, fill=black]{2.0pt}) at the jet nozzle and the zero-dimensional simulation results for a variation of the absorbed power. $(b)$ the spatially resolved measurements in gas flow direction and the pseudo-one-dimensional model calculations at an absorbed power of 0.4 W. The COST-Jet is fed with 1000 sccm He + $0.5$\:\% synthetic air mixture. The solid lines (\textcolor{black}{\full}) represent the simulation results with the reference chemical kinetics. The dashed lines (\textcolor{OliveGreen}{\dashed}) denote the simulation results with a rate coefficient of $1 \times 10^{-17}$ $ \mathrm{m^3/s}$ for reaction (\ref{eqn:N2v13O3p_NON4s}) instead of the reference value $1 \times 10^{-19}$ $ \mathrm{m^3/s}$ in set (1). The dashed-dotted lines (\textcolor{blue}{\chain}) denote those with the same rate coefficient $1 \times 10^{-17}$ $ \mathrm{m^3/s}$ as well as an additional wall reaction $\rm N(^4S) + wall \rightarrow NO$ with a probability of $1$ in set (2). The dotted lines (\textcolor{red}{\dotted}) show the simulation results with a rate coefficient value of $7 \times 10^{-15}$ $ \mathrm{m^3/s}$ for reaction (\ref{eqn:N2aO3p_NON2d}) instead of the reference value $7 \times 10^{-18}$ $ \mathrm{m^3/s}$ in set (3). The star lines (\textcolor{red}{$\ast\!\cdot\!\cdot\!\ast\!\cdot\!\cdot\!\ast$}) show those with the same rate coefficient $7 \times 10^{-15}$ $ \mathrm{m^3/s}$ as well as an additional reaction (\ref{eqn:N2bO3p_NON2d}) with a rate coefficient value of $7 \times 10^{-15}$ $ \mathrm{m^3/s}$ in set (4).}\label{fig:B2_OinJet}
\end{figure}

The $\mathrm{O(^3 P)}$ density measurements in the COST-Jet fed with $1000$ sccm He + 0.5\:\% synthetic air mixture are shown in figure \ref{fig:B2_OinJet}$(a)$: data at the jet nozzle for a variation of the absorbed power, \ref{fig:B2_OinJet}$(b)$: one dimensional data in gas flow direction at an absorbed power of 0.4 W. The experimental details to obtain these measurements are summarized in an investigation of a helium oxygen mixture by Steuer $et$ $al.$ \cite{Ste21}. The zero-dimensional and the pseudo-one-dimensional $\mathrm{O(^3 P)}$ simulation results of the aforementioned distinct sets of NO formation mechanisms are presented, accordingly. A pressure of 98000 Pa and a gas temperature of 345 K is used in the simulation following the sameness of the operation conditions as those of Preissing $et$ $al.$ \cite{Pre20}. Similar values and trends are observed between the measured and simulated $\mathrm{O(^3 P)}$ densities for all the chemical sets, however the closest agreement is obtained via set (4). It should be emphasized that the reactive quenching of the species $ \mathrm{N_2(A^3 \Sigma,B ^3\Pi;}\:v>0) $ and multiple higher $ \mathrm{N_2} $ electronically excited states in set (4) better explains the $\mathrm{O(^3 P)}$ density besides its underlying role in the NO production at low absorbed power (see figure \ref{fig:B2_air_flow_power}).

A well predictive capability of the developed models to the plasma properties in the considered atmospheric-pressure plasma jets is discussed by the above-shown benchmark and characterization results. A computational insight into the detailed vibrational kinetics and the sensitivity analyses is presented for the COST-Jet in the following sections.

\subsection{Population of the vibrationally excited molecules}
\label{sec:VDF}

\begin{figure}[tp]
        \centering
	\subfigure{\includegraphics[scale=0.25,clip=true,trim=0cm 0cm 0cm 0cm,]{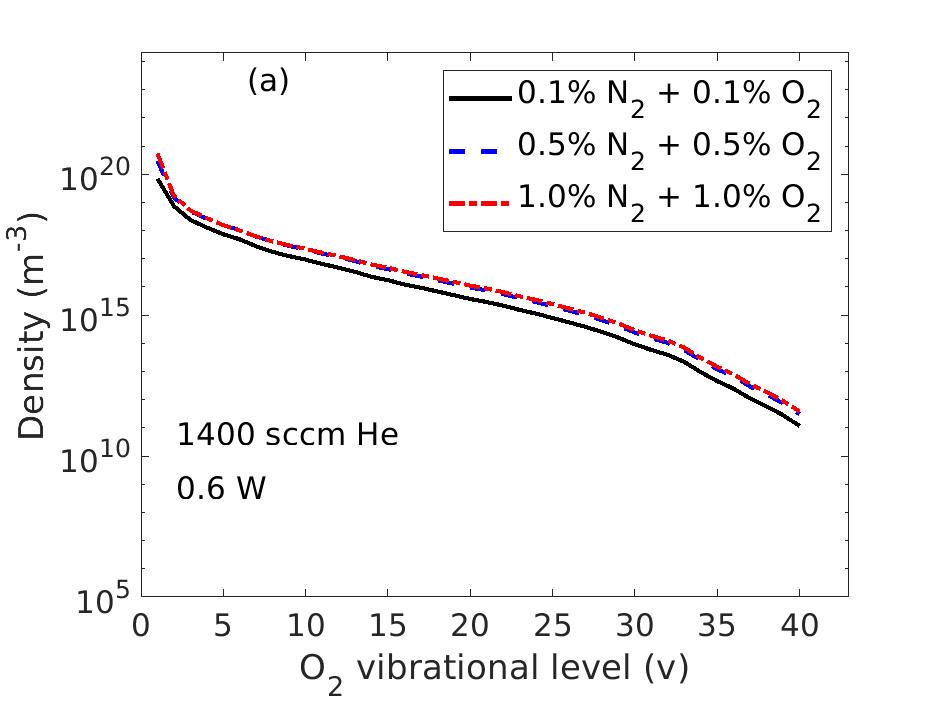}} 
	\subfigure{\includegraphics[scale=0.25,clip=true,trim=0cm 0cm 0cm 0cm,]{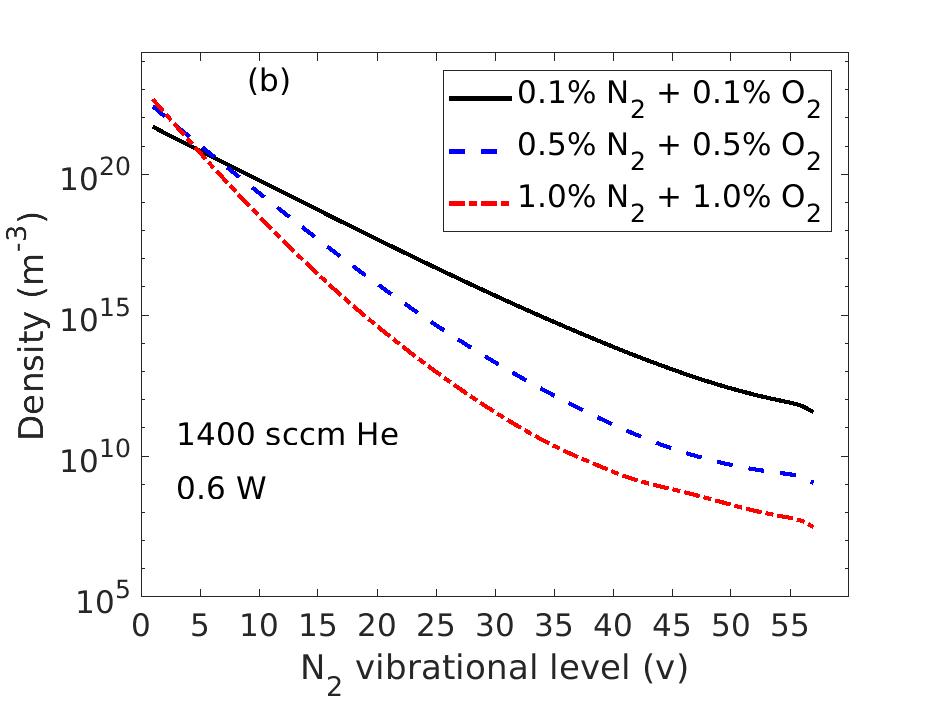}} \\
	\subfigure{\includegraphics[scale=0.25,clip=true,trim=0cm 0cm 0cm 0cm,]{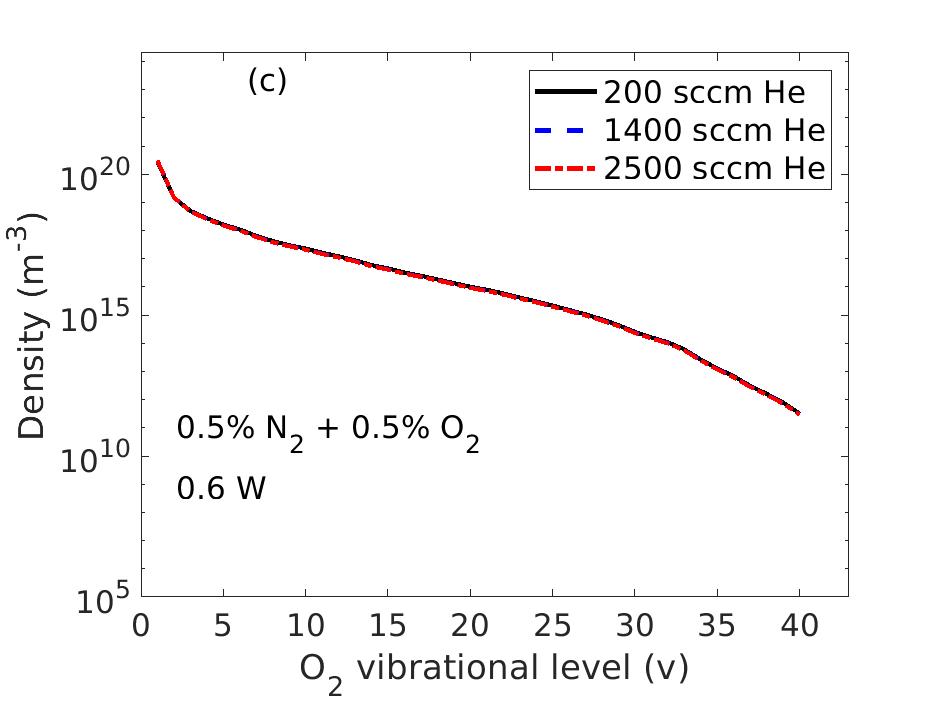}} 
	\subfigure{\includegraphics[scale=0.25,clip=true,trim=0cm 0cm 0cm 0cm,]{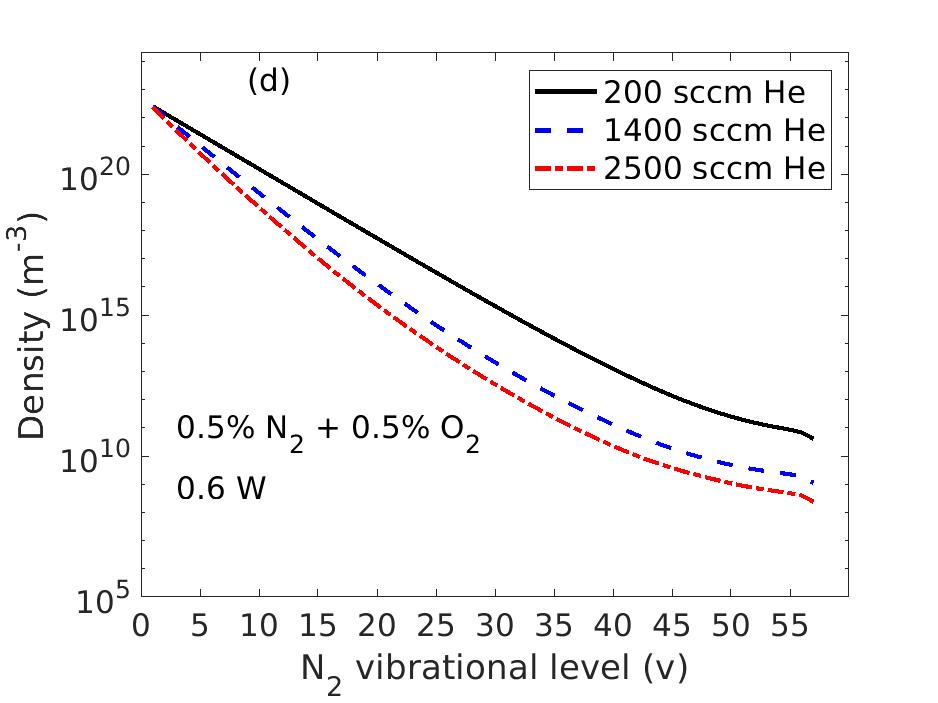}} \\
	\subfigure{\includegraphics[scale=0.25,clip=true,trim=0cm 0cm 0cm 0cm,]{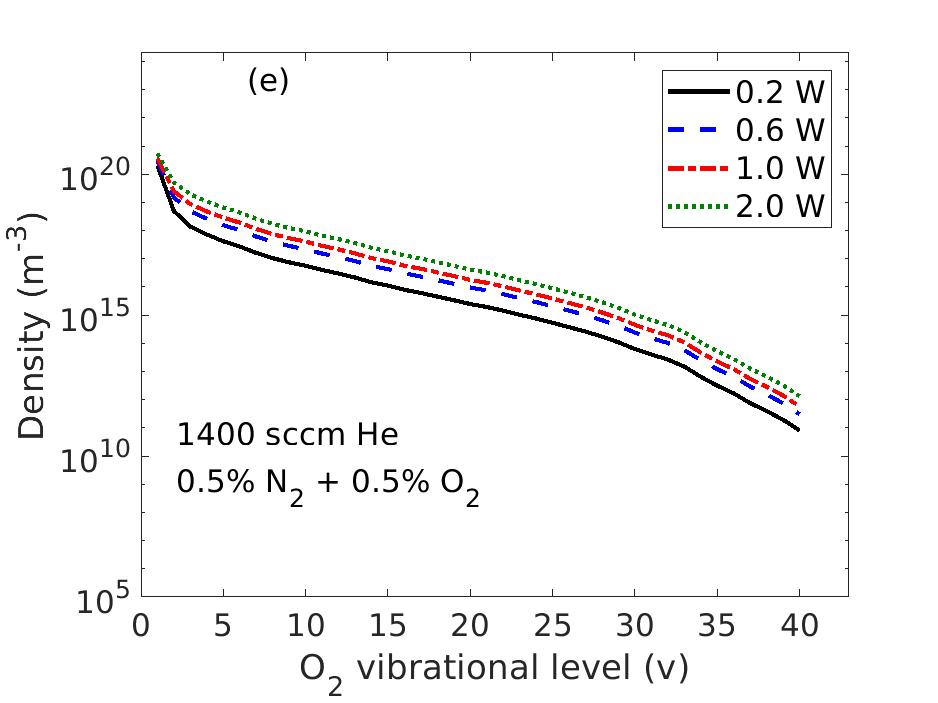}} 
	\subfigure{\includegraphics[scale=0.25,clip=true,trim=0cm 0cm 0cm 0cm,]{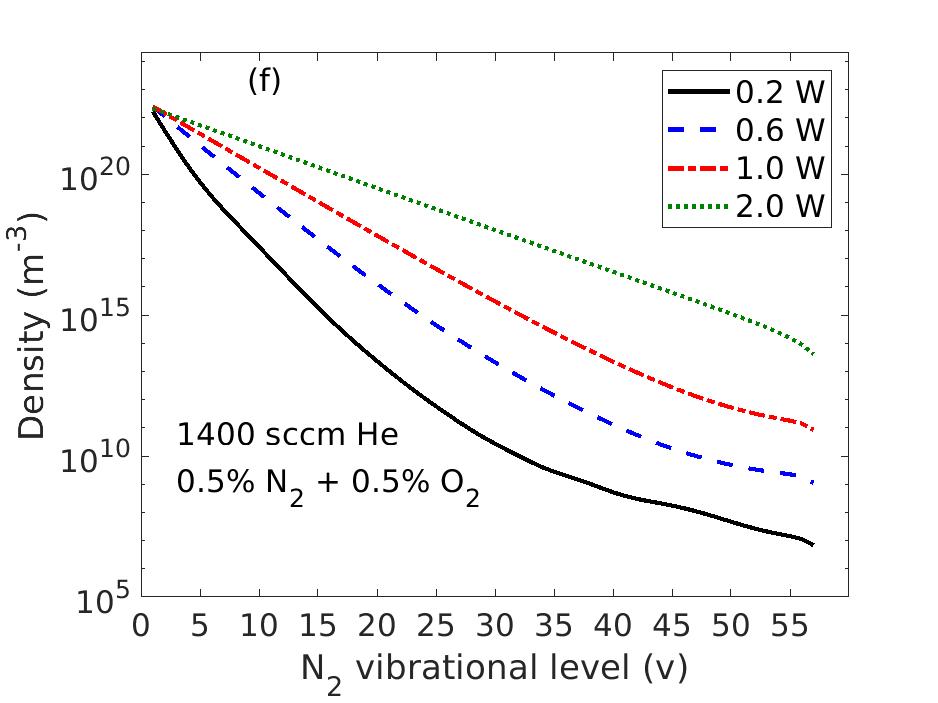}}
	\caption{The zero-dimensional simulation results of the $ \mathrm{O_2} $ and $ \mathrm{N_2} $ VDFs in the COST-Jet for a variation of $(a)-(b)$ the mixture ratio (0.1\:-\:1.0\:\% $ \mathrm{N_2} $ + 0.1\:-\:1.0\:\% $ \mathrm{O_2} $), $(c)-(d)$ the helium flow rate (200\:-\:2500 sccm) and $(e)-(f)$ the absorbed power (0.2\:-\:2.0 W) at a pressure of 101325 Pa and a gas temperature of 345 K. 
}\label{fig:E_VDF_HeN2O2}
\end{figure}

The vibrational distribution functions in a He/$\mathrm{N_2}$/$\mathrm{O_2}$ COST-Jet for a variation of the gas mixture ratio (0.1\:-\:1.0\:\% $ \mathrm{N_2} $ + 0.1\:-\:1.0\:\% $ \mathrm{O_2} $), the helium flow rate (200\:-\:2500 sccm) and the absorbed power (0.2\:-\:2.0 W) are shown in figure \ref{fig:E_VDF_HeN2O2}. A close similarity is obtained between the distribution functions of the gas mixtures He/$ \mathrm{N_2} $/$ \mathrm{O_2} $, He/$ \mathrm{O_2} $ and He/$ \mathrm{N_2} $, therefore only those of He/$ \mathrm{N_2} $/$ \mathrm{O_2} $ are addressed here. Three distinct regions are located in the oxygen distribution function: (1) fast depletion at low, (2) plateau area at intermediate and (3) Boltzmann tail at high vibrational levels, similar to low-pressure oxygen plasma \cite{Kem16,Pie20}. The synergistic role of the whole vibrational chemical kinetics (i.e., e-V, V-V and V-T reactions) is important for the profile at low quantum numbers, whereas the V-T mechanism is more influential on the intermediate and high vibrational levels. A Maxwellian-like distribution function at low and intermediate vibrational quanta is obtained for nitrogen molecules, similar to those at low-pressure \cite{Coc16,Pin16,Gue19_rev}. The tail of the distribution function is slightly elevated except at a high power of $2.0$ W. The vibrational population is initially driven by e-V energy transfer, however the quasi-steady distribution function is mainly shaped by the V-V collisions. The negligible role of the electron-impact vibrational excitation at low quantum numbers is attributed to much higher densities of vibrationally excited nitrogen than that of electron. On the other hand, a more important influence of the e-V mechanisms on high quantum region is observed at lower absorbed power due to the significantly reduced vibrational population. Additionally, the increasing influence of the e-V mechanisms with decreasing power is the main mechanism responsible for the elevated tail at the low power operation (see figure \ref{fig:E_VDF_HeN2O2}$(f)$).

The $ \mathrm{N_2} $ VDF shows a substantial dependency on the operation parameters relative to a weak response of the oxygen vibrational population. A likely reason is that the higher efficiency of nitrogen up-pumping promotes a larger variation with respect to the operation parameters. Furthermore, nitrogen levels store more internal energy with a reinforcement at higher quantum number. Hence, these levels are far more influential on the EEDF as well as on the resultant chemical kinetics \cite{Gue19_rev}. For a high mixture ratio in figure \ref{fig:E_VDF_HeN2O2}$(a)$ and $(b)$, the collisions significantly re-distribute the vibrational energy stored in nitrogen and hence reduce the $ \mathrm{N_2} $ VDF. The re-distributed energy is partially transferred to a slightly enhanced oxygen vibrational population. The role of the gas flow rate in both oxygen and nitrogen vibrational populations in figure \ref{fig:E_VDF_HeN2O2}$(c)$ and $(d)$ is less important than those of the mixture ratio and the absorbed power due to the much less influence of the flow on the electron density and temperature. Both VDFs are enhanced with higher absorbed power shown in figure \ref{fig:E_VDF_HeN2O2}$(e)$ and $(f)$, as a result of the sufficient energy input to the e-V and V-V transfer for climbing the vibrational ladder.

\begin{figure}[tp]
        \centering
	\subfigure{\includegraphics[scale=0.24,clip=true,trim=0cm 0cm 0cm 0cm,]{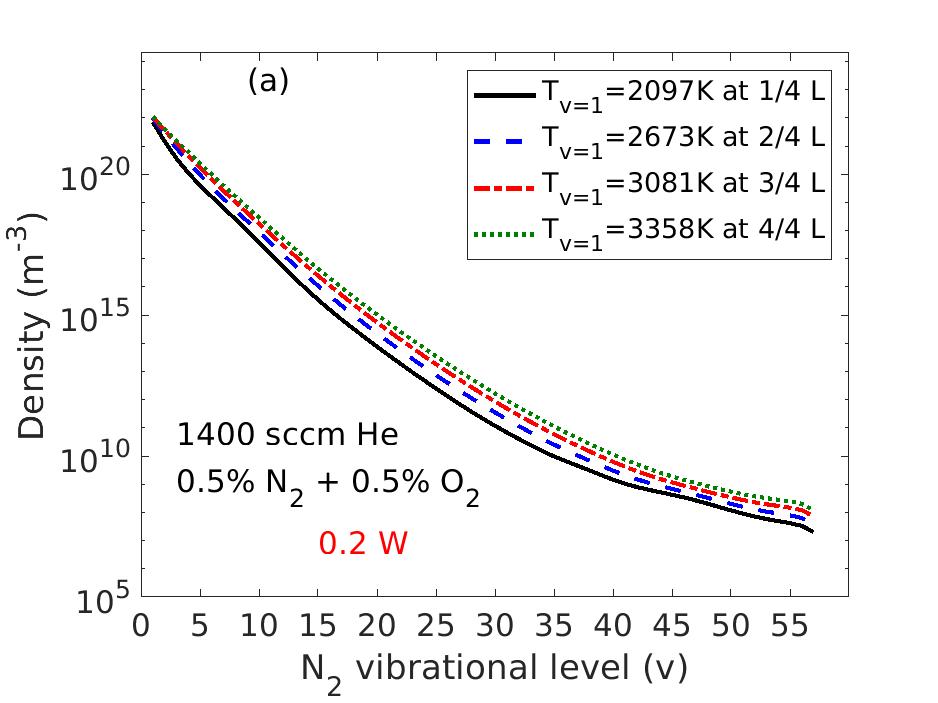}} 
	\subfigure{\includegraphics[scale=0.24,clip=true,trim=0cm 0cm 0cm 0cm,]{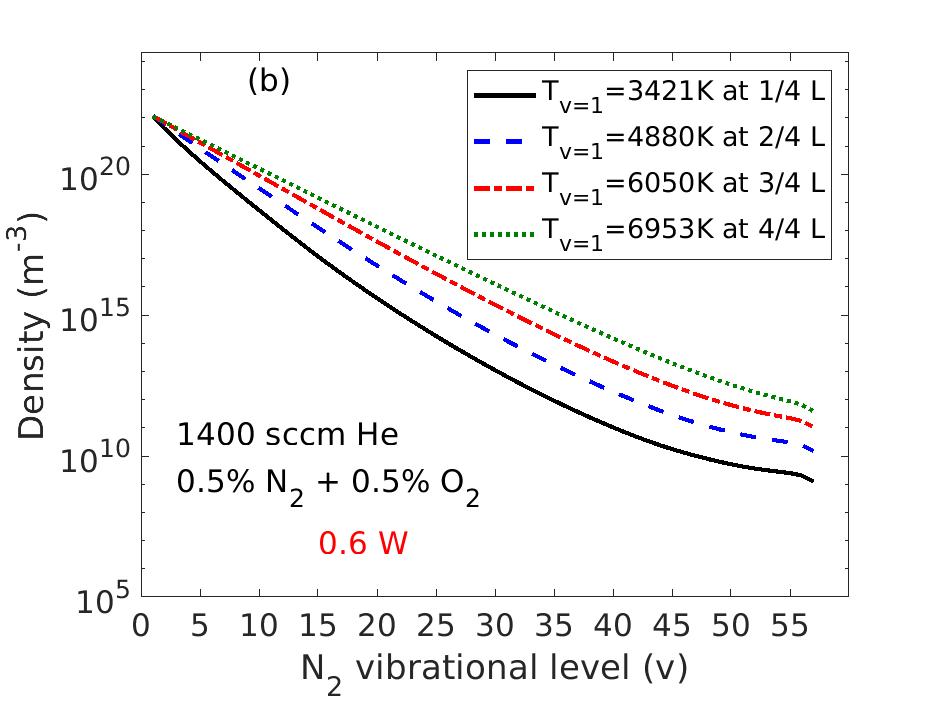}} \\
	\subfigure{\includegraphics[scale=0.24,clip=true,trim=0cm 0cm 0cm 0cm,]{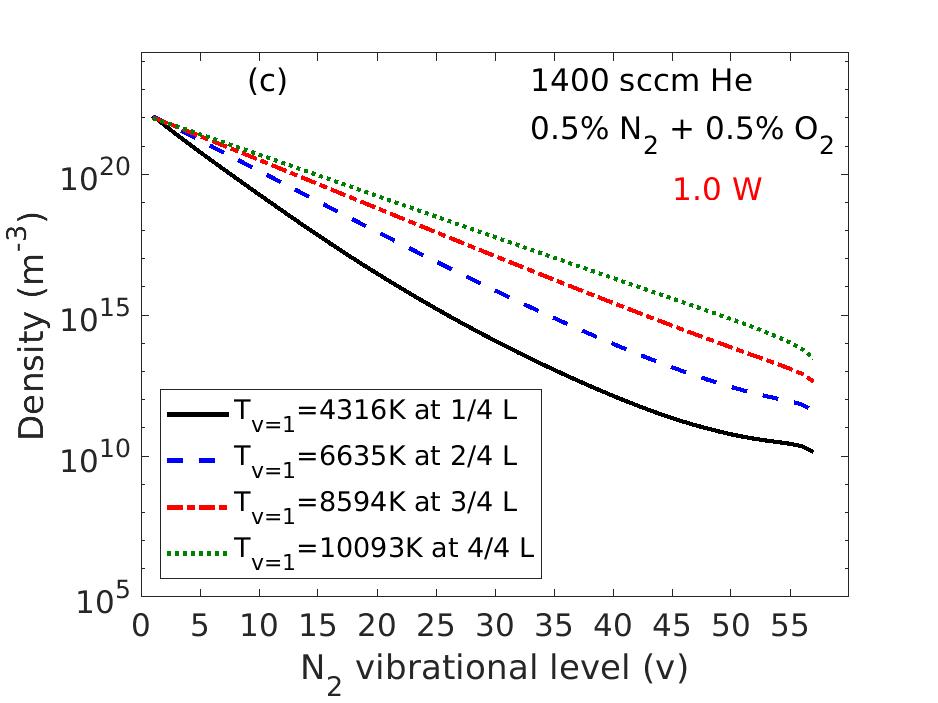}} 
	\subfigure{\includegraphics[scale=0.24,clip=true,trim=0cm 0cm 0cm 0cm,]{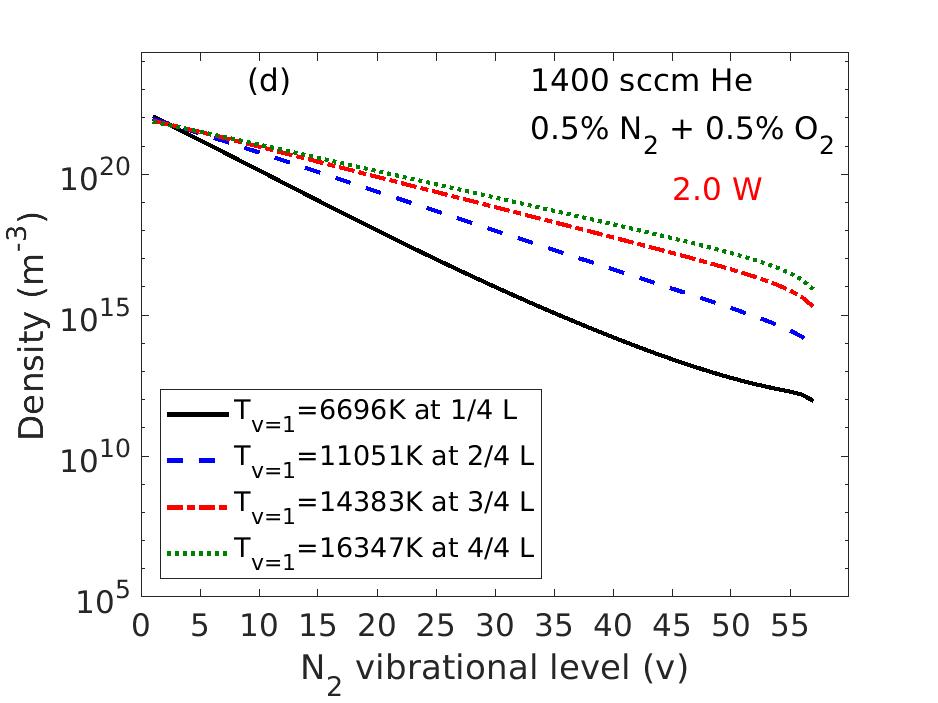}}
	\caption{The spatial evolution of the $ \mathrm{N_2} $ VDFs from entrance (0/4 L) to exit (4/4 L) of the plasma chamber. The COST-Jet is fed with 1400 sccm He + 0.5\:\% $ \mathrm{N_2} $ + 0.5\:\% $ \mathrm{O_2} $ at a pressure of 101325 Pa and a gas temperature of 345 K for absorbed powers of $(a)$ 0.2 W, $(b)$ 0.6 W, $(c)$ 1.0 W and $(d)$ 2.0 W.
}\label{fig:E_VDF_HeN2O2_plug_N2}
\end{figure}

The spatial evolution of the $ \mathrm{N_2} $ VDF in gas flow direction of the He/$ \mathrm{N_2} $/$ \mathrm{O_2}$ COST-Jet is given in figure \ref{fig:E_VDF_HeN2O2_plug_N2} for absorbed powers of $(a)$ 0.2 W, $(b)$ 0.6 W, $(c)$ 1.0 W and $(d)$ 2.0 W. The oxygen vibrational population is already saturated at one fourth of the plasma chamber length (1/4 L) and therefore it is not depicted in the figure. A significant growth of the $\mathrm{N_2}$ VDF is observed in the direction of the flow. It is amplified with an increase of the absorbed power specifically at the tail region. The vibrational temperature $\mathrm{T_{v=1}}$ increases from 2097 K to 3358 K along the electrodes at an absorbed power of 0.2 W, whereas it raises from 6696 K to 16347 K at 2.0 W.

\subsection{Sensitivity analyses}
\label{sec:Sensitivity}

\begin{figure}[t!]
        \centering
	\subfigure{\includegraphics[scale=0.25,clip=true,trim=0cm 0cm 0cm 0cm,]{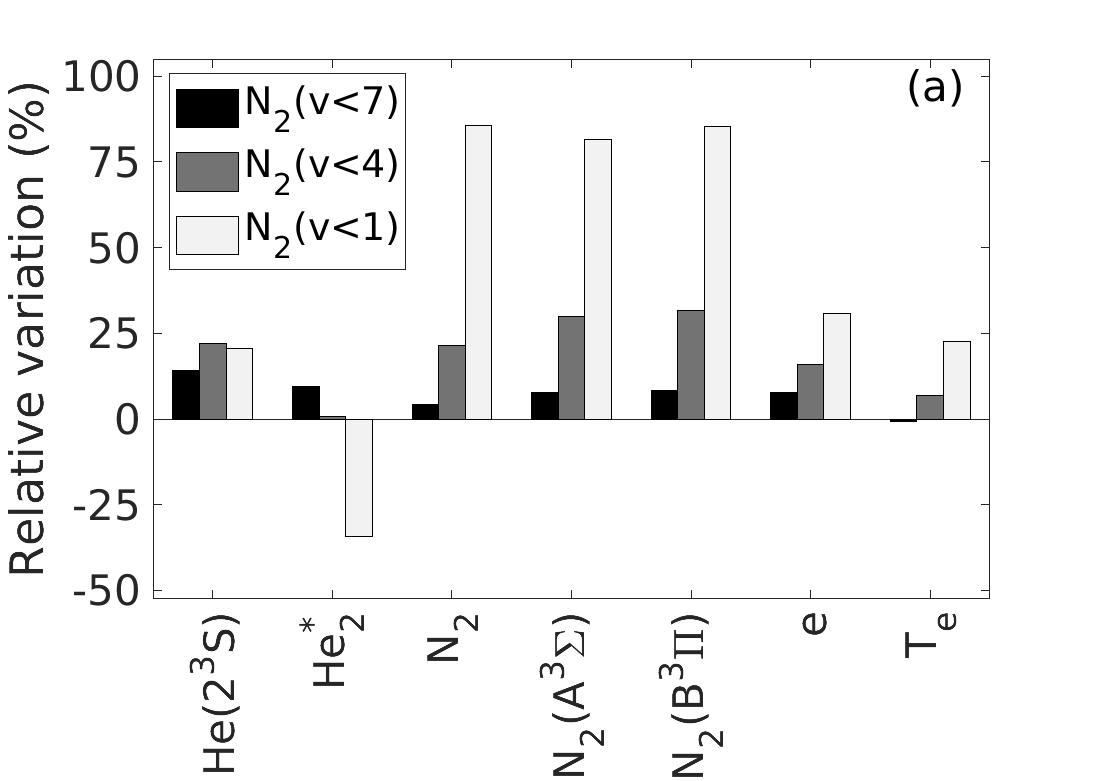}} \\
	\subfigure{\includegraphics[scale=0.25,clip=true,trim=0cm 0cm 0cm 0cm,]{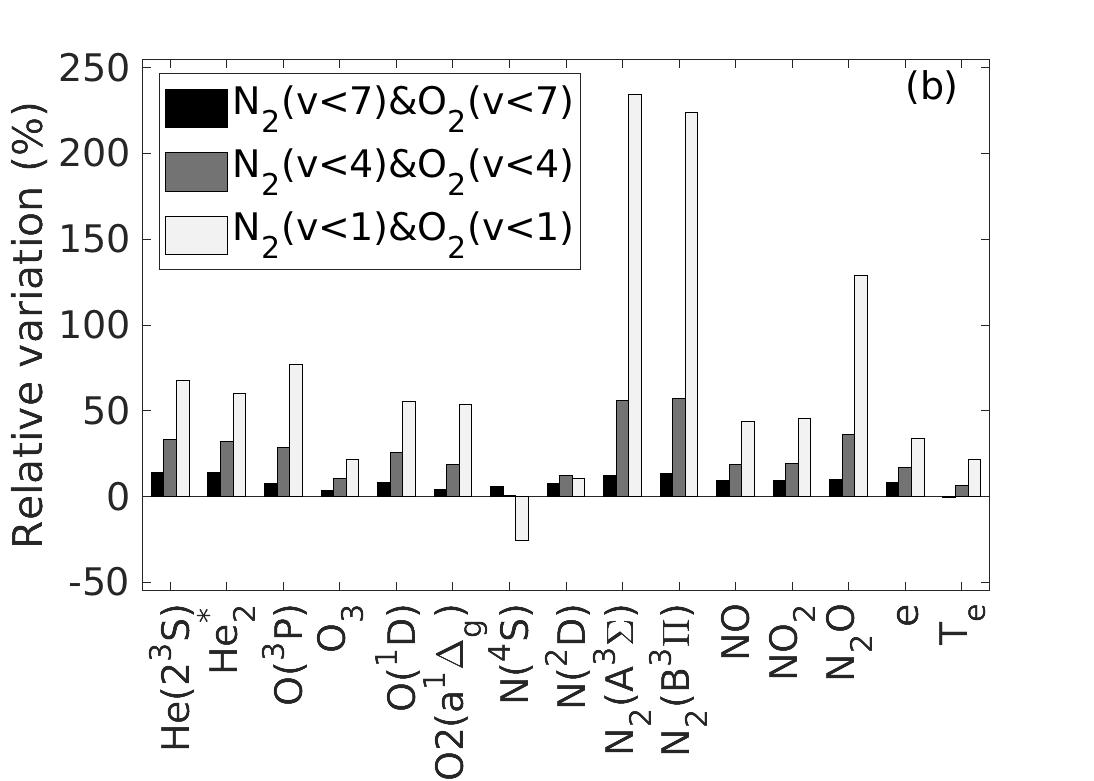}}
	\caption{The variation of the simulated plasma properties at different resolution of VDFs relative to those at a detailed resolution with  N$_2(v<58)$\:\&\:O$_2(v<41)$. The COST-Jet is sustained by an absorbed power of $0.6$ W at a pressure of $101325$ Pa and a gas temperature of $345$ K, and fed with $(a)$ 1400 sccm He + 0.5\:\% $\mathrm{N_2}$ and $(b)$ 1400 sccm He + 0.5\:\% $\mathrm{O_2}$ + 0.5\:\% $\mathrm{N_2}$. 
}\label{fig:VibStates_HeN2O2}
\end{figure}

The variation of the simulated COST-Jet plasma properties for varying numbers of vibrationally excited levels considered in the model is given in figure \ref{fig:VibStates_HeN2O2} relative to those including the whole set, N$_2(v<58)$\:\&\:O$_2(v<41)$. Those of the He/$\mathrm{O_2}$ plasma jet are not shown here, since they are below 10\:\%. A virtually negligible deviation of the simulation results is observed between the cases of N$_2(v<7)$\:\&\:O$_2(v<7)$ and N$_2(v<58)$\:\&\:O$_2(v<41)$. However, the deviation substantially increases with a decreasing number of the vibrationally excited levels in the model. For a feeding gas mixture of He/$\mathrm{N_2}$, the densities of $\mathrm{N_2}$, $\mathrm{N_2(A ^3\Sigma)}$ and $\mathrm{N_2(B ^3\Pi)}$ are altered significantly as shown in figure \ref{fig:VibStates_HeN2O2}$(a)$. The electron and helium metastable densities as well as the electron temperature are varied up to about $35\:\%$ when the vibrational levels are completely ignored. The variation is remarkably larger for a feeding gas mixture of He/$\mathrm{N_2}$/$\mathrm{O_2}$, up to about $234\:\%$ presented in figure \ref{fig:VibStates_HeN2O2}$(b)$. The plasma properties are strongly altered by an inclusion of N$_2(v<4)$\:\&\:O$_2(v<4)$, particularly on the densities of $\mathrm{N_2(A ^3\Sigma)}$ and $\mathrm{N_2(B ^3\Pi)}$. It should be emphasized that the quantitative analysis here is only valid for the considered operation conditions and the variation is reinforced with increasing absorbed power (not shown here).

\begin{ThreePartTable}
{\scriptsize
\begin{longtable}[h]{p{1.5cm}p{2.8cm}p{2.4cm}p{2.8cm}p{2.4cm}p{2.5cm}}
%\begin{longtable}[h]{p{1.5cm}b{2.8cm}b{2.4cm}b{2.8cm}b{2.4cm}b{2.5cm}}
\caption{The variation of the simulated plasma properties for a wall reaction probability value of 1 relative to that of 0 is presented with a percentage of increase (``$\textcolor{blue}{\uparrow}$") or decrease (``$\textcolor{red}{\downarrow}$"). Those smaller than 1\:\% are set to 0\:\% for the sake of clarity. For the wall reactions labeled by a symbol ``\textcolor{OliveGreen}{\textbf{*}}" under ``$\textcolor{black}{\rightarrow}$", the corresponding reactions are only included in this table for a sensitivity analysis, unless stated otherwise (see section \ref{sec:Chem_kin} and table \ref{tab:wrHeN2O2}). The COST-Jet is sustained by an absorbed power of $0.6$ W at a pressure of $101325$ Pa and a gas temperature of $345$ K for a feeding gas mixture of $1400$ sccm He + 0.5\:\% $\mathrm{O_2}$, $1400$ sccm He + 0.5\:\% $\mathrm{N_2}$ or $1400$ sccm He + 0.5\:\% $\mathrm{N_2}$ + 0.5\:\% $\mathrm{O_2}$. 
}
%----------------------------
\\\hline
\\[\dimexpr-\normalbaselineskip+3pt]
\\[\dimexpr-\normalbaselineskip+3pt]
%$\frac{(n_{\gamma=1}-n_{\gamma=0})}{n_{\gamma=0}}$
 & %    $ \mathrm{N(^2D)} \xrightarrow[\text{}]{\text{w}} \mathrm{N(^4S)}  $    &
$ \mathrm{N(^4S,\:^2D)} \xrightarrow[\text{}]{\text{w}} 1/2 \mathrm{N}_2  $     &    $ \mathrm{O_2(a^1 \Delta_g)} \xrightarrow[\text{}]{\text{w}} \mathrm{O}_2  $     &    $ \mathrm{O(^3P,\:^1D)} \xrightarrow[\textcolor{OliveGreen}{\textbf{*}}]{\text{w}} 1/2\mathrm{O}_2  $     &    $ \mathrm{O(^3P)} \xrightarrow[\textcolor{OliveGreen}{\textbf{*}}]{\text{w}} 1/3\mathrm{O}_3  $     &    $ \mathrm{N}_2(v) \xrightarrow[\text{}]{\text{w}} \mathrm{N}_2(v-1)  $         \\ 
\\[\dimexpr-\normalbaselineskip+3pt]
%& $[0 1]$     &    $[0 1]$     &    $[0 1]$     &    $[0 1]$     &    $[0 1]$         \\ 
\hline
\\[\dimexpr-\normalbaselineskip+3pt]
\endfirsthead
%----------------------------

%-----------------------------------------------------------------------------------------------------------------------------------------------------
\\[\dimexpr-\normalbaselineskip+3pt]
\\[\dimexpr-\normalbaselineskip+7pt]
%\textbf{Input Gas $\mathrm{He/O_2}$}       &        &        &        &        &        &        \\ 
\textbf{{\underline{He/O$_2$}}}       %&
&        &        &        &        &        \\ 
%\hline
\\[\dimexpr-\normalbaselineskip+3pt] 
   $   \mathrm{O(^3P)}    $   %&    $\;\;\;\;\;\;\;\;\;\;\;\;-$    
   &    $\;\;\;\;\;\;\;\;\;\;\;\;-$    &   $\;\;\;\;\;\;\;\;\textcolor{red}{\downarrow} \;\:\:3\:\%$   &     $\;\;\;\;\;\;\;\;\textcolor{red}{\downarrow} \;\:\:58\:\%$   &     $\;\;\;\;\;\;\;\;\textcolor{red}{\downarrow} \;51\:\%$      &    $\;\;\;\;\;\;\;\;\;\;\;\;-$      \\   %6
\\[\dimexpr-\normalbaselineskip+3pt] 
   $   \mathrm{O_3}    $   %&    $\;\;\;\;\;\;\;\;\;\;\;\;-$    
   &    $\;\;\;\;\;\;\;\;\;\;\;\;-$    &     $\;\;\;\;\;\;\;\;\textcolor{blue}{\uparrow} \;56\:\%$   &     $\;\;\;\;\;\;\;\;\textcolor{red}{\downarrow} \;\:\:74\:\%$   &   $\;\;\;\;\;\;\;\;\textcolor{blue}{\uparrow} \;16\:\%$      &    $\;\;\;\;\;\;\;\;\;\;\;\;-$      \\   %8
\\[\dimexpr-\normalbaselineskip+3pt] 
   $   \mathrm{O(^1D)}    $   %&    $\;\;\;\;\;\;\;\;\;\;\;\;-$    
   &    $\;\;\;\;\;\;\;\;\;\;\;\;-$    &   $\;\;\;\;\;\;\;\;\textcolor{red}{\downarrow} \;\:\:8\:\%$   &     $\;\;\;\;\;\;\;\;\textcolor{red}{\downarrow} \;\:\:29\:\%$   &     $\;\;\;\;\;\;\;\;\textcolor{red}{\downarrow} \;\:\:3\:\%$      &    $\;\;\;\;\;\;\;\;\;\;\;\;-$      \\   %9
\\[\dimexpr-\normalbaselineskip+3pt] 
   $   \mathrm{O_2 (a^1 \Delta_g)}    $   %&    $\;\;\;\;\;\;\;\;\;\;\;\;-$    
   &    $\;\;\;\;\;\;\;\;\;\;\;\;-$    &   $\;\;\;\;\;\;\;\;\textcolor{red}{\downarrow} \;44\:\%$   &   $\;\;\;\;\;\;\;\;\textcolor{blue}{\uparrow} \;106\:\%$   &     $\;\;\;\;\;\;\;\;\textcolor{red}{\downarrow} \;10\:\%$      &    $\;\;\;\;\;\;\;\;\;\;\;\;-$      \\   %10

%-----------------------------------------------------------------------------------------------------------------------------------------------------
\\[\dimexpr-\normalbaselineskip+3pt]
\\[\dimexpr-\normalbaselineskip+7pt]
%\textbf{Input Gas $\mathrm{He/N_2}$}       &        &        &        &        &        &        \\ 
\textbf{\underline{{He/N$_2$}}}      % &        
&        &        &        &        &        \\ 
%\hline
%\\[\dimexpr-\normalbaselineskip+3pt] 
%   $   \mathrm{He_2^*}    \:\%$   &    $0\:\%$   &    $0\:\%$   &    $\;\;\;\;\;\;\;\;\;\;\;\;-$    &    $\;\;\;\;\;\;\;\;\;\;\;\;-$    &    $\;\;\;\;\;\;\;\;\;\;\;\;-$    &   $\;\;\;\;\;\;\;\;\textcolor{blue}{\uparrow} \;0.25\:\%$     \\   %4
\\[\dimexpr-\normalbaselineskip+3pt] 
   $   \mathrm{N(^4S)}    $  % &     $\;\;\;\;\;\;\;\;\textcolor{red}{\downarrow} \;0.10$   
   &     $\;\;\;\;\;\;\;\;\textcolor{red}{\downarrow} \;70\:\%$   &    $\;\;\;\;\;\;\;\;\;\;\;\;-$    &    $\;\;\;\;\;\;\;\;\;\;\;\;-$    &    $\;\;\;\;\;\;\;\;\;\;\;\;-$    &   $\;\;\;\;\;\;\;\;\textcolor{red}{\downarrow} \;13\:\%$     \\   %7
\\[\dimexpr-\normalbaselineskip+3pt] 
   $   \mathrm{N(^2D)}    $  % &   $\;\;\;\;\;\;\;\;\textcolor{blue}{\uparrow} \;0.90$   
   &     $\;\;\;\;\;\;\;\;\textcolor{red}{\downarrow} \;50\:\%$   &    $\;\;\;\;\;\;\;\;\;\;\;\;-$    &    $\;\;\;\;\;\;\;\;\;\;\;\;-$    &    $\;\;\;\;\;\;\;\;\;\;\;\;-$    &   $\;\;\;\;\;\;\;\;\textcolor{red}{\downarrow} \;25\:\%$     \\   %8
\\[\dimexpr-\normalbaselineskip+3pt] 
   $   \mathrm{N_2}(v=3)    $   %&    $0$   
   &    $\;\;\;\;\;\;\;\;\:\:\:\:\:\,0\:\%$   &    $\;\;\;\;\;\;\;\;\;\;\;\;-$    &    $\;\;\;\;\;\;\;\;\;\;\;\;-$    &    $\;\;\;\;\;\;\;\;\;\;\;\;-$    &   $\;\;\;\;\;\;\;\;\textcolor{red}{\downarrow} \;32\:\%$     \\   %18
\\[\dimexpr-\normalbaselineskip+3pt] 
   $   \mathrm{N_2}(v=35)    $   %&    $0$   
   &    $\;\;\;\;\;\;\;\;\:\:\:\:\:\,0\:\%$   &    $\;\;\;\;\;\;\;\;\;\;\;\;-$    &    $\;\;\;\;\;\;\;\;\;\;\;\;-$    &    $\;\;\;\;\;\;\;\;\;\;\;\;-$    &   $\;\;\;\;\;\;\;\;\textcolor{red}{\downarrow} \;98\:\%$     \\   %50

%-----------------------------------------------------------------------------------------------------------------------------------------------------
\\[\dimexpr-\normalbaselineskip+3pt]
\\[\dimexpr-\normalbaselineskip+7pt]
\textbf{\underline{{He/N$_2$/O$_2$}}}       %&        
&        &        &        &        &        \\ 
%\hline
\\[\dimexpr-\normalbaselineskip+3pt] 
   $   \mathrm{N(^4S)}    $   %&    $0$   
   &     $\;\;\;\;\;\;\;\;\textcolor{red}{\downarrow} \;56\:\%$   &     $\;\;\;\;\;\;\;\;\textcolor{blue}{\uparrow} \;\:\:1\:\%$   &     $\;\;\;\;\;\;\;\;\textcolor{red}{\downarrow} \;\:\:\:\:7\:\%$   &     $\;\;\;\;\;\;\;\;\textcolor{red}{\downarrow} \;\:\:2\:\%$   &   $\;\;\;\;\;\;\;\;\textcolor{red}{\downarrow} \;47\:\%$     \\   %7
\\[\dimexpr-\normalbaselineskip+3pt] 
   $   \mathrm{N(^2D)}    $   %&    $0$   
   &     $\;\;\;\;\;\;\;\;\textcolor{red}{\downarrow} \;\:\:6\:\%$   &     $\;\;\;\;\;\;\;\;\textcolor{blue}{\uparrow} \;\:\:2\:\%$   &     $\;\;\;\;\;\;\;\;\textcolor{red}{\downarrow} \;\:\:\:\:9\:\%$   &     $\;\;\;\;\;\;\;\;\textcolor{red}{\downarrow} \;\:\:1\:\%$   &   $\;\;\;\;\;\;\;\;\textcolor{red}{\downarrow} \;28\:\%$     \\   %8
\\[\dimexpr-\normalbaselineskip+3pt] 
   $   \mathrm{O(^3P)}    $   %&    $0$   
   &    $\;\;\;\;\;\;\;\;\:\:\:\:\:\,0\:\%$   &   $\;\;\;\;\;\;\;\;\textcolor{red}{\downarrow} \;\:\:2\:\%$   &     $\;\;\;\;\;\;\;\;\textcolor{red}{\downarrow} \;\:\:60\:\%$   &     $\;\;\;\;\;\;\;\;\textcolor{red}{\downarrow} \;53\:\%$   &   $\;\;\;\;\;\;\;\;\textcolor{red}{\downarrow} \;\:\:1\:\%$     \\   %16
\\[\dimexpr-\normalbaselineskip+3pt] 
   $   \mathrm{O_3}    $   %&    $0$   
   &    $\;\;\;\;\;\;\;\;\:\:\:\:\:\,0\:\%$   &     $\;\;\;\;\;\;\;\;\textcolor{blue}{\uparrow} \;35\:\%$   &     $\;\;\;\;\;\;\;\;\textcolor{red}{\downarrow} \;\:\:75\:\%$   &   $\;\;\;\;\;\;\;\;\textcolor{blue}{\uparrow} \;\:\:3\:\%$   &     $\;\;\;\;\;\;\;\;\textcolor{blue}{\uparrow} \;\:\:1\:\%$     \\   %18
\\[\dimexpr-\normalbaselineskip+3pt] 
   $   \mathrm{O(^1D)}    $   %&    $0$   
   &    $\;\;\;\;\;\;\;\;\:\:\:\:\:\,0\:\%$   &   $\;\;\;\;\;\;\;\;\textcolor{red}{\downarrow} \;\:\:7\:\%$   &     $\;\;\;\;\;\;\;\;\textcolor{red}{\downarrow} \;\:\:30\:\%$   &     $\;\;\;\;\;\;\;\;\textcolor{red}{\downarrow} \;\:\:2\:\%$   &   $\;\;\;\;\;\;\;\;\textcolor{red}{\downarrow} \;\:\:9\:\%$     \\   %19
\\[\dimexpr-\normalbaselineskip+3pt] 
   $   \mathrm{O_2 (a^1 \Delta_g)}    $  % &    $0$   
   &    $\;\;\;\;\;\;\;\;\:\:\:\:\:\,0\:\%$   &   $\;\;\;\;\;\;\;\;\textcolor{red}{\downarrow} \;36\:\%$   &   $\;\;\;\;\;\;\;\;\textcolor{blue}{\uparrow} \;109\:\%$   &     $\;\;\;\;\;\;\;\;\textcolor{red}{\downarrow} \;\:\:1\:\%$   &   $\;\;\;\;\;\;\;\;\textcolor{red}{\downarrow} \;\:\:3\:\%$     \\   %20
\\[\dimexpr-\normalbaselineskip+3pt] 
   $   \mathrm{NO}    $   %&    $0$  
   &   $\;\;\;\;\;\;\;\;\textcolor{blue}{\uparrow} \;\:\:8\:\%$   &     $\;\;\;\;\;\;\;\;\textcolor{blue}{\uparrow} \;\:\:1\:\%$   &     $\;\;\;\;\;\;\;\;\textcolor{red}{\downarrow} \;\:\:12\:\%$   &     $\;\;\;\;\;\;\;\;\textcolor{red}{\downarrow} \;\:\:6\:\%$   &   $\;\;\;\;\;\;\;\;\textcolor{red}{\downarrow} \;46\:\%$     \\   %28
\\[\dimexpr-\normalbaselineskip+3pt] 
   $   \mathrm{NO_2}    $   %&    $0$   
   &   $\;\;\;\;\;\;\;\;\textcolor{blue}{\uparrow} \;\:\:8\:\%$   &     $\;\;\;\;\;\;\;\;\textcolor{blue}{\uparrow} \;\:\:1\:\%$   &     $\;\;\;\;\;\;\;\;\textcolor{red}{\downarrow} \;\:\:16\:\%$   &     $\;\;\;\;\;\;\;\;\textcolor{red}{\downarrow} \;\:\:9\:\%$   &   $\;\;\;\;\;\;\;\;\textcolor{red}{\downarrow} \;45\:\%$     \\   %31
\\[\dimexpr-\normalbaselineskip+3pt] 
   $   \mathrm{NO_3}    $   %&    $0$   
   &   $\;\;\;\;\;\;\;\;\textcolor{blue}{\uparrow} \;\:\:8\:\%$   &     $\;\;\;\;\;\;\;\;\textcolor{blue}{\uparrow} \;\:\:1\:\%$   &     $\;\;\;\;\;\;\;\;\textcolor{red}{\downarrow} \;\:\:18\:\%$   &     $\;\;\;\;\;\;\;\;\textcolor{red}{\downarrow} \;11\:\%$   &   $\;\;\;\;\;\;\;\;\textcolor{red}{\downarrow} \;45\:\%$     \\   %34
\\[\dimexpr-\normalbaselineskip+3pt] 
   $   \mathrm{N_2O_3}    $   %&    $0$   
   &   $\;\;\;\;\;\;\;\;\textcolor{blue}{\uparrow} \;17\:\%$   &     $\;\;\;\;\;\;\;\;\textcolor{blue}{\uparrow} \;\:\:2\:\%$   &     $\;\;\;\;\;\;\;\;\textcolor{red}{\downarrow} \;\:\:27\:\%$   &     $\;\;\;\;\;\;\;\;\textcolor{red}{\downarrow} \;16\:\%$   &   $\;\;\;\;\;\;\;\;\textcolor{red}{\downarrow} \;70\:\%$     \\   %39
\\[\dimexpr-\normalbaselineskip+3pt] 
   $   \mathrm{N_2O_4}    $  % &    $0$   
   &   $\;\;\;\;\;\;\;\;\textcolor{blue}{\uparrow} \;17\:\%$   &     $\;\;\;\;\;\;\;\;\textcolor{blue}{\uparrow} \;\:\:2\:\%$   &     $\;\;\;\;\;\;\;\;\textcolor{red}{\downarrow} \;\:\:30\:\%$   &     $\;\;\;\;\;\;\;\;\textcolor{red}{\downarrow} \;18\:\%$   &   $\;\;\;\;\;\;\;\;\textcolor{red}{\downarrow} \;70\:\%$     \\   %40
\\[\dimexpr-\normalbaselineskip+3pt] 
   $   \mathrm{N_2O_5}    $   %&    $0$   
   &   $\;\;\;\;\;\;\;\;\textcolor{blue}{\uparrow} \;17\:\%$   &     $\;\;\;\;\;\;\;\;\textcolor{blue}{\uparrow} \;\:\:2\:\%$   &     $\;\;\;\;\;\;\;\;\textcolor{red}{\downarrow} \;\:\:31\:\%$   &     $\;\;\;\;\;\;\;\;\textcolor{red}{\downarrow} \;19\:\%$   &   $\;\;\;\;\;\;\;\;\textcolor{red}{\downarrow} \;70\:\%$     \\   %41
\\[\dimexpr-\normalbaselineskip+3pt] 
   $   \mathrm{N_2}(v=3)    $   %&    $0$   
   &    $\;\;\;\;\;\;\;\;\:\:\:\:\:\,0\:\%$   &    $\;\;\;\;\;\;\;\;\:\:\:\:\:\,0\:\%$   &   $\;\;\;\;\;\;\;\;\textcolor{blue}{\uparrow} \;\:\:\:\:1\:\%$   &    $\;\;\;\;\;\;\;\;\:\:\:\:\:\,0\:\%$   &   $\;\;\;\;\;\;\;\;\textcolor{red}{\downarrow} \;36\:\%$     \\   %44
\\[\dimexpr-\normalbaselineskip+3pt] 
   $   \mathrm{N_2}(v=35)    $   %&    $0$   
   &    $\;\;\;\;\;\;\;\;\:\:\:\:\:\,0\:\%$   &   $\;\;\;\;\;\;\;\;\textcolor{red}{\downarrow} \;\:\:4\:\%$   &   $\;\;\;\;\;\;\;\;\textcolor{blue}{\uparrow} \;\:\:11\:\%$   &     $\;\;\;\;\;\;\;\;\:\:\:\:\:\,0\:\%$   &   $\;\;\;\;\;\;\;\;\textcolor{red}{\downarrow} \;98\:\%$     \\   %76
\\[\dimexpr-\normalbaselineskip+3pt] 
\hline
\label{tab:roleOfWallReaction}
\end{longtable}
}
\end{ThreePartTable}

The modification of the simulated plasma properties in the COST-Jet to the changes of the wall reaction probabilities from 0 to 1 is summarized in table \ref{tab:roleOfWallReaction} for the gas mixtures He/$\mathrm{O_2}$, He/$\mathrm{N_2}$ and He/$\mathrm{N_2}$/$\mathrm{O_2}$. The COST-Jet He plasma is not altered by these changes and it is excluded from the table. The electron density and temperature are negligibly affected by the neutral wall reaction mechanisms within the considered gas mixtures. Additionally, it is observed that all the plasma properties are insensitive to the wall quenching of $ \mathrm{He(2 ^3S)}$, $\mathrm{He_2^*}$, $\mathrm{N_2(A^3 \Sigma)}$, $\mathrm{N_2(B ^3\Pi)}$, $\mathrm{O(^1 D)}$ and $\mathrm{O_2}(v<41)$. The quenching of $\mathrm{N(^2 D)}$ is of importance merely for its concentration in a He/$\mathrm{N_2}$ jet (not shown here). In the He/O$_2$ and He/$\mathrm{N_2}$/$\mathrm{O_2}$ mixtures, an increase of the $\mathrm{O_2 (a^1 \Delta_g)}$ quenching probability reduces the density of $\mathrm{O_2 (a^1 \Delta_g)}$ and raises that of $\mathrm{O_3}$. The oxygen atom wall recombination and ozone wall formation probabilities have a significant influence on the $\mathrm{O(^3 P)}$ density. Moreover, the recombination probability plays an important role in the concentrations of $\mathrm{O_3}$ and $\mathrm{O_2 (a^1 \Delta_g)}$. In the He/$\mathrm{N_2}$ and He/$\mathrm{N_2}$/$\mathrm{O_2}$ mixtures, a rise of the nitrogen atom wall recombination probability reduces the densities of $\mathrm{N(^4 S)}$ and $\mathrm{N(^2 D)}$. The wall quenching of N$_2(v<58)$ has an impact on the vibrationally excited nitrogen molecule densities, particularly on those with high quantum numbers. This quenching additionally alters the concentrations of nitrogen atom and nitrogen oxides for a feeding gas mixture of He/$\mathrm{N_2}$/$\mathrm{O_2}$.

\begin{figure}[tp]
        \centering
	\subfigure{\includegraphics[scale=0.25,clip=true,trim=0cm 0cm 0cm 0cm,]{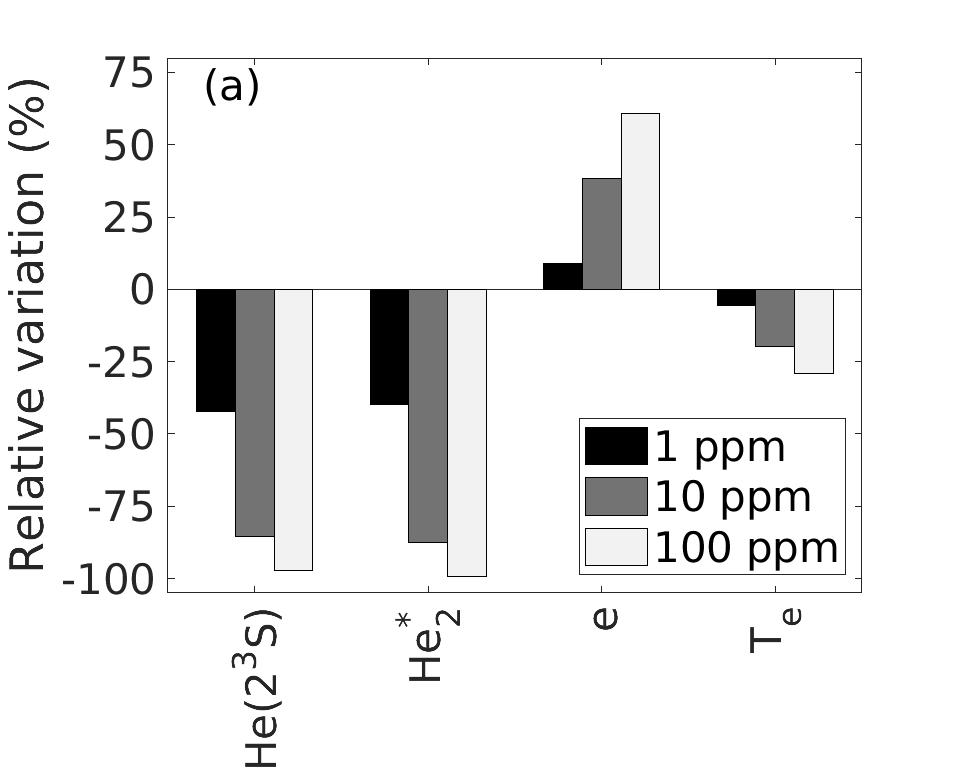}} \\
	\subfigure{\includegraphics[scale=0.25,clip=true,trim=0cm 0cm 0cm 0cm,]{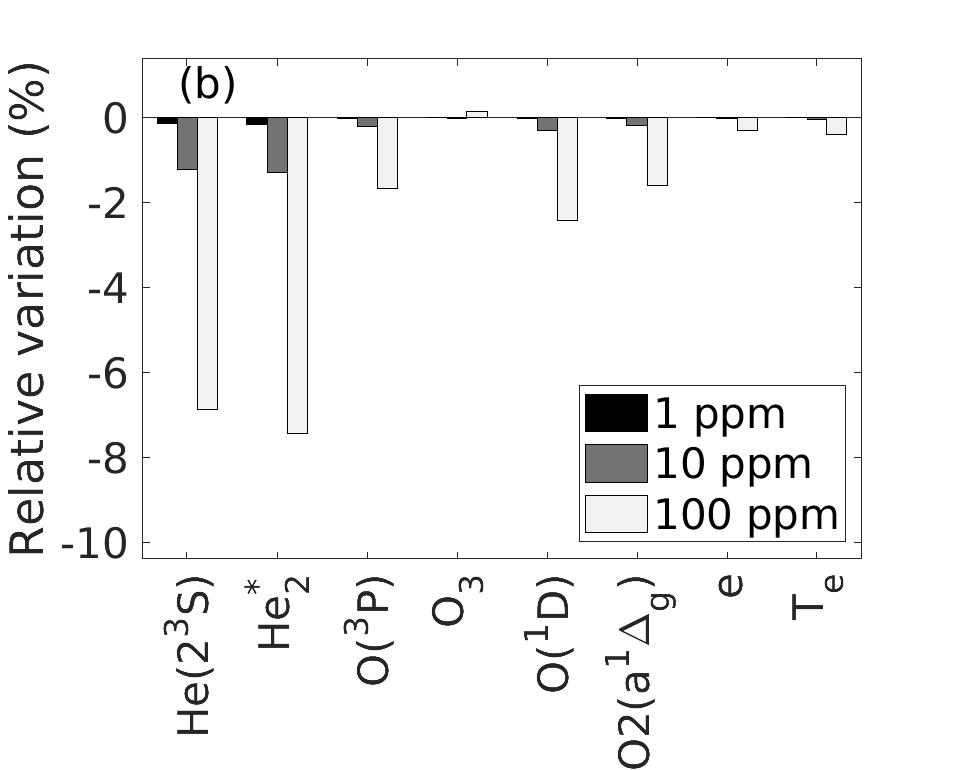}} \\
	\subfigure{\includegraphics[scale=0.25,clip=true,trim=0cm 0cm 0cm 0cm,]{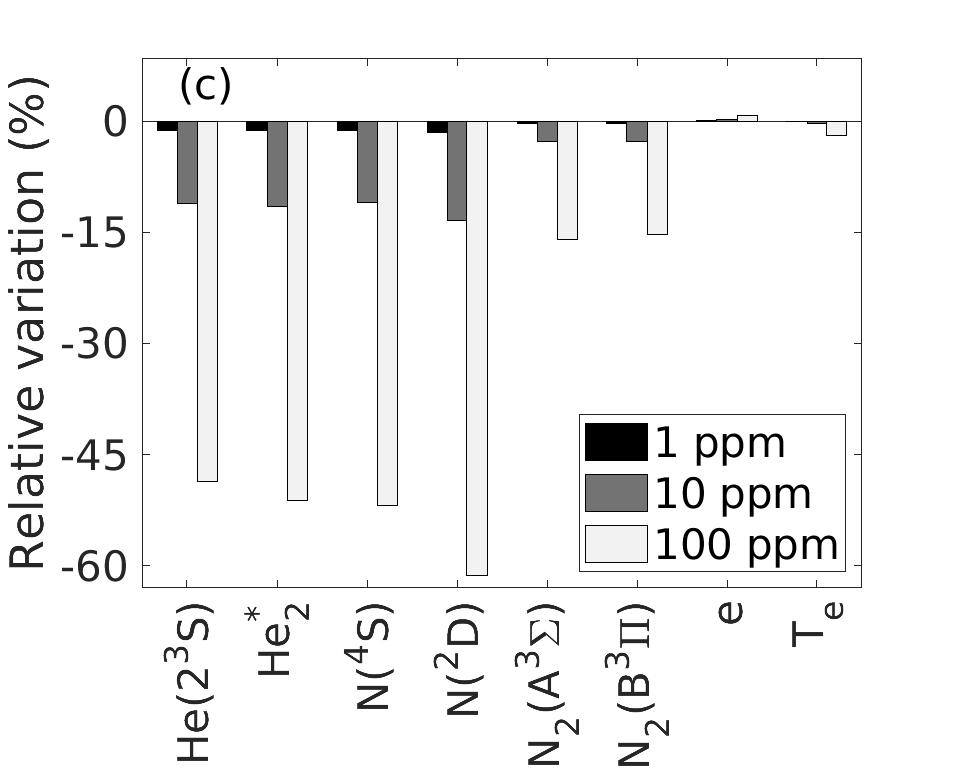}}
	\caption{The variation of the simulated plasma properties at varying orders of synthetic air impurity relative to those with a pure feeding gas. The COST-Jet is sustained by an absorbed power of $0.6$ W at a pressure of $101325$ Pa and a gas temperature of $345$ K for feeding gases of $(a)$ 1400 sccm He, $(b)$ 1400 sccm He + 0.5\:\% $\mathrm{O_2}$ and $(c)$ 1400 sccm He + 0.5\:\% $\mathrm{N_2}$. 
}\label{fig:Imputrity_HeN2O2}
\end{figure}

The variation of the simulated plasma properties at varying orders of synthetic air impurity relative to those with a pure feeding gas in the COST-Jet is shown in figure \ref{fig:Imputrity_HeN2O2} for gases $(a)$ He, $(b)$ He/O$_2$ and $(c)$ He/N$_2$. The simulations confirm the substantial influence of the impurity level on the plasma properties. The electron density increases with increasing amount of synthetic air in He plasma, whereas the electron temperature as well as the densities of $ \mathrm{He(2 ^3S)}$ and $\mathrm{He_2^*}$ decrease. The metastable densities experience the largest variation due to the efficient Penning ionization by nitrogen and oxygen species \cite{Sun19}. The role of the synthetic air impurity in He/$\mathrm{O_2}$ plasma is limited and produces below $10\:\%$ variation at a value of $100$ ppm. The He/$\mathrm{N_2}$ mixture is comparatively more sensitive to this impurity level, and up to $60\:\%$ reduction of the concentrations is observed.

\section{Conclusion}
\label{sec:con}

Atmospheric-pressure plasma jet devices of planar electrode configuration fed with He, He/$\mathrm{O_2}$, He/$\mathrm{N_2}$ and He/$\mathrm{N_2}$/$\mathrm{O_2}$ mixtures are investigated by a zero-dimensional (volume-averaged) and a pseudo-one-dimensional (plug-flow) modelling approach. The models are developed with a focus on the vibrational kinetics and self-consistently coupled with a Boltzmann solver (LoKI-B) under the two-term approximation to properly address the electron kinetics. A good agreement is obtained between the model calculations and the spatially resolved simulation results as well as diverse measurements available from literature, including the electron density, the electron temperature, and the concentrations of helium metastable, helium excimer, ozone, oxygen atom and nitrogen atom in distinct operation conditions.

Nitric oxide is thoroughly characterized for a variation of the gas mixture ratio, helium flow rate and absorbed power. The measurements at low power are better described by introducing the reactive quenching of $\mathrm{N_2(A^3 \Sigma, B ^3\Pi;}\:v>0)$ and of multiple $\mathrm{N_2}$ electronically excited states with a hypothetical $``$effective$"$ rate coefficient value in the simulations, that forms underlying NO production mechanisms as well as significant $\mathrm{O(^3 P)}$ loss channels. A continuous growth of the NO spatial density profile is obtained in gas flow direction, and the saturation is merely observed at a relatively high absorbed power of about 2.2 W. Within the range of the considered operation conditions, the NO loss is governed by its flow-out rate, the $\mathrm{O(^3 P)} + \mathrm{NO} + \mathrm{M} \rightarrow \mathrm{NO_2} + \mathrm{M}$ and $\mathrm{N(^4 S)} + \mathrm{NO} \rightarrow \mathrm{N_2} + \mathrm{O(^3 P)}$ reactions, while the formation is dominated by the $\mathrm{O(^3 P)} + \mathrm{NO_2} \rightarrow \mathrm{NO} + \mathrm{O_2}$ and $\mathrm{N(^2 D)} + \mathrm{O_2} \rightarrow \mathrm{NO} + \mathrm{O(^1 D)}$ channels. The contribution of the vibrationally excited nitrogen molecules $\mathrm{N_2}(v \geqslant 13)$ in the net NO formation is enhanced with increasing power.

The vibrationally excited levels of $\mathrm{O_2}(v<41)$ and $\mathrm{N_2}(v<58)$ are analysed. A larger response of the N$_2$ VDF to the variation of the gas mixture ratio, helium flow rate and absorbed power is observed relative to that of the O$_2$ VDF. A continuous spatial growth of the N$_2$ vibrational population is obtained in gas flow direction (i.e., the population is not yet saturated at the jet nozzle). On the contrary, a fast spatial equilibrium of the O$_2$ vibrational population is acquired at the beginning of the plasma chamber. The simulation results are affected by the total number of vibrationally excited nitrogen molecules considered in the chemical model. However, the influence of $\mathrm{N_2}(v<58)$ on the overall plasma behaviour is captured by a limited set of vibrational levels depending on the applied power value.

The sensitivity of the simulation results to a variation of the wall reaction probabilities and the  synthetic air impurity levels is provided. The concentrations of nitrogen atom and nitrogen oxides are influenced by the wall quenching of vibrationally excited nitrogen molecules. A dramatic sensitivity of the atomic oxygen and ozone densities to the oxygen atom wall recombination is observed. The plasma properties of pure He gas are significantly altered in the presence of the impurity due to the efficient Penning ionization mechanism. An impurity level up to 100 ppm is not influential on the simulation results of He/$\mathrm{O_2}$ mixture, but has a strong impact on those of He/$\mathrm{N_2}$ plasma. 

A further analysis of the rate coefficients of the termolecular Penning ionization $ \mathrm{He(2 ^3S)} + \mathrm{N_2} + \mathrm{He} \rightarrow e + \mathrm{N_2^+} + 2 \mathrm{He}  $, the metastable quenching via vibrationally excited nitrogen molecules $ {\rm He(2 ^3S)} + {\rm N}_2(v>0) \rightarrow e + {\rm He} + {\rm N}_2^+ $ and the nitric oxide formation channels $  \mathrm{N_2(X,A ^3\Sigma,B ^3\Pi,...;}\:v) + {\rm O(^3P)} \rightarrow {\rm NO} + {\rm N(^2D)} $ is of importance for a more accurate model prediction.

\section*{Acknowledgements}

Funded by the Deutsche Forschungsgemeinschaft (DFG, German Research Foundation) - Project-ID 327886311 (SFB 1316: simulations by A9 \& A8 and measurements by B2 \& A4).
The authors are in depth with Antonio Tejero del Caz and Luis L. Alves for discussions on the self-consistent LoKI-B adaptation, with Peter Bruggeman and Nader Sadeghi for discussions on the pulse-modulated plasma jet, and with Judith Golda for  discussions on the experimental details of the He COST-Jet. Vasco Guerra was partially funded by the Portuguese FCT-Fundacao para a Ciencia e a Tecnologia, under projects UIDB/50010/2020 and UIDP/50010/2020.

\section*{ORCID iDs}
Youfan He  \href{target}{https://orcid.org/0000-0003-1275-7695}

Patrick Preissing  \href{target}{https://orcid.org/0000-0001-7375-3984}

David Steuer  \href{target}{https://orcid.org/0000-0003-3005-0829}

Maximilian Klich  \href{target}{https://orcid.org/0000-0002-3913-1783}

Volker Schulz-von der Gathen  \href{target}{https://orcid.org/0000-0002-7182-3253}

Marc B{\"o}ke  \href{target}{https://orcid.org/0000-0003-1062-5808}

Ihor Korolov  \href{target}{https://orcid.org/0000-0003-2384-1243}

Julian Schulze  \href{target}{https://orcid.org/0000-0001-7929-5734}

Vasco Guerra  \href{target}{https://orcid.org/0000-0002-6878-6850}

Ralf Peter Brinkmann  \href{target}{https://orcid.org/0000-0002-2581-9894}

Efe Kemaneci  \href{target}{https://orcid.org/0000-0002-5540-0947}
\clearpage
\newpage
{\bf \LARGE Appendix}
\appendix
\section{Chemical kinetics}
\label{sec:chemkin}

\begin{table}[hbt!] \footnotesize
\centering
\caption{The distinct species in the model of $\mathrm{He}$, $\mathrm{He/O_2}$, $\mathrm{He/N_2}$ or $\mathrm{He/N_2/O_2}$ are selected from the table below.}
\begin{tabular}{l}
\\[\dimexpr-\normalbaselineskip+3pt]
\hline
\\[\dimexpr-\normalbaselineskip+3pt]
 Plasma species  \\ \hline
\\[\dimexpr-\normalbaselineskip+3pt]
 $ \mathrm{He}$,
 $ \mathrm{He(2 ^3S)}$,
  $\mathrm{He_2^*}$,
 $\mathrm{He^+}$,
 $\mathrm{He_2^+}$,\\
 \\[\dimexpr-\normalbaselineskip+3pt]
 $\mathrm{N(^4 S)} $,
 $\mathrm{N_2}$,
 $\mathrm{N(^2 D)}$,
 $\mathrm{N_2(A ^3\Sigma)}$,
 $\mathrm{N_2(B ^3\Pi)}$,
 $\mathrm{N^+}$,
 $\mathrm{N_2^+}$,
 $\mathrm{N_3^+}$,
 $\mathrm{N_4^+}$,\\
\\[\dimexpr-\normalbaselineskip+3pt]
 $\mathrm{O(^3 P)} $,
 $\mathrm{O_2}$,
 $\mathrm{O_3}$,
 $\mathrm{O(^1 D)}$,
 $\mathrm{O_2 (a^1 \Delta_g)}$,
 $\mathrm{O^+}$,
 $\mathrm{O_2^+}$,
 $\mathrm{O_4^+}$,
 $\mathrm{O^-}$,
 $\mathrm{O_2^-}$,
 $\mathrm{O_3^-}$,
 $\mathrm{O_4^-}$,\\
 \\[\dimexpr-\normalbaselineskip+3pt]
 $\mathrm{NO} $,
 $\mathrm{NO_2} $,
 $\mathrm{NO_3} $,
 $\mathrm{N_2O} $,
 $\mathrm{N_2O_3} $,
 $\mathrm{N_2O_4} $,
 $\mathrm{N_2O_5} $,
 $\mathrm{NO^+}$,
 $\mathrm{NO_2^+}$,
 $\mathrm{N_2O^+}$,
 $\mathrm{NO^-}$,
 $\mathrm{NO_2^-}$,
 $\mathrm{NO_3^-}$,
 $\mathrm{N_2O^-}$,\\
\\[\dimexpr-\normalbaselineskip+3pt]
 $\mathrm{N_2}$(0$<v<$58),
 $\mathrm{O_2}$(0$<v<$41),
 $e$\\
\\[\dimexpr-\normalbaselineskip+3pt]
\hline
\end{tabular}
\label{tab:SpeciesHeN2O2}
\end{table}

\noindent 
\begin{ThreePartTable}
{\scriptsize
\setlength{\tabcolsep}{8pt}
\begin{longtable}[h]{p{0.7cm}p{6.5cm}p{6.0cm}p{1.1cm}p{0.2cm}}
\caption{The volume reactions included in the $ \mathrm{He} $ model. The rate coefficient units are given in s$^{-1}$, m$^3$\:s$^{-1}$ and m$^6$\:s$^{-1}$ for one-, two- and three-body reactions, respectively. $\rm T_e$ is in eV and $\rm T_g$ in K, if not stated otherwise. The rate coefficient $ f(\epsilon) $ is taken from a look-up-table calculated via the referred cross-section self-consistently coupled to the EEDF \cite{Tej19}. The reverse reaction rate coefficient of the electron-impact excitation labeled with a symbol ``*" near the number is calculated via the principle of {\em detailed balancing} \cite{LieBook2005}. 
}
\label{tab:ReactionListHe} \\

\hline
\# &Reaction     & Rate Coefficient & Ref & \\ \hline
\\[\dimexpr-\normalbaselineskip+3pt]
\endfirsthead

\multicolumn{4}{c}%
%{{\bfseries \tablename\ \thetable{} -- $ Continued $ $ from $ $ previous $ $ page $}} \\
{{ \tablename\ \thetable{} -- $ Continued $ $ from $ $ previous $ $ page $}} \\
\hline
\# &Reaction     & Rate Coefficient & Ref & \\ 
\hline
\\[\dimexpr-\normalbaselineskip+3pt]

\endhead

\hline \multicolumn{4}{r@{}}{{$ Continued $ $ on $ $ next $ $ page $}} \\
\\[\dimexpr-\normalbaselineskip+3pt]
\endfoot

%\hline
\\[\dimexpr-\normalbaselineskip+3pt]
\endlastfoot

%\\[\dimexpr-\normalbaselineskip+3pt]
%\hline
%\\[\dimexpr-\normalbaselineskip+3pt]

%\\[\dimexpr-\normalbaselineskip+4pt]
%%\hline
% & Electron - Neutral     &        &     & \\ 
%\\[\dimexpr-\normalbaselineskip+4pt]

1   &   $ e + \mathrm{He} \rightarrow 2 e + \mathrm{He^+}  $   &   $ f(\epsilon) $   &     \cite{IST-Lisbon-LXCat}     &      \\
\\[\dimexpr-\normalbaselineskip+3pt]
2$^*$   &   $ e + \mathrm{He} \rightarrow e + \mathrm{He(2 ^3S)}  $   &   $ f(\epsilon) $   &     \cite{IST-Lisbon-LXCat}     &      \\
\\[\dimexpr-\normalbaselineskip+3pt]
3   &   $ e + \mathrm{He(2 ^3S)} \rightarrow 2 e + \mathrm{He^+}  $   &   $ f(\epsilon) $   &     \cite{TRINITI-LXCat}     &      \\
\\[\dimexpr-\normalbaselineskip+3pt]
4   &   $ e + \mathrm{He_2^*} \rightarrow e + 2 \mathrm{He}  $   &   $ 3.8 \times 10^{-15} $   &     \cite{Sch18_PCCP}     &      \\
\\
\hline
\end{longtable}}

\end{ThreePartTable}

\newpage
\noindent 
\begin{ThreePartTable}
{\scriptsize
\setlength{\tabcolsep}{8pt}
\begin{longtable}[h]{p{0.7cm}p{6.5cm}p{5.0cm}p{1.1cm}p{0.2cm}}
\caption{The oxygen volume reactions in the $ \mathrm{He/O_2} $ model. M is the background gas helium. The rate coefficient units are given in m$^3$\:s$^{-1}$ and m$^6$\:s$^{-1}$ for two- and three-body reactions, respectively. $\rm T_e$ is in eV and $\rm T_g$ in K, if not stated otherwise. The rate coefficient $ f(\epsilon) $ is taken from a look-up-table calculated via the referred cross-section self-consistently coupled to the EEDF \cite{Tej19}. The reverse reaction rate coefficients of the electron-impact excitation labeled with a symbol ``*" near the number are calculated via the principle of {\em detailed balancing} \cite{LieBook2005}.
}
\label{tab:ReactionListHeO2} \\

\hline
\# &Reaction     & Rate Coefficient & Ref & \\ \hline
\\[\dimexpr-\normalbaselineskip+3pt]
\endfirsthead

\multicolumn{4}{c}%
%{{\bfseries \tablename\ \thetable{} -- $ Continued $ $ from $ $ previous $ $ page $}} \\
{{ \tablename\ \thetable{} -- $ Continued $ $ from $ $ previous $ $ page $}} \\
\hline
\# &Reaction     & Rate Coefficient & Ref & \\ 
\hline
\\[\dimexpr-\normalbaselineskip+3pt]

\endhead

\hline \multicolumn{4}{r@{}}{{$ Continued $ $ on $ $ next $ $ page $}} \\
\\[\dimexpr-\normalbaselineskip+3pt]
\endfoot

%\hline
\\[\dimexpr-\normalbaselineskip+3pt]
\endlastfoot

%\\[\dimexpr-\normalbaselineskip+3pt]
%\hline
%\\[\dimexpr-\normalbaselineskip+3pt]

%%\hline
%\# & Electron - Neutral Particle Collision     &        &     & \\ 
\\[\dimexpr-\normalbaselineskip+3pt]

%1   &   $ e + \mathrm{O(^3 P)} + \mathrm{O_2} \rightarrow \mathrm{O^-} + \mathrm{O_2}  $   &   $ 1.0 \times 10^{-43} $   &     \cite{CapBook}     &      \\
1   &   $ e + \mathrm{O(^3 P)} + \mathrm{O_2} \rightarrow \mathrm{O^-} + \mathrm{O_2}  $   &   $ 1 \times 10^{-43} $   &     \cite{CapBook}     &      \\
\\[\dimexpr-\normalbaselineskip+3pt]
%2   &   $ e + \mathrm{O(^3 P)} + \mathrm{O_2} \rightarrow \mathrm{O_2^-} + \mathrm{O(^3 P)}  $   &   $ 1.0 \times 10^{-43} $   &     \cite{CapBook}     &      \\
2   &   $ e + \mathrm{O(^3 P)} + \mathrm{O_2} \rightarrow \mathrm{O_2^-} + \mathrm{O(^3 P)}  $   &   $ 1 \times 10^{-43} $   &     \cite{CapBook}     &      \\
\\[\dimexpr-\normalbaselineskip+3pt]
3$^*$   &   $ e + \mathrm{O(^3 P)} \rightarrow e + \mathrm{O(^1 D)}  $   &   $ f(\epsilon) $   &     \cite{IST-Lisbon-LXCat}     &      \\
\\[\dimexpr-\normalbaselineskip+3pt]
4   &   $ e + \mathrm{O(^3 P)} \rightarrow 2 e + \mathrm{O^+}  $   &   $ f(\epsilon) $   &     \cite{IST-Lisbon-LXCat}     &      \\
\hline
\end{longtable}}

\end{ThreePartTable}

\newpage
\noindent 
\begin{ThreePartTable}
{\scriptsize
\setlength{\tabcolsep}{8pt}
\begin{longtable}[h]{p{0.2cm}p{5.5cm}p{5.5cm}p{1.9cm}p{0.1cm}}
\caption{The nitrogen volume reactions in the $ \mathrm{He/N_2} $ model. M is the background gas helium. The rate coefficient units are given in s$^{-1}$, m$^3$\:s$^{-1}$ and m$^6$\:s$^{-1}$ for one-, two- and three-body reactions, respectively. $\rm T_e$ is in eV and $\rm T_g$ in K, if not stated otherwise. The rate coefficient $ f(\epsilon) $ is taken from a look-up-table calculated via the referred cross-section self-consistently coupled to the EEDF \cite{Tej19}. The reverse reaction rate coefficients of the electron-impact excitation labeled with a symbol ``*" near the number are calculated via the principle of {\em detailed balancing} \cite{LieBook2005}. }
\label{tab:ReactionListHeN2} \\

\hline
%\# &Reaction     & Rate Coefficient & Ref & \\ \hline
\# &Reaction     & Rate Coefficient & Ref & \\ \hline
\\[\dimexpr-\normalbaselineskip+3pt]
\endfirsthead

\multicolumn{4}{c}%
%{{\bfseries \tablename\ \thetable{} -- $ Continued $ $ from $ $ previous $ $ page $}} \\
{{ \tablename\ \thetable{} -- $ Continued $ $ from $ $ previous $ $ page $}} \\
\hline
\# &Reaction     & Rate Coefficient & Ref & \\ 
\hline
\\[\dimexpr-\normalbaselineskip+3pt]

\endhead

\hline \multicolumn{4}{r@{}}{{$ Continued $ $ on $ $ next $ $ page $}} \\
\\[\dimexpr-\normalbaselineskip+3pt]
\endfoot

%\hline
\\[\dimexpr-\normalbaselineskip+3pt]
\endlastfoot

%\\[\dimexpr-\normalbaselineskip+3pt]
%\hline
%\\[\dimexpr-\normalbaselineskip+3pt]

%%\hline
%\# & Electron - Neutral Particle Collision     &        &     & \\ 
%\\[\dimexpr-\normalbaselineskip+3pt]

1$^*$   &   $ e + \mathrm{N(^4S)} \rightarrow e + \mathrm{N(^2D)}  $   &   $ f(\epsilon) $   &     \cite{IST-Lisbon-LXCat}     &      \\
\\[\dimexpr-\normalbaselineskip+3pt]
2   &   $ e + \mathrm{N(^4S)} \rightarrow 2 e + \mathrm{N^+}  $   &   $ f(\epsilon) $   &     \cite{IST-Lisbon-LXCat}     &      \\
\\[\dimexpr-\normalbaselineskip+3pt]
3   &   $ e + \mathrm{N(^2D)} \rightarrow 2 e + \mathrm{N^+}  $   &   $ 1.67 \times 10^{-14} \: T_{\rm e}^{0.50} \: \exp(-13.07/T_{\rm e}) $   &     \cite{Tho08}     &      \\
\\[\dimexpr-\normalbaselineskip+3pt]
4   &   $ e + \mathrm{N_2} \rightarrow e + \mathrm{N(^4S)} + \mathrm{N(^4S)}  $   &   $ f(\epsilon) $   &     \cite{Phys4Entry}     &      \\
\\
\hline
\end{longtable}}

\end{ThreePartTable}

\newpage
\noindent 
\begin{ThreePartTable}
{\scriptsize
\setlength{\tabcolsep}{8pt}
\begin{longtable}[h]{p{0.7cm}p{6.5cm}p{5.0cm}p{1.9cm}p{0.3cm}}
\caption{ The oxygen and nitrogen volume reactions in the $ \mathrm{He/N_2/O_2} $ model. M is the background gas helium. The rate coefficient units are given in m$^3$\:s$^{-1}$ and m$^6$\:s$^{-1}$ for two- and three-body reactions, respectively. $\rm T_e$ is in eV and $\rm T_g$ in K, if not stated otherwise. The rate coefficient $ f(\epsilon) $ is taken from a look-up-table calculated via the referred cross-section self-consistently coupled to the EEDF \cite{Tej19}. 
}
\label{tab:ReactionListHeNxOy} \\

\hline
\# &Reaction     & Rate Coefficient & Ref & \\ \hline
\\[\dimexpr-\normalbaselineskip+3pt]
\endfirsthead

\multicolumn{4}{c}%
%{{\bfseries \tablename\ \thetable{} -- $ Continued $ $ from $ $ previous $ $ page $}} \\
{{ \tablename\ \thetable{} -- $ Continued $ $ from $ $ previous $ $ page $}} \\
\hline
\# &Reaction     & Rate Coefficient & Ref & \\ 
\hline
\\[\dimexpr-\normalbaselineskip+3pt]

\endhead

\hline \multicolumn{4}{r@{}}{{$ Continued $ $ on $ $ next $ $ page $}} \\
\\[\dimexpr-\normalbaselineskip+3pt]
\endfoot

%\hline
\\[\dimexpr-\normalbaselineskip+3pt]
\endlastfoot

%\\[\dimexpr-\normalbaselineskip+3pt]
%\hline
%\\[\dimexpr-\normalbaselineskip+3pt]

%%\hline
%\# & Electron - Neutral Particle Collision     &        &     & \\ 
\\[\dimexpr-\normalbaselineskip+3pt]

%1   &   $ e + \mathrm{O_2} + \mathrm{N_2} \rightarrow \mathrm{O_2^-} + \mathrm{N_2}  $   &   $ 1.1 \times 10^(-43). \times ((T_{\rm g}./(T_{\rm e}. \times 1.1604505e4)).^2). \times (e^(-70/T_{\rm g})). \times (e^(1500 \times ((T_{\rm e}. \times 1.1604505e4)-T_{\rm g})./((T_{\rm e}. \times 1.1604505e4) \times T_{\rm g}))) $   &     \cite{Cha62}     &      \\
1   &   $ e + \mathrm{O_2} + \mathrm{N_2} \rightarrow \mathrm{O_2^-} + \mathrm{N_2}  $   &   $ 1.1 \times 10^{-43} \: (T_{\rm g}/T{\rm _e[K]})^2 \: \exp(-70/T_{\rm g}) \times \exp\{1500 (T{\rm _e[K]}-T_{\rm g})/(T{\rm _e[K]} T_{\rm g})\} $   &     \cite{Cha62}     &      \\
\\[\dimexpr-\normalbaselineskip+3pt]
2   &   $ e + \mathrm{NO} \rightarrow \mathrm{NO^-}  $   &   $ f(\epsilon) $   &     \cite{Hayashi-LXCat}     &      \\
\\[\dimexpr-\normalbaselineskip+3pt]
%3   &   $ e + \mathrm{NO} + M \rightarrow \mathrm{NO^-} + M  $   &   $ (8. \times 10.^(-43)) $   &     \cite{CapBook}     &      \\
3   &   $ e + \mathrm{NO} + \mathrm{M} \rightarrow \mathrm{NO^-} + \mathrm{M}  $   &   $ 8 \times 10^{-43} $   &     \cite{CapBook}     &      \\
\\[\dimexpr-\normalbaselineskip+3pt]
4   &   $ e + \mathrm{NO} \rightarrow \mathrm{O^-} + \mathrm{N(^4S)}  $   &   $ f(\epsilon) $   &     \cite{Itikawa-LXCat}     &      \\
\hline
\end{longtable}}

\end{ThreePartTable}

\newpage
\begin{table}[hbt!]\scriptsize
\setlength{\tabcolsep}{14pt}
\centering
\caption{The oxygen vibrational kinetics in the $\mathrm{He/O_2}$ model. The letters $v$ and $w$ represent the vibrationally excited levels of oxygen molecule. The unit of the rate coefficients is m$^3$\:s$^{-1}$, and that of $\rm T_g$ is K. The rate coefficient $ f(\epsilon) $ is taken from a look-up-table calculated via the referred cross-section. Only the cross-sections of first six vibrational levels are used in the solution to the Boltzmann equation \cite{Tej19}, while those of the higher levels are directly evaluated to the rate coefficients according to the established EEDF. The reverse reaction rate coefficient of the electron-impact excitation labeled with a symbol ``*" near the number is calculated via the principle of {\em detailed balancing} \cite{LieBook2005}. The reverse reaction rate coefficients of the V-T mechanism labeled with a symbol "+" near the number are calculated by {\em detailed balance} \cite{Gue19_rev}.
}
\begin{tabular}{lllll}
\\[\dimexpr-\normalbaselineskip+3pt]
\hline
\\[\dimexpr-\normalbaselineskip+3pt]
\# &Reaction     & Rate Coefficient & Ref & \\ \hline
%\\[\dimexpr-\normalbaselineskip+3pt]
%\# & E-V Kinetics     &        &     & \\ 
\\[\dimexpr-\normalbaselineskip+3pt]
1$^*$  & $ e + {\rm O}_2(v\geqslant0) \rightarrow e + {\rm O}_2(v<w) $ & $ f(\epsilon) $  &  \cite{IST-Lisbon-LXCat,Lap13} &  \\ 
\\[\dimexpr-\normalbaselineskip+3pt]
2  & $ e + {\rm O}_2(v>0) \rightarrow e + {\rm O(^3P)} + {\rm O(^3P)}   $ & $ f(\epsilon) $ &  \cite{Phys4Entry,Lap15} &  \\ 
\\[\dimexpr-\normalbaselineskip+3pt]
3  & $ e + {\rm O}_2(v>0) \rightarrow {\rm O(^3P)} + {\rm O}^-  $ & $ f(\epsilon) $ &  \cite{Phys4Entry,Lap15} &  \\ 
\\[\dimexpr-\normalbaselineskip+3pt]
4  & $ e + {\rm O}_2(v=1-32) \rightarrow e + e + {\rm O_2}^+  $ & $ f(\epsilon) $ &  \cite{Phys4Entry} &  \\ 
%\\[\dimexpr-\normalbaselineskip+3pt]
%\# & V-T Kinetics     &        &     & \\ 
\\[\dimexpr-\normalbaselineskip+3pt]
%5*  & $ {\rm O}_2(v\geq1) + {\rm O} \rightarrow {\rm O}_2(v-1) + {\rm O}    $ & paper differ with code &  \cite{Cac78} &   \\  % Cac78 is old reaction rate coefficient in Efe's 2016 paper, that's why differ.
5$^+$  & $ {\rm O}_2(v>0) + {\rm O(^3P)} \rightarrow {\rm O}_2(v-1) + {\rm O(^3P)}    $ &  &  \cite{Esp08}$ ^{a} $ &   \\ 
\\[\dimexpr-\normalbaselineskip+3pt]
6$^+$  & $ {\rm O}_2(v>0) + {\rm O}_2 \rightarrow {\rm O}_2(v-1) + {\rm O}_2    $ &  &  \cite{Cap78}$ ^{b} $ &  \\ % figure 1 in "Cap78" (data plot digitizer)
\\[\dimexpr-\normalbaselineskip+3pt]
7$^+$  & $ {\rm O}_2(v>0) + {\rm He} \rightarrow {\rm O}_2(v-1) + {\rm He}    $ &  &  \cite{CapBook,Koz14}$ ^{c} $ &  \\ 
%\\[\dimexpr-\normalbaselineskip+3pt]
%\# & V-V Kinetics     &        &     & \\ 
\\[\dimexpr-\normalbaselineskip+3pt]
8  & $ {\rm O}_2(v \geqslant 0) + {\rm O}_2(w=1) \rightarrow {\rm O}_2(v+1) + {\rm O}_2(w=0)    $ &   &  \cite{Cap78}$ ^{b} $ &  \\ 
\\[\dimexpr-\normalbaselineskip+3pt]
9  & $ {\rm O}_2(v \geqslant 0) + {\rm O}_2(w>1) \rightarrow {\rm O}_2(v+1) + {\rm O}_2(w-1)    $ & $(v+1)w 5.68 \times 10^{-23} T_{\mathrm{g}}^{3/2} $  &  \cite{Kem16,Kie72} &  \\ % the original reference "Kie72" can not found, so also refer to "Kem16" (However, similar to k10 of Capbook: 9e-23Tg^1.5)
%\\[\dimexpr-\normalbaselineskip+3pt]
%\# & Other Kinetics     &        &     & \\ 
\\[\dimexpr-\normalbaselineskip+3pt]
10  & $ {\rm He^+} + {\rm O}_2(v>0) \rightarrow {\rm He} + {\rm O}_2^+    $ & $ 3.3 \times 10^{-17}(300/T_{\mathrm{g}})^{-1/2} $ &  \cite{Sta04} & \\ 
\\[\dimexpr-\normalbaselineskip+3pt]
11  & $ {\rm He^+} + {\rm O}_2(v>0) \rightarrow {\rm He} + {\rm O(^3P)} + {\rm O}^+    $ & $ 1.07 \times 10^{-15} (300/T_{\mathrm{g}})^{-1/2} $ &  \cite{Sta04} & \\ 
\\[\dimexpr-\normalbaselineskip+3pt]
12  & $ {\rm He(2 ^3S)} + {\rm O}_2(v>0) \rightarrow e + {\rm He} + {\rm O}_2^+    $ & $ 2.54 \times 10^{-16} (300/T_{\mathrm{g}})^{-1/2} $ &  \cite{Sta04} & \\ 
\\[\dimexpr-\normalbaselineskip+3pt]
13  & $ {\rm O(^3P)} + {\rm O}_2(v>0) \rightarrow 3 {\rm O(^3P)}     $ & $ 1 \times 10^{-6} \times 10^{\sum_j a_j v^j} $ &  \cite{Esp02} &   \\  % rate is when Tg = 500K, not 300K (and there is no "Tg" in rate like that in "Kem16": $\sum_j a_j(T_{\mathrm{g}}) v^j$) 
\\[\dimexpr-\normalbaselineskip+3pt]
\hline 
\end{tabular}

\begin{tablenotes}
\tiny
\item[1]$ ^{a} $ The rate coefficients are calculated from the formula in \cite{Esp08}.
\item[2]$ ^{b} $ The rate coefficients are adopted from the figures in \cite{Cap78}.
\item[3]$ ^{c} $ The rate coefficients are calculated by a product between the rate coefficient of $v=1$ in \cite{CapBook} and the scaling law in \cite{Koz14}. 
\end{tablenotes}

\label{tab:ReactionListHeO2Vib}
\end{table}

\begin{table}[H]\scriptsize
\setlength{\tabcolsep}{14pt}
\centering
\caption{The stepwise dissociation mechanism in the $ \mathrm{He/O_2} $ model by a vibrational quantum at the pseudo level $v'$=41. The letters $v$ and $w$ represent the vibrational quantum numbers.
}
\begin{tabular}{lllll}
\hline
\\[\dimexpr-\normalbaselineskip+3pt]
SD-1  & $ e + {\rm O}_2(v\geqslant0) \rightarrow {\rm O}_2(v') + e \rightarrow 2 {\rm O(^3P)} + e         $  &  \cite{Lap13,IST-Lisbon-LXCat} &  \\ 

\\[\dimexpr-\normalbaselineskip+3pt]
SD-2  & $ {\rm O(^3P)} + {\rm O}_2(v'-1) \rightarrow {\rm O(^3P)} + {\rm O}_2(v') \rightarrow 3 {\rm O(^3P)}    $  &  \cite{Kem16} &   \\    % the rate in code not from Esp08, but from the what Kem16 refered to.
\\[\dimexpr-\normalbaselineskip+3pt]
SD-3  & $ {\rm O}_2 + {\rm O}_2(v'-1) \rightarrow {\rm O}_2 + {\rm O}_2(v') \rightarrow {\rm O}_2  + 2 {\rm O(^3P)}    $   &  \cite{Cap78} &  \\ 
\\[\dimexpr-\normalbaselineskip+3pt]
SD-4  & $ {\rm He} + {\rm O}_2(v'-1) \rightarrow {\rm He} + {\rm O}_2(v') \rightarrow {\rm He}  + 2 {\rm O(^3P)}    $   &  \cite{CapBook,Koz14} &  \\ 
\\[\dimexpr-\normalbaselineskip+3pt]
SD-5  & $ {\rm O}_2(v'-1) + {\rm O}_2(w>0) \rightarrow {\rm O}_2(v') + {\rm O}_2(w-1) \rightarrow  2 {\rm O(^3P)} + {\rm O}_2(w-1)    $  &  \cite{Kem16,Kie72} &  \\ 
\\[\dimexpr-\normalbaselineskip+3pt]
\hline 
\end{tabular}

\label{tab:ReactionListHeO2vprime}
\end{table}

\newpage
\begin{table}[hbt!]\scriptsize
\setlength{\tabcolsep}{14pt}
\centering
\caption{The nitrogen vibrational kinetics in the $\mathrm{He/N_2}$ model. The letters $v$ and $w$ represent the vibrationally excited levels of nitrogen molecule. The unit of the rate coefficients is m$^3$\:s$^{-1}$, and that of $\rm T_g$ is K. The rate coefficient $ f(\epsilon) $ is taken from a look-up-table calculated via the referred cross-section. Only the cross-sections of first six vibrational levels are used in the solution to the Boltzmann equation \cite{Tej19}, while those of the higher levels are directly evaluated to the rate coefficients according to the established EEDF. The reverse reaction rate coefficient of the electron-impact excitation labeled with a symbol ``*" near the number is calculated via the principle of {\em detailed balancing} \cite{LieBook2005}. The reverse reaction rate coefficients of the V-T mechanism labeled with a symbol "+" near the number are calculated by {\em detailed balance} \cite{Gue19_rev}.
}
\begin{tabular}{p{0.3cm}p{6.0cm}p{6.0cm}p{1.1cm}p{0.2cm}}
%\begin{tabular}{lllll}
\\[\dimexpr-\normalbaselineskip+3pt]
\hline
\\[\dimexpr-\normalbaselineskip+3pt]
\# &Reaction     & Rate Coefficient & Ref & \\ \hline
\\[\dimexpr-\normalbaselineskip+3pt]
%\# & E-V Kinetics     &        &     & \\ 
%\\[\dimexpr-\normalbaselineskip+3pt]
1$^*$  & $ e + {\rm N}_2(v\geqslant0) \rightarrow e + {\rm N}_2(v<w)         $ & $ f(\epsilon) $ &  \cite{Phys4Entry,Lap14} &  \\ 
\\[\dimexpr-\normalbaselineskip+3pt]
2  & $ e + {\rm N}_2(v>0) \rightarrow e + {\rm N(^4S)} + {\rm N(^4S)}         $ & $ f(\epsilon) $ &  \cite{Phys4Entry,Lap14} &  \\ \\[\dimexpr-\normalbaselineskip+3pt]
3  & $ e + {\rm N}_2(v=1-35) \rightarrow e + {\rm N(^4S)} + {\rm N(^2 D)}         $ & $ f(\epsilon) $ &  \cite{Phys4Entry} &  \\ 
\\[\dimexpr-\normalbaselineskip+3pt]
4  & $ e + {\rm N}_2(v=1-40) \rightarrow e + e + {\rm N_2^+ }         $ & $ f(\epsilon) $ &  \cite{Phys4Entry} &  \\ 
%\\[\dimexpr-\normalbaselineskip+3pt]
%\# & V-T Kinetics     &        &     & \\ 
\\[\dimexpr-\normalbaselineskip+3pt]
5$^+$  & $ {\rm N}_2(v>0) + {\rm N(^4S)} \rightarrow {\rm N}_2(v-1) + {\rm N(^4S)}    $ &  $ v 4.0 \times 10^{-16} (T_{\mathrm{g}}/300)^{1/2} \exp(-7062.76/T_{\mathrm{g}})  $  &  \cite{CapBook}$ ^{a} $ &  \\ % $ v 4.0 \times 10^{-16} (T_{\mathrm{g}}/300)^{1/2} {\rm exp}(-7280/T_{\mathrm{g}} + 0.065 \times 0.288 \times 1.16e4 / T_{\mathrm{g}} )  $
\\[\dimexpr-\normalbaselineskip+3pt]
6$^+$  & $ {\rm N}_2(v>0) + {\rm N}_2 \rightarrow {\rm N}_2(v-1) + {\rm N}_2    $ & $ v 7.8 \times 10^{-18} T_{\mathrm{g}}  {\rm exp}(-218 / T_{\mathrm{g}}^{1/3} + 690 / T_{\mathrm{g}} ) \times {[1 - {\rm exp}( -3340.8 / T_{\mathrm{g}})]}^{-1}  $ &  \cite{CapBook}$ ^{a} $  & \\ %  % $ v 7.8 \times 10^{-18} T_{\mathrm{g}} \exp(-218 / T_{\mathrm{g}}^{1/3} + 690 / T_{\mathrm{g}} ) \times {(1 - e^{-0.288 \times 1.16e4 / T_{\mathrm{g}}})}^{-1}  $
\\[\dimexpr-\normalbaselineskip+3pt]
7$^+$  & $ {\rm N}_2(v>0) + {\rm He} \rightarrow {\rm N}_2(v-1) + {\rm He}    $ & $ v 3.0 \times 10^{-14} T_{\mathrm{g}}^{1/3} {\rm exp}(-196 / T_{\mathrm{g}}^{1/3} + 1680 / T_{\mathrm{g}} ) $ &  \cite{CapBook}$ ^{a} $ & \\ 
\\[\dimexpr-\normalbaselineskip+3pt]
%\# & V-V Kinetics     &        &     & \\ 
%\\[\dimexpr-\normalbaselineskip+3pt]
8  & $ {\rm N}_2(v \geqslant 0) + {\rm N}_2(w>0) \rightarrow {\rm N}_2(v+1) + {\rm N}_2(w-1)    $ & $(v+1)w 6.35 \times 10^{-23} T_{\mathrm{g}}^{3/2} $  &  \cite{Gue19_rev}$ ^{b} $ &  \\ 
\\[\dimexpr-\normalbaselineskip+3pt]
%\# & Other Kinetics     &        &     & \\ 
%\\[\dimexpr-\normalbaselineskip+3pt]
9  & $ {\rm He^+} + {\rm N}_2(v>0) \rightarrow {\rm He} + {\rm N}_2^+    $ & $ 5.0 \times 10^{-16} $ &  \cite{Som92_2} & \\ 
\\[\dimexpr-\normalbaselineskip+3pt]
10  & $ {\rm He^+} + {\rm N}_2(v>0) \rightarrow {\rm He} + {\rm N(^4S)} + {\rm N}^+    $ & $ 7.0 \times 10^{-16} $ &  \cite{Som92_2} & \\ 
\\[\dimexpr-\normalbaselineskip+3pt]
11  & $ {\rm He(2 ^3S)} + {\rm N}_2(v>0) \rightarrow e + {\rm He} + {\rm N}_2^+    $ & $ 7.0 \times 10^{-17} $ &  \cite{Som92_2} & \\ 
\\[\dimexpr-\normalbaselineskip+3pt]
\hline 
\end{tabular}
%\end{longtable}

\begin{tablenotes}
\tiny
\item[1]$ ^{a} $ The rate coefficients are calculated by a product between the rate coefficient of $v=1$ in \cite{CapBook} and the scaling law in \cite{Gue19_rev}.
\item[1]$ ^{b} $ The rate coefficients are calculated by a product between the rate coefficient of $v=0,w=1$ in \cite{Gue19_rev} and the scaling law in \cite{Gue19_rev}.
\end{tablenotes}

\label{tab:ReactionListHeN2Vib}
\end{table}

\begin{table}[H]\scriptsize
\setlength{\tabcolsep}{14pt}
\centering
\caption{The stepwise dissociation mechanism in the $ \mathrm{He/N_2} $ model by a vibrational quantum at the pseudo level $v'$=58. The letters $v$ and $w$ represent the vibrational quantum numbers.
}
\begin{tabular}{llccl}
\hline
\\[\dimexpr-\normalbaselineskip+3pt]
SD-1  & $ e + {\rm N}_2(v\geqslant0) \rightarrow {\rm N}_2(v') + e \rightarrow 2 {\rm N(^4S)} + e         $  &  \cite{Phys4Entry} &  \\ 

\\[\dimexpr-\normalbaselineskip+3pt]
SD-2  & $ {\rm N(^4S)} + {\rm N}_2(v'-1) \rightarrow {\rm N(^4S)} + {\rm N}_2(v') \rightarrow 3 {\rm N(^4S)}    $  &  \cite{CapBook} &   \\ 
\\[\dimexpr-\normalbaselineskip+3pt]
SD-3  & $ {\rm N}_2 + {\rm N}_2(v'-1) \rightarrow {\rm N}_2 + {\rm N}_2(v') \rightarrow {\rm N}_2  + 2 {\rm N(^4S)}    $   &  \cite{CapBook} &  \\ 
\\[\dimexpr-\normalbaselineskip+3pt]
SD-4  & $ {\rm He} + {\rm N}_2(v'-1) \rightarrow {\rm He} + {\rm N}_2(v') \rightarrow {\rm He}  + 2 {\rm N(^4S)}    $   &  \cite{CapBook} &  \\ 

\\[\dimexpr-\normalbaselineskip+3pt]
SD-5  & $ {\rm N}_2(v'-1) + {\rm N}_2(w>0) \rightarrow {\rm N}_2(v') + {\rm N}_2(w-1) \rightarrow  2 {\rm N(^4S)} + {\rm N}_2(w-1)    $  &  \cite{Gue19_rev} &  \\ 
\\[\dimexpr-\normalbaselineskip+3pt]
\hline 
\end{tabular}

\label{tab:ReactionListHeN2vprime}
\end{table}

\begin{table}[hbt!]\scriptsize
\setlength{\tabcolsep}{14pt}
\centering
\caption{The vibrational kinetics between oxygen and nitrogen in the $ \mathrm{He/N_2/O_2} $ model. The letters $v$ and $w$ represent the vibrational quantum numbers. The unit of the rate coefficients is m$^3$\:s$^{-1}$, and that of $\rm T_g$ is K. 
}
\begin{tabular}{p{0.3cm}p{6.0cm}p{6.0cm}p{1.1cm}p{0.2cm}}
\\[\dimexpr-\normalbaselineskip+3pt]
\hline
\\[\dimexpr-\normalbaselineskip+3pt]
\# &Reaction     & Rate Coefficient & Ref & \\ \hline
\\[\dimexpr-\normalbaselineskip+3pt]
1   & $ {\rm N}_2(v > 0) + {\rm O(^3P)} \rightarrow {\rm N}_2(v-1) + {\rm O(^3P)} $ & $ v [2.3 \times 10^{-19} {\rm \exp}(-1280 / T_{\mathrm{g}}) + 2.7 \times 10^{-17} {\rm \exp}(-10840 / T_{\mathrm{g}}) ] $ &  \cite{CapBook,Gue19_rev}$^a$ &   \\ % $ v (2.3 \times 10^{-19} e^{-1280 / T_{\mathrm{g}}} + 2.7 \times 10^{-17} e^{-10840 / T_{\mathrm{g}}} ) $
\\[\dimexpr-\normalbaselineskip+3pt]
2   & $ {\rm N}_2(v > 0) + {\rm O}_2 \rightarrow {\rm N}_2(v-1) + {\rm O}_2 $ &  &  \cite{Gue19_rev}$^b$ &  \\ 
\\[\dimexpr-\normalbaselineskip+3pt]
3   & $ {\rm O}_2(v > 0) + {\rm N}_2 \rightarrow {\rm O}_2(v-1) + {\rm N}_2 $ &  &  \cite{Gue19_rev}$^b$ &  \\
\\[\dimexpr-\normalbaselineskip+3pt]
4   & $ {\rm O}_2(v \geqslant 0) + {\rm N}_2(w>1) \rightarrow {\rm O}_2(v+1) + {\rm N}_2(w-1)    $ &  $ (v+1)w3.69 \times 10^{-18}(T_{\mathrm{g}}/300) {\rm \exp}(-104 / T_{\mathrm{g}}^{1/3}) $  &  \cite{CapBook,Gue19_rev}$^c$ &  \\ % $ (v+1)w3.69 \times 10^{-18}(T_{\mathrm{g}}/300) e^{-104 / T_{\mathrm{g}}^{1/3}} $
\\[\dimexpr-\normalbaselineskip+3pt]
5   & $ {\rm N}_2(v \geqslant 13) + {\rm O(^3P)} \rightarrow {\rm NO} + {\rm N(^4S)}  $ & $ 1 \times 10^{-19} $ &  \cite{Gue95,Gue19_rev} &  \\ 
\\[\dimexpr-\normalbaselineskip+3pt]
6   & $ {\rm N(^4S)} + {\rm NO} \rightarrow {\rm N}_2(v=3) + {\rm O(^3P)} $ & $ 1.05 \times 10^{-18} \: T_{\mathrm{g}}^{1/2} $ &  \cite{Gue95,Gue19_rev} &  \\ 
\\[\dimexpr-\normalbaselineskip+3pt]
\hline 
\end{tabular}

\begin{tablenotes}
\tiny
\item[1]$^a$ The rate coefficients are calculated by a product between the rate coefficient of $v=1$ in \cite{CapBook} and the scaling law in \cite{Gue19_rev}.
\item[2]$^b$ The rate coefficients of N$_2(v>0)$-O$_2$ and O$_2(v>0)$-N$_2$ V-T reactions are estimated by those of N$_2(v>0)$-N$_2$ and O$_2(v>0)$-O$_2$, respectively. The estimation is based on the equation (27) in \cite{Gue19_rev}.
\item[3]$^c$ The rate coefficients are calculated by a product between the rate coefficient of $v=0,w=1$ in \cite{CapBook} and the scaling law in \cite{Gue19_rev}.
\end{tablenotes}

\label{tab:ReactionListN2O2Vib}
\end{table}

\begin{table}[H]\scriptsize
\setlength{\tabcolsep}{14pt}
\centering
\caption{The neutral wall reactions in the model of $\mathrm{He}$, $\mathrm{He/O_2}$, $\mathrm{He/N_2}$ or $\mathrm{He/N_2/O_2}$ are selected from the table below. 
}
\begin{tabular}{lllll}
\\[\dimexpr-\normalbaselineskip+3pt]
\hline
\\[\dimexpr-\normalbaselineskip+3pt]
\# &Reaction     &  Probability($\gamma$) &  Ref &  \\ \hline
\\[\dimexpr-\normalbaselineskip+3pt]
1  & $ \mathrm{He(2 ^3S)} + \mathrm{wall} \rightarrow \mathrm{He}  $ & $1$ &    \cite{Liu10,Yan16} &    \\ 
\\[\dimexpr-\normalbaselineskip+3pt]
2  & $ \mathrm{He_2^*} + \mathrm{wall} \rightarrow 2 \mathrm{He}  $ & $1$ &    \cite{Liu10,Yan16} &    \\ 
\\[\dimexpr-\normalbaselineskip+3pt]
3  & $ \mathrm{N(^2D)} + \mathrm{wall} \rightarrow \mathrm{N(^4S)}  $ & $0.93$  &    \cite{Tho09}   & \\ 
\\[\dimexpr-\normalbaselineskip+3pt]
4  & $ \mathrm{N_2(A^3 \Sigma)} + \mathrm{wall} \rightarrow \mathrm{N}_2  $ & $1$ &  \cite{Tho09}   &   \\ 
\\[\dimexpr-\normalbaselineskip+3pt]
5  & $ \mathrm{N_2(B^3 \Pi)} + \mathrm{wall} \rightarrow \mathrm{N}_2  $ & $1$ &  \cite{Tho09}   &   \\ 
\\[\dimexpr-\normalbaselineskip+3pt]
6  & $ \mathrm{N(^4S,\:^2D)} + \mathrm{wall} \rightarrow 1/2\mathrm{N}_2  $ & $10^{-6}$  &    \cite{Tal17}  &  \\ 
\\[\dimexpr-\normalbaselineskip+3pt]
7  & $ \mathrm{O(^1D)} + \mathrm{wall} \rightarrow \mathrm{O(^3P)}  $ & $0.1$  &    \cite{Tho2010}   & \\ 
\\[\dimexpr-\normalbaselineskip+3pt]
8  & $ \mathrm{O_2(a^1 \Delta_g)} + \mathrm{wall} \rightarrow \mathrm{O}_2  $ & $0.007$ &   \cite{Sha89}  &   \\ 
\\[\dimexpr-\normalbaselineskip+3pt]
9  & $ \mathrm{O(^3P,\:^1D)} + \mathrm{wall} \rightarrow 1/2\mathrm{O}_2  $ & $0$  &   $^a$   &  \\ 
\\[\dimexpr-\normalbaselineskip+3pt]
10  & $ \mathrm{O(^3P)} + \mathrm{wall} \rightarrow 1/3\mathrm{O}_3  $ & $0$ &   $^a$   &  \\

\\[\dimexpr-\normalbaselineskip+3pt]

11  & $ \mathrm{N}_2(v) + \mathrm{wall} \rightarrow \mathrm{N}_2(v-1)  $ & $1$  &   \cite{Tho09}   &  \\ 
\\[\dimexpr-\normalbaselineskip+3pt]
12  & $ \mathrm{O}_2(v) + \mathrm{wall} \rightarrow \mathrm{O}_2(v-1)  $ & $1$  &    \cite{Kem16}  &  \\ 
\\[\dimexpr-\normalbaselineskip+3pt]
13  & $ \mathrm{N(^4S)} + \mathrm{wall} \rightarrow \mathrm{NO}  $ & 0 %1
&   $^b$   &  \\

\\[\dimexpr-\normalbaselineskip+3pt]
\hline
\end{tabular}

\begin{tablenotes}
\tiny
\item[1]$^a$ Both reactions are only included in section \ref{sec:Sensitivity}, except reaction 9 is additionally considered in figure \ref{fig:was10}$(b)$ for a test of sensitivity.  
\item[2]$ ^b $ The reaction is only considered in section \ref{sec:NOproduction} for a test of sensitivity.
\end{tablenotes}

\label{tab:wrHeN2O2}
\end{table}

\begin{table}[H]\scriptsize
\setlength{\tabcolsep}{14pt}
\centering
\caption{The ion wall reactions in the model of $\mathrm{He}$, $\mathrm{He/O_2}$, $\mathrm{He/N_2}$ or $\mathrm{He/N_2/O_2}$ are selected from the table below. 
}
\begin{tabular}{ll}
\\[\dimexpr-\normalbaselineskip+3pt]
\hline
\\[\dimexpr-\normalbaselineskip+3pt]
\# &Reaction       \\ \hline
\\[\dimexpr-\normalbaselineskip+3pt]
1  & $ \mathrm{He^+} + \mathrm{wall} \rightarrow \mathrm{He}  $    \\ 
\\[\dimexpr-\normalbaselineskip+3pt]
2  & $ \mathrm{He_2^+} + \mathrm{wall} \rightarrow 2 \mathrm{He}  $    \\ 
\\[\dimexpr-\normalbaselineskip+3pt]
3  & $ \mathrm{N^+} + \mathrm{wall} \rightarrow \mathrm{N(^4S)}  $  \\ 
\\[\dimexpr-\normalbaselineskip+3pt]
4  & $ \mathrm{N_2^+} + \mathrm{wall} \rightarrow \mathrm{N}_2  $    \\ 
\\[\dimexpr-\normalbaselineskip+3pt]
5  & $ \mathrm{N_3^+} + \mathrm{wall} \rightarrow \mathrm{N(^4S)} + \mathrm{N}_2  $    \\ 
\\[\dimexpr-\normalbaselineskip+3pt]
6  & $ \mathrm{N_4^+} + \mathrm{wall} \rightarrow 2 \mathrm{N_2}  $  \\ 
\\[\dimexpr-\normalbaselineskip+3pt]
7  & $ \mathrm{O^+} + \mathrm{wall} \rightarrow \mathrm{O(^3P)}  $    \\ 
\\[\dimexpr-\normalbaselineskip+3pt]
8  & $ \mathrm{O_2^+} + \mathrm{wall} \rightarrow \mathrm{O}_2  $   \\ 
\\[\dimexpr-\normalbaselineskip+3pt]
9  & $ \mathrm{O_4^+} + \mathrm{wall} \rightarrow 2 \mathrm{O}_2  $   \\ 
\\[\dimexpr-\normalbaselineskip+3pt]
10  & $ \mathrm{NO^+} + \mathrm{wall} \rightarrow \mathrm{NO}  $   \\ 
\\[\dimexpr-\normalbaselineskip+3pt]
11  & $ \mathrm{NO_2^+} + \mathrm{wall} \rightarrow \mathrm{NO}_2  $   \\ 
\\[\dimexpr-\normalbaselineskip+3pt]
12  & $ \mathrm{N_2O^+} + \mathrm{wall} \rightarrow \mathrm{N_2O}  $   \\ 
\\[\dimexpr-\normalbaselineskip+3pt]

\hline
\end{tabular}

\label{tab:iwrHeN2O2}
\end{table}

\begin{table}[hbt!]\scriptsize
\setlength{\tabcolsep}{25pt}
\centering
\caption{The elastic electron-impact collisions in the model of $\mathrm{He}$, $\mathrm{He/O_2}$, $\mathrm{He/N_2}$ or $\mathrm{He/N_2/O_2}$ are selected from the table below. 
}
\begin{tabular}{lllll}
\\[\dimexpr-\normalbaselineskip+3pt]
\hline
\\[\dimexpr-\normalbaselineskip+3pt]
\# &Collision     &  Ref  & \\ \hline
\\[\dimexpr-\normalbaselineskip+3pt]
1  & $ e + \mathrm{He} \rightarrow e + \mathrm{He} $ &   \cite{IST-Lisbon-LXCat} &      \\ 
\\[\dimexpr-\normalbaselineskip+3pt]
2  & $  e+ \mathrm{N(^4 S)}  \rightarrow e + \mathrm{N(^4 S)}  $ &   \cite{IST-Lisbon-LXCat}   &     \\ 
\\[\dimexpr-\normalbaselineskip+3pt]
3  & $  e+ \mathrm{N_2}  \rightarrow e + \mathrm{N_2}  $ &   \cite{Biagi-LXCat} &      \\
\\[\dimexpr-\normalbaselineskip+3pt]
4  & $  e+ \mathrm{O(^3 P)}  \rightarrow e + \mathrm{O(^3 P)}  $ &   \cite{IST-Lisbon-LXCat}   &     \\ 
\\[\dimexpr-\normalbaselineskip+3pt]
5  & $  e+ \mathrm{O_2}  \rightarrow e + \mathrm{O_2}  $ &   \cite{Biagi-LXCat} &      \\ 
\\[\dimexpr-\normalbaselineskip+3pt]
6  & $  e+ \mathrm{O_3}  \rightarrow e + \mathrm{O_3}  $ &   \cite{Morgan-LXCat} &      \\ 
\\[\dimexpr-\normalbaselineskip+3pt]
7  & $ e+ \mathrm{NO} \rightarrow e + \mathrm{NO}    $ &   \cite{Itikawa-LXCat} &      \\ 
\\[\dimexpr-\normalbaselineskip+3pt]
8  & $ e+ \mathrm{NO_2} \rightarrow e + \mathrm{NO_2}    $ &   \cite{Son19} &      \\ 
\\[\dimexpr-\normalbaselineskip+3pt]
9  & $ e+ \mathrm{N_2O} \rightarrow e + \mathrm{N_2O}    $ &   \cite{Hayashi-LXCat} &      \\ 
\\[\dimexpr-\normalbaselineskip+3pt]

\hline 
\end{tabular}

\label{tab:elaHeN2O2}
\end{table}

\newpage
\clearpage
\bibliographystyle{unsrt}
\bibliography{ref}

\end{document}